\documentclass[useAMS,usenatbib,psfig,eps,graphics]{mn2e}
\usepackage{natbib,psfig,graphicx, epsfig}

\def\grtsim{\mathrel{\hbox{\rlap{\hbox{\lower2pt\hbox{$\sim$}}}\raise2pt\hbox{$>$}}}} 
\def\lesssim{\mathrel{\hbox{\rlap{\hbox{\lower2pt\hbox{$\sim$}}}\raise2pt\hbox{$<$}}}}

\def\degree{\nobreak\ifmmode{^\circ}\else{$^\circ$}\fi}

\newcommand{\whzsr}{W~Hz$^{-1}$~sr$^{-1}$}

\newcommand{\kms}{km~s$^{-1}$}

\newcommand{\nh}{$N_{\rm H}$}

\newcommand{\lx}{$L_{\rm X}$}

\newcommand{\qjet}{$Q_{\rm jet}$}

\def\msol{M$_{\odot}$}
\def\msolyr{M$_{\odot}$~yr$^{-1}$}

\def\amean{$\langle \hat{a} (z) \rangle$}
\def\sp{$\hat{a}$}

\newcommand{\mbh}{$m_{\bullet}$}

\newcommand{\aap}{A\&A}
\newcommand{\aapr}{A\&AR}
\newcommand{\aaps}{A\&AS}

\newcommand{\aj}{AJ}
\newcommand{\apj}{ApJ}
\newcommand{\apjl}{ApJL}
\newcommand{\apjs}{ApJS}

\newcommand{\araa}{ARA\&A}
\newcommand{\mnras}{MNRAS}
\newcommand{\nat}{Nat}

\newcommand{\pasj}{PASJ}

\newcommand{\prd}{Phys. Rev. D}

 \newcommand{\myemail}{alejo.martinez-sansigre@port.ac.uk}

\begin{document}
\topmargin -0.5in %this is only for astro-ph, uncomment when submitting paper

\title[Observational constraints on the spin of SMBHs]{Observational constraints on the spin of the most massive black holes from radio observations}
\author[A. Mart\'\i nez-Sansigre \& S. Rawlings]{Alejo Mart\'\i nez-Sansigre$^{1,2,3}$\thanks{\myemail}, Steve Rawlings$^{2}$\\
$^{1}$Institute of Cosmology and Gravitation, University of Portsmouth, Dennis Sciama Building, Burnaby Road, Portsmouth, \\ PO1 3FX, United Kingdom \\
$^{2}$Astrophysics, Department of Physics, University of Oxford, Keble Road, Oxford OX1 3RH, United Kingdom \\
$^{3}$SEP{\it net}, South-East Physics network \\
 }

\date{}

\pagerange{\pageref{firstpage}--\pageref{lastpage}} \pubyear{}

\maketitle

\label{firstpage}
\vspace{-0.5 cm}

\begin{abstract}  
 We use recent progress in simulating the production of
 magnetohydrodynamic jets around black holes to derive the cosmic spin
 history of the most massive black holes. Our work focusses on black
 holes with masses $\grtsim10^{8}$~\msol. Under the assumption that
 the efficiency of jet production is a function of spin \sp,
 as given by the simulations, we can approximately reproduce the
 observed `radio loudness' of quasars and the local radio luminosity
 function.

 Using the X-ray luminosity function and the local mass function of supermassive black holes, SMBHs,  we can reproduce the individual radio luminosity functions of radio sources showing high- and low-excitation narrow emission lines. We find that the data favour spin distributions that are bimodal, with one component around spin zero and the other close to maximal spin. The `typical' spin is therefore really the expectation value, lying between the two peaks.  In the low-excitation galaxies, the two components have similar amplitudes, meaning approximately half of the sources have very high spins, and the other have very low spins. For the high-excitation galaxies, the amplitude of the high-spin peak is typically much smaller than that of the low-spin peak, so that most of the sources have low spins. However, a small population of near maximally spinning high-accretion rate objects is inferred.  A bimodality should be seen in the radio loudness of quasars, although there are  a variety of physical and selection effects that may obfuscate this feature.

We predict that the low-excitation galaxies are dominated by SMBHs with masses $\grtsim$$10^{8}$~\msol, down to radio luminosity densities  $\sim$$10^{21}$ \whzsr\, at 1.4~GHz. Under reasonable assumptions, our model is also able to predict the radio luminosity function at $z$$=$1, and predicts it to be dominated by radio sources with high-excitation narrow emission lines above luminosity densities $\grtsim$$10^{26}$ \whzsr\, at 1.4~GHz,  and this is in full agreement with the observations.

 From our parametrisation of the spin distributions of the high- and low-accretion rate SMBHs, we derive an estimate of the spin history of SMBHs, which shows a weak evolution between $z$$=$1 and 0.    A larger fraction of low-redshift SMBHs have high spins compared to high-redshift SMBHs. Using the best fitting jet efficiencies 
there is marginal evidence for evolution in spin: the mean spin increases slightly from $\langle \hat{a} \rangle$$\sim$0.25 at $z$$=$1 to $\langle \hat{a} \rangle$$\sim$0.35 at $z$$=$0, and 
  the fraction of SMBHs with \sp$\geq$0.5  increases from  0.16$\pm$0.03 at $z$$=$1 
 to     0.24$\pm$0.09 at $z$$=$0. 
  Our inferred spin history of  SMBHs  is in excellent agreement with constraints from the mean radiative efficiency of quasars, as well as the results from recent simulations of growing SMBHs. We discuss the implications in terms of accretion and SMBH mergers. We also discuss other work related to the spin of SMBHs as well as  work discussing the spin of galactic black holes and their jet powers.  
  \end{abstract} 
 
 
\section{Introduction} \label{sec:intro}

Astrophysical black holes are described by two parameters, mass and
spin, which define the structure of space-time around the black
hole. Distant observers can infer black hole properties from the
dynamics of material orbiting the black hole or more indirectly when
matter is being accreted, from the radiation and jets produced.

Active galactic nuclei (AGN) are believed to be powered by accretion
onto supermassive black holes \citep[SMBHs,
  e.g.][]{1969Natur.223..690L,1984ARA&A..22..471R}. Matter accreted
onto the SMBH radiates via thermal and non-thermal processes,
dominating the observed spectral energy distribution (SED) between
X-ray energies and far-infrared wavelengths \citep[see
  e.g.][]{1999agnc.book.....K}. Multiwavelength observations can
therefore be used to measure the bolometric luminosity of the AGN,
although in practice reasonable estimates of this quantity can be
obtained from observations in a single band
\citep[e.g.][]{1994ApJS...95....1E,2007ApJ...654..731H}.

Accretion onto SMBHs can also produce jets.  The bulk of the jet power
is used to fill and expand a cocoon of low-density relativistic
magnetised plasma that pushes back the intergalactic medium
\citep[IGM, see
  e.g.][]{1973MNRAS.164..243L,1974MNRAS.166..513S,1984RvMP...56..255B}. These
lobes are sources of synchrotron radiation emitted at radio
wavelengths.  However, the power is radiated at a much lower rate than
is supplied to the lobes via the jets, at least while the jets are
active, so that energy is stored in the lobes. Making assumptions
about the contents of the lobes (charged particles and magnetic
fields), and how the energy density is distributed between them, as
well as assumptions on the typical jet advance speeds and IGM density
profiles, the jet power can be estimated from the low-frequency
monochromatic radio luminosity
\citep[e.g.][]{1993MNRAS.263..425M,1999MNRAS.309.1017W}.

Observations of AGN require an extra `hidden variable' to explain the
variations in radio luminosity, reflecting variations in jet
power. For a given radiative power, AGN can show radio luminosities
that vary by more than two orders of magnitude
\citep[e.g.][]{1989AJ.....98.1195K,1993MNRAS.263..425M,2007ApJ...658..815S,2011ApJ...727...39M}. This
huge scatter in `radio-loudness' is not dependent on black hole mass
or accretion rate \citep[e.g.][]{2007ApJ...658..815S}, and the X-ray
to far-infrared SEDs of radio-loud and radio-quiet AGN do not show
significant differences \citep[e.g.][]{1994ApJS...95....1E}.

Conversely, observational studies of AGN selected at radio frequencies
have, for a long time, realised that the AGN showed two types of
optical spectra. Some show high-excitation narrow emission lines
(known as `high-excitation galaxies' or HEGs) indicative of high
accretion rates, while the others show only low-excitation narrow
lines (LEGs) suggesting low accretion rates
\citep[e.g.][]{1979MNRAS.188..111H,1983MNRAS.204..151L,1994ASPC...54..201L,1997MNRAS.286..241J}. This
confirms that SMBHs producing jets are observed to accrete both at
high and low rates.

Comparison of the jet powers and bolometric luminosities of AGN
suggests that the conversion of accreted energy into jets must occur
extremely efficiently in some sources, since the inferred jet power is
comparable or sometimes even greater than the inferred accretion power $\dot{m}c^{2}$
\citep{2007MNRAS.374L..10P,2010arXiv1012.1910P,2011ApJ...727...39M,2010arXiv1010.0691F}.  These results
still carry considerable uncertainty in their estimates of jet power
and accretion rate, however they do show that an extremely efficient
mechanism must be powering the jets.

In theory, spinning black holes
can reach nominal jet efficiencies greater than one, since as well as
drawing energy from the accreted mass, the jet draws energy from the
rotation of the black hole. Hence, spin is an attractive candidate to explain the radio loudness
of AGN \citep[e.g. ][]{1977MNRAS.179..433B,1984ARA&A..22..471R,1990ApJ...354..583P,1992ersf.meet..368S,1995ApJ...438...62W,2002Sci...295.1688K,2007ApJ...658..815S}. It is attractive because, in principle, it is capable of high efficiencies and in addition it provides a physical explanation for a hidden variable. This is usually referred to as the `spin paradigm'.

The jets are most likely
powered by differential rotation of poloidal magnetic fields in the accretion disc or near
the black hole. Theoretically, the spin of the black hole is therefore a promising candidate since frame dragging around spinning black holes would provide particularly strong magnetic fields and particularly fast rotation \citep[e.g. ][]{1977MNRAS.179..433B,1990ApJ...354..583P,2001ApJ...548L...9M,2002Sci...295.1688K}. 

However, whether the jets are really tapping the rotational energy of the black hole is still a matter of debate. Some authors argue that the power extracted in this way is at  most comparable to the power generated by the inner regions of the accretion disc \citep{1997MNRAS.292..887G,1999ApJ...512..100L}. Others argue that in the case of  geometrically thick accretion discs,  powerful 
jets can be be driven by the black hole spin \citep[e.g.][]{2001ApJ...548L...9M}.

In galactic black holes (GBHs), the presence of steady jets occurs typically during low accretion rates (where by low we mean an Eddington rate\footnote{The Eddington rate $\lambda$ is defined here as $\lambda \equiv {\dot{m} \over \dot{m_{\rm E}}}$, where $\dot{m}$ is the accretion rate and $\dot{m_{\rm E}}$ is the accretion rate corresponding to  the Eddington-limit. 
} $\lambda$$\lesssim$0.01), or during high accretion rates \citep[$\lambda$$\grtsim$0.3, e.g. GRS 1915+105, see e.g.][]{2004MNRAS.355.1105F}.  The jets can be steady, transient or absent at $\lambda$$\grtsim$0.01, depending on the accretion state.  This is consistent with  steady jets requiring geometrically-thick accretion flows, either ADAFs \citep[advection-dominated accretion flows][]{1995ApJ...452..710N} or slim discs \citep{1988ApJ...332..646A}.   However, these differences in the accretion flow are observable at X-ray energies, while as explained in the preceding paragraphs, the radio-loudness of AGN requires an extra variable parameter that is not readily observable. 

Finally, we note that comparison of individual spin measurements for GBHs find no correlation between radio loudness and individual spin estimates \citep{2010MNRAS.406.1425F}. We discuss these results in detail later in this paper, in Section~\ref{sec:lack}, and argue that they are not robust enough to  invalidate the spin paradigm.

In this work we begin by spelling out our assumptions in Section~\ref{sec:assum}, where we also introduce the different jet efficiencies, drawn from the literature, that we will use.  The most 
important assumption is that the
efficiency with which jets are produced is an increasing function of
spin.  In Section~\ref{sec:loud} we test the spin paradigm and  show that under our assumptions we can
explain the observed range in `radio-loudness' of quasars. 

In Section~\ref{sec:fit_rlf}, we model the  contribution to the local radio luminosity function (LF) of  HEGs and LEGs as being composed of high-accretion rate objects (quasars or QSOs) and low-accretion rate ones (ADAFs). We allow these populations to have several different spin distributions, parametrised by three different models. Fitting to the data,
and using a bayesian model selection scheme, in Section~\ref{sec:res} we derive the spin distribution of SMBHs of mass \mbh$\geq$$10^{8}$~\msol\, 
in the local Universe, and make a prediction for the $z$$=$1 radio LF. 
In Section~\ref{sec:spin_his}, we present an estimate of the cosmic spin history of the most
massive black holes from our modelling of the radio LF. Section~\ref{sec:other} compares and discusses our
inferrence about the spins of black holes to other results in the
literature, using different methods. It includes a discussion about the lack of correlation between spin measurements and radio loudness found in GBHs.  Finally, in Section~\ref{sec:disc} we
discuss the implications of our result in terms of SMBH growth through
accretion and mergers.

We adopt a $\Lambda$CDM cosmology with the following parameters: $h =
H_{0} / (100 ~ \rm km ~ s^{-1} ~ Mpc^{-1}) = 0.7$; $\Omega_{\rm m} =
0.3$; $\Omega_{\Lambda} = 0.7$.

\section{Assumptions}\label{sec:assum}

\subsection{Radiative efficiency}\label{sec:rad_eff}

We assume that the bolometric luminosity (due to radiation), $L_{\rm
  bol}$ is given by:

\begin{equation}
L_{\rm bol} = \epsilon  \dot{m} c^{2} 
\label{eq:lbol}
\end{equation}

\noindent where $\dot{m}$ is the accretion rate and $c$ the speed of
light. The term $\epsilon$ is the radiative efficiency.  In the
classical disc picture of \citet[][ hereafter
  NT73]{1973blho.conf..343N}, $\epsilon$ is given by (one minus) the
binding energy per unit mass of the ISCO, which is itself a function
of the black hole spin.

Some simulations have shown that the radiative efficiency estimated by
NT73 to slightly underestimate the radiative efficiency for low spins
\citep{2008MNRAS.390...21B,2009ApJ...692..411N}.  On the other hand,
other simulations suggest that the radiative efficiency is very close
to that predicted by the NT73 model
\citep{2004ApJ...611..977M,2008ApJ...687L..25S,2010MNRAS.408..752P}.

\begin{figure} 
\begin{center}
\psfig{file=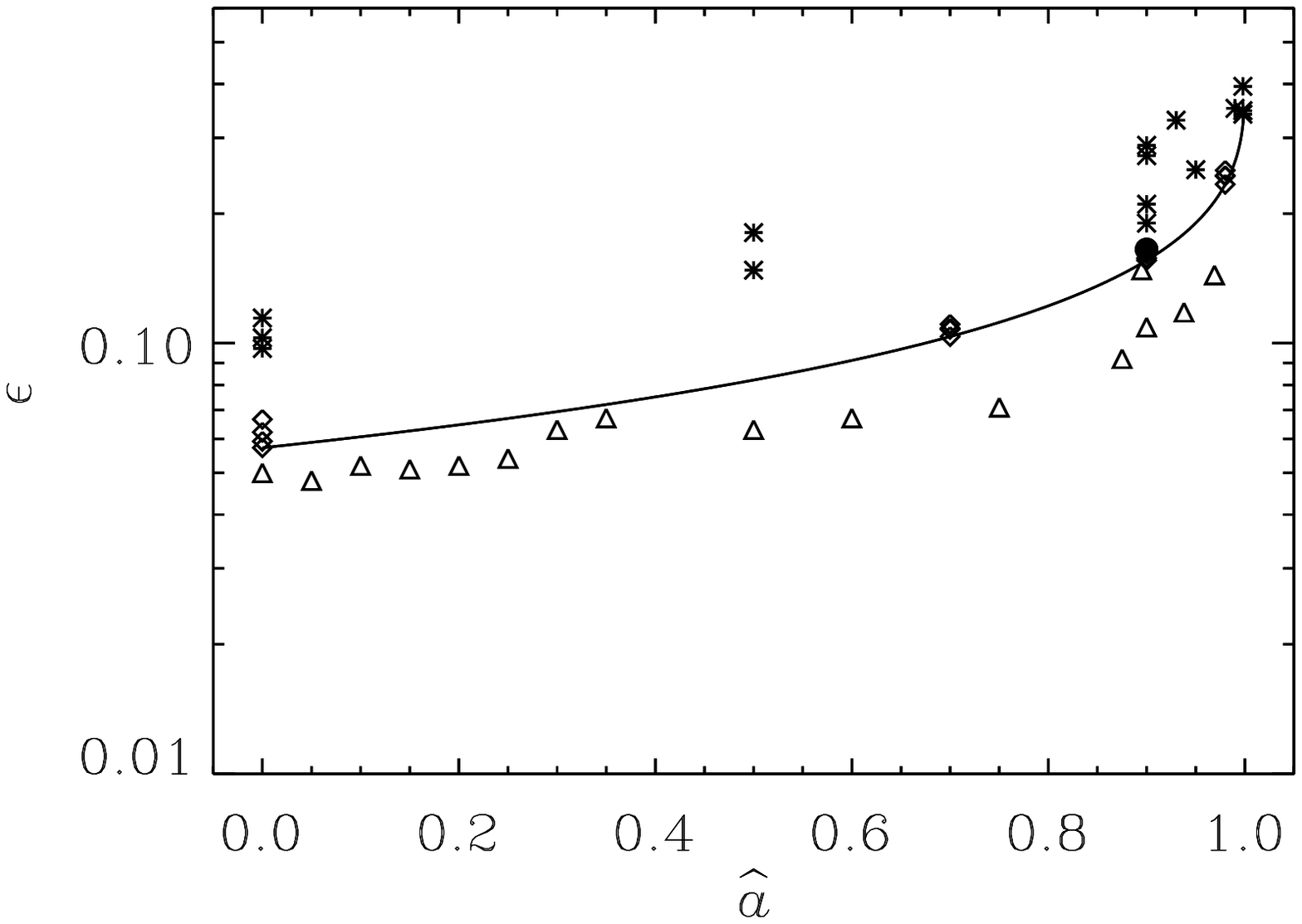, width=8cm, angle=0}
\caption{\noindent The radiative efficiency as a function of spin. The
  figure compares the analytical model of NT73 (solid line) to the
  results from different GRMHD simulations. The simulation data are
  from \citet [empty triangles]{2004ApJ...611..977M},
  \citet[stars]{2008MNRAS.390...21B}, \citet[black
    circle at \sp$=$0.9]{2009ApJ...692..411N} and \citet[empty
    diamonds]{2010MNRAS.408..752P}.  }
\label{fig:eps}
\end{center}
\end{figure}

Figure~\ref{fig:eps} compares the radiative efficiency as a function
of spin from the NT73 model to the results of simulations by different
groups.  In some cases, the results from the simulations have higher
efficiencies than the NT73 model, in others they have lower
efficiencies. Overall, and given the range of results from the
simulations, the NT73 model is not obviously wrong.

For our work we require analytic expressions for $\epsilon(\hat{a})$
and the jet efficiency $\eta(\hat{a})$, and we will use the NT73
approximation for $\epsilon(\hat{a})$. Figure~\ref{fig:eps} suggests this is an acceptable
approximation.  This paper is concerned with variations in jet powers
that span orders of magnitudes, and the $\sim$2 uncertainty in
$\epsilon$ has only a small effect on our results.  Indeed the uncertainties
in the jet efficiencies discussed in Section~\ref{sec:jet_eff} are far larger
than this.

We are also neglecting the fraction of energy advected into the hole
(in excess of the energy of the ISCO). This is a reasonable assumption
for SMBHs with  Eddington rates in the range  $10^{-2}$$\lesssim \lambda\lesssim$1. Accretion rates with higher or lower Eddington ratios will tend to have a large fraction of the energy advected into the hole
\citep{1995ApJ...452..710N,1998MNRAS.297..739B}. Throughout this paper we will
only require to use the radiative efficiency for quasars, not for
ADAFs. Quasars have typically 0.1$\lesssim$$\lambda$$\lesssim$1.0 \citep{2004MNRAS.352.1390M}, and hence neglecting the advected energy is a reasonable approximation.

\begin{figure*}
\hbox{ \psfig{file=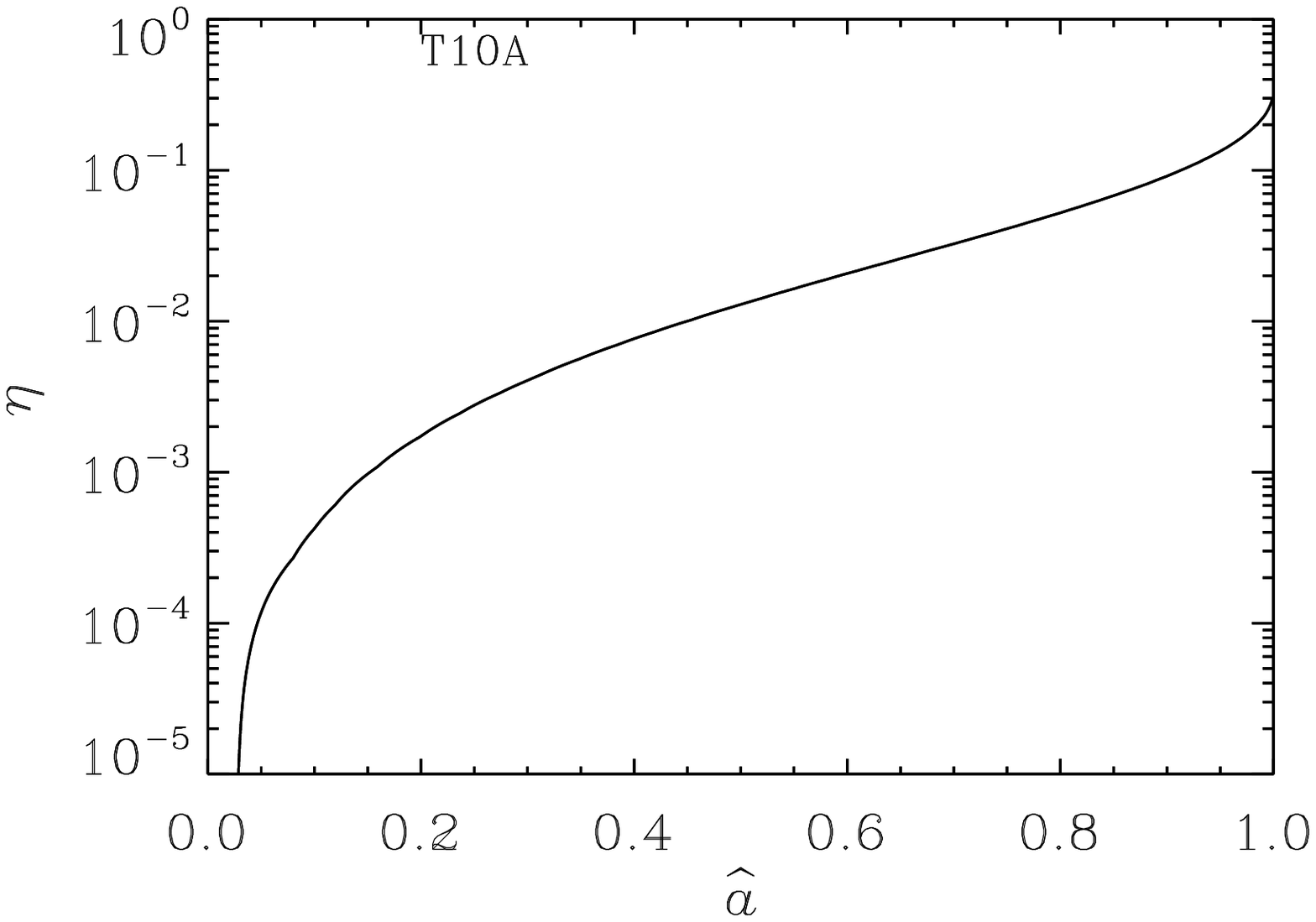, width=8cm, angle=0} \psfig{file=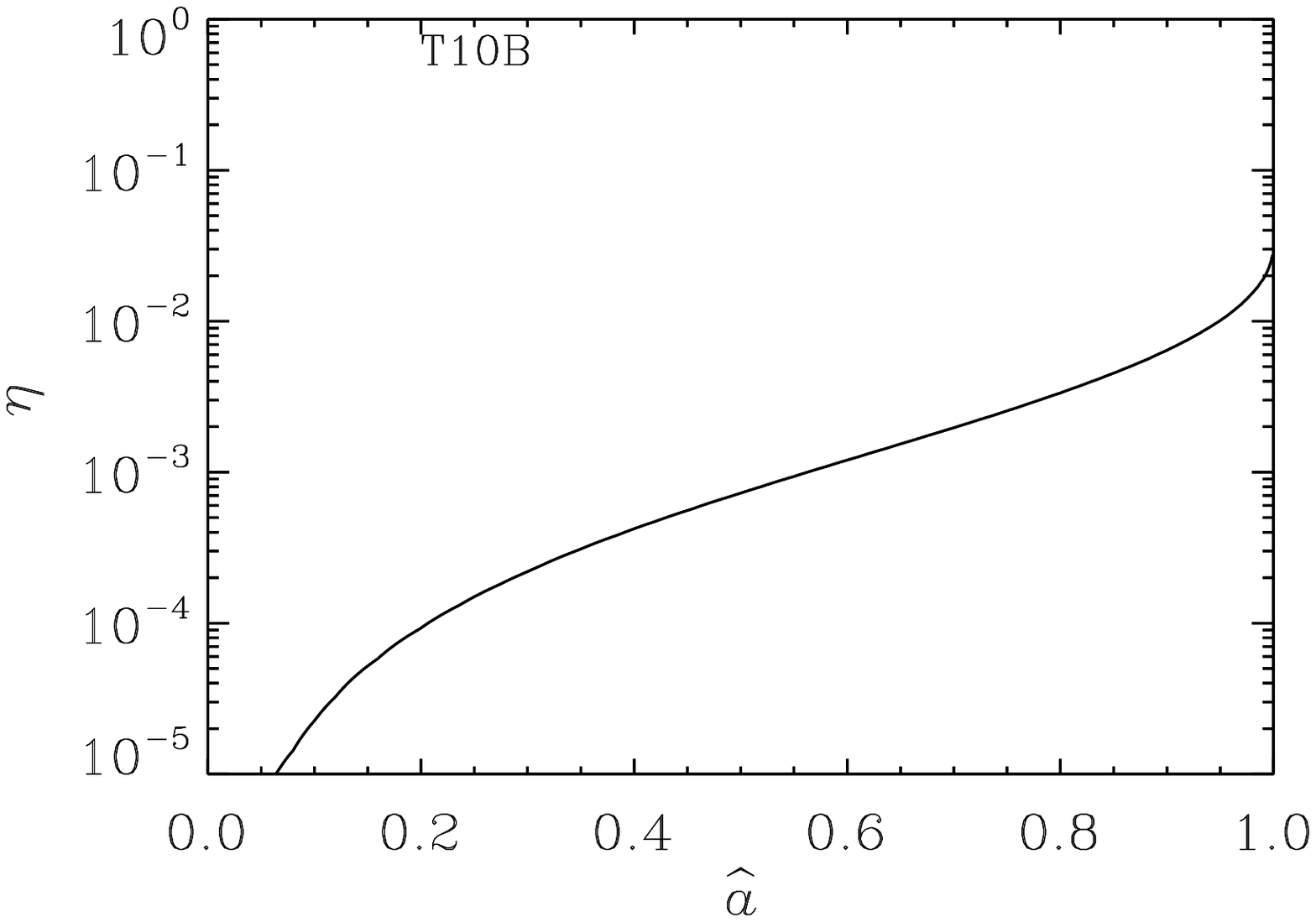, width=8cm, angle=0} }
\hbox{ \psfig{file=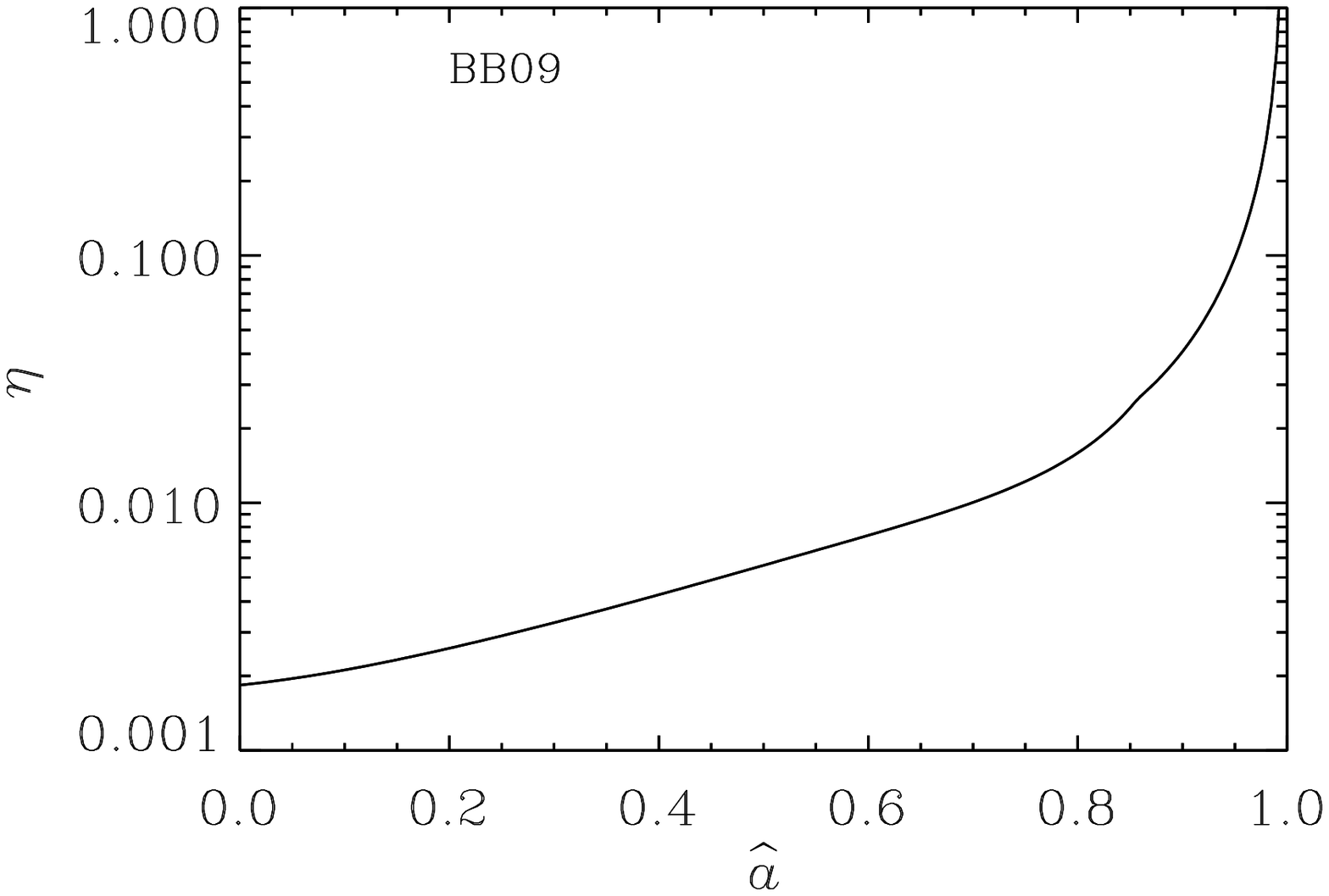, width=8cm, angle=0} \psfig{file=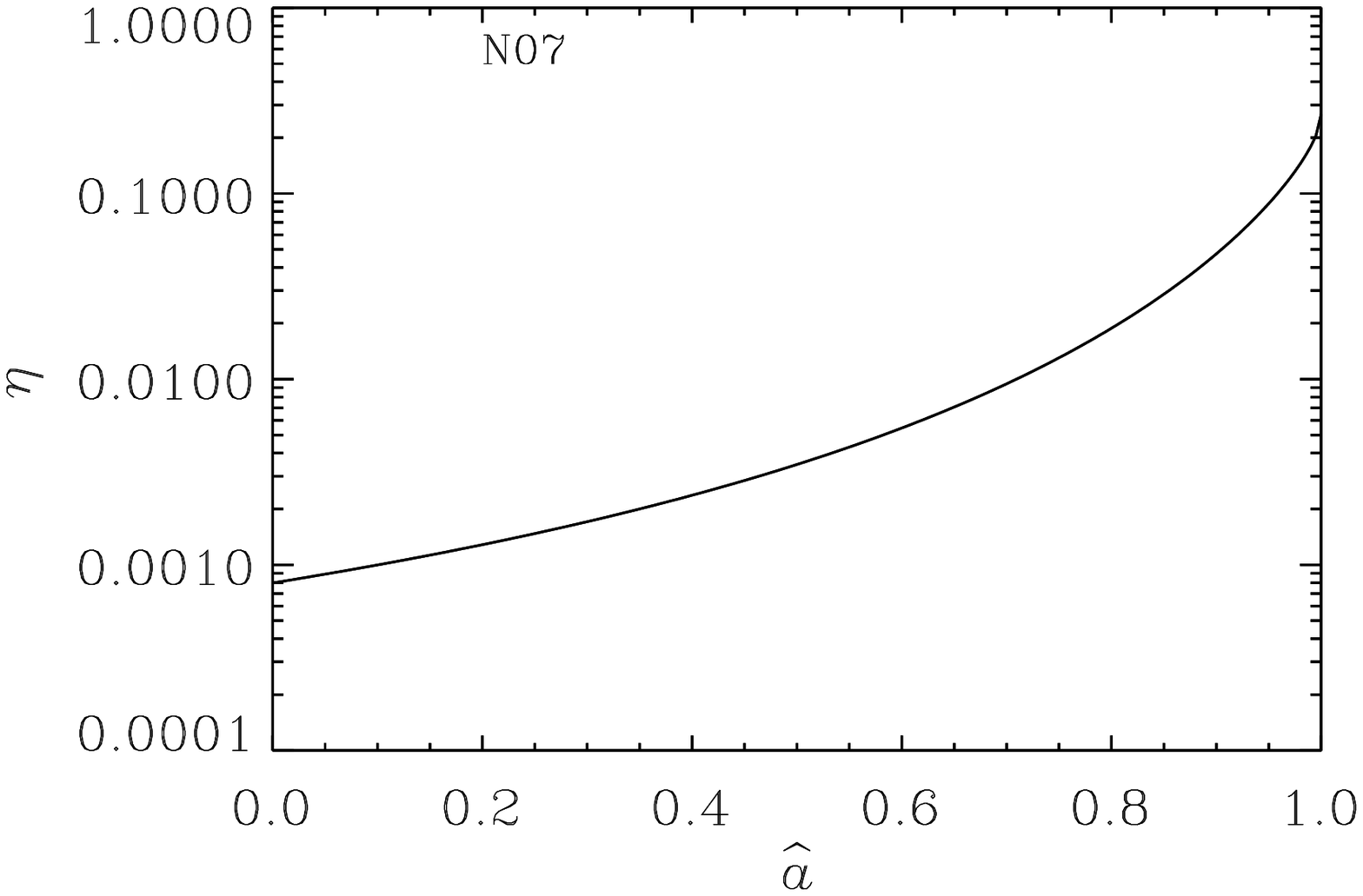, width=8cm, angle=0} }
\hbox{ \psfig{file=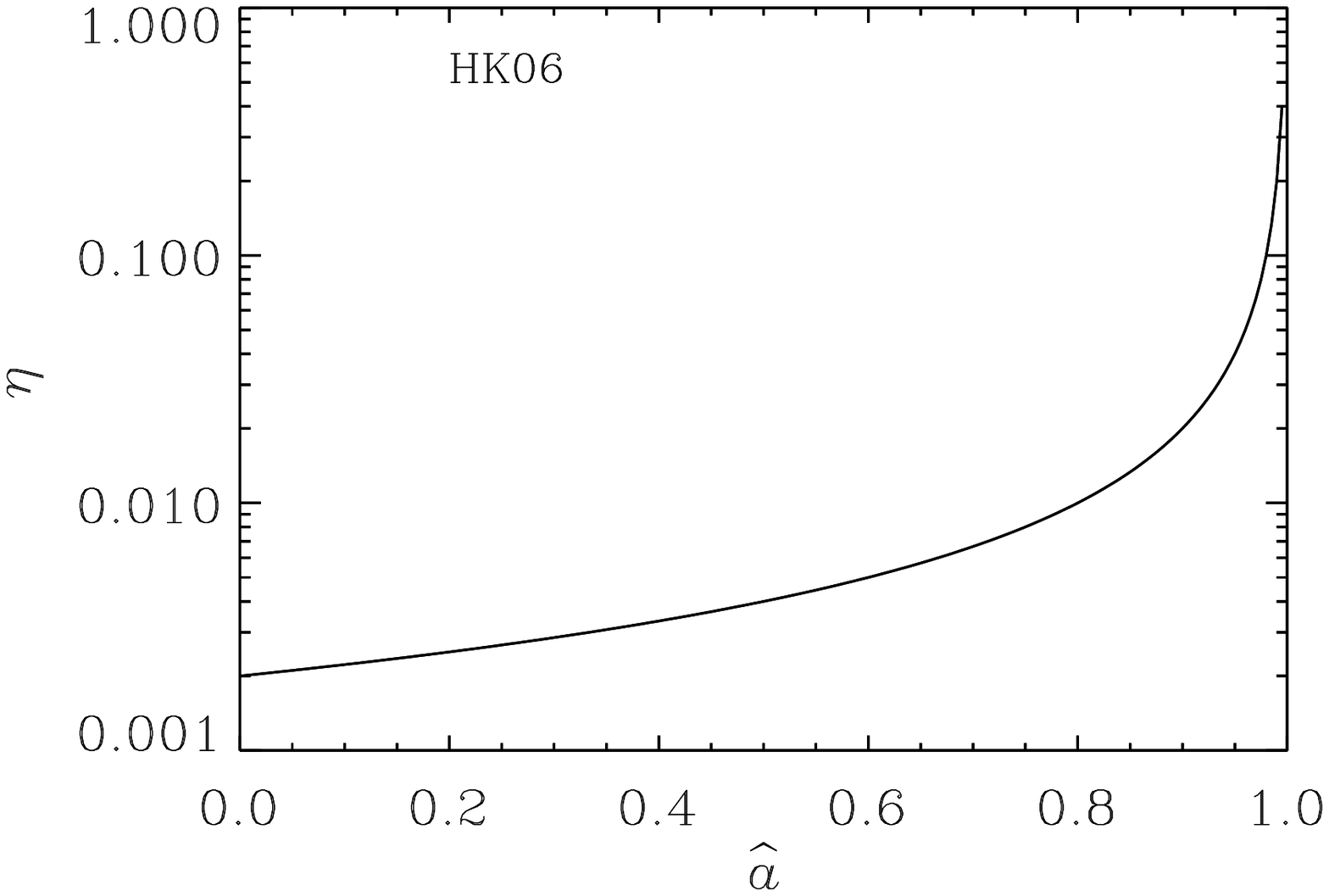, width=8cm, angle=0} \psfig{file=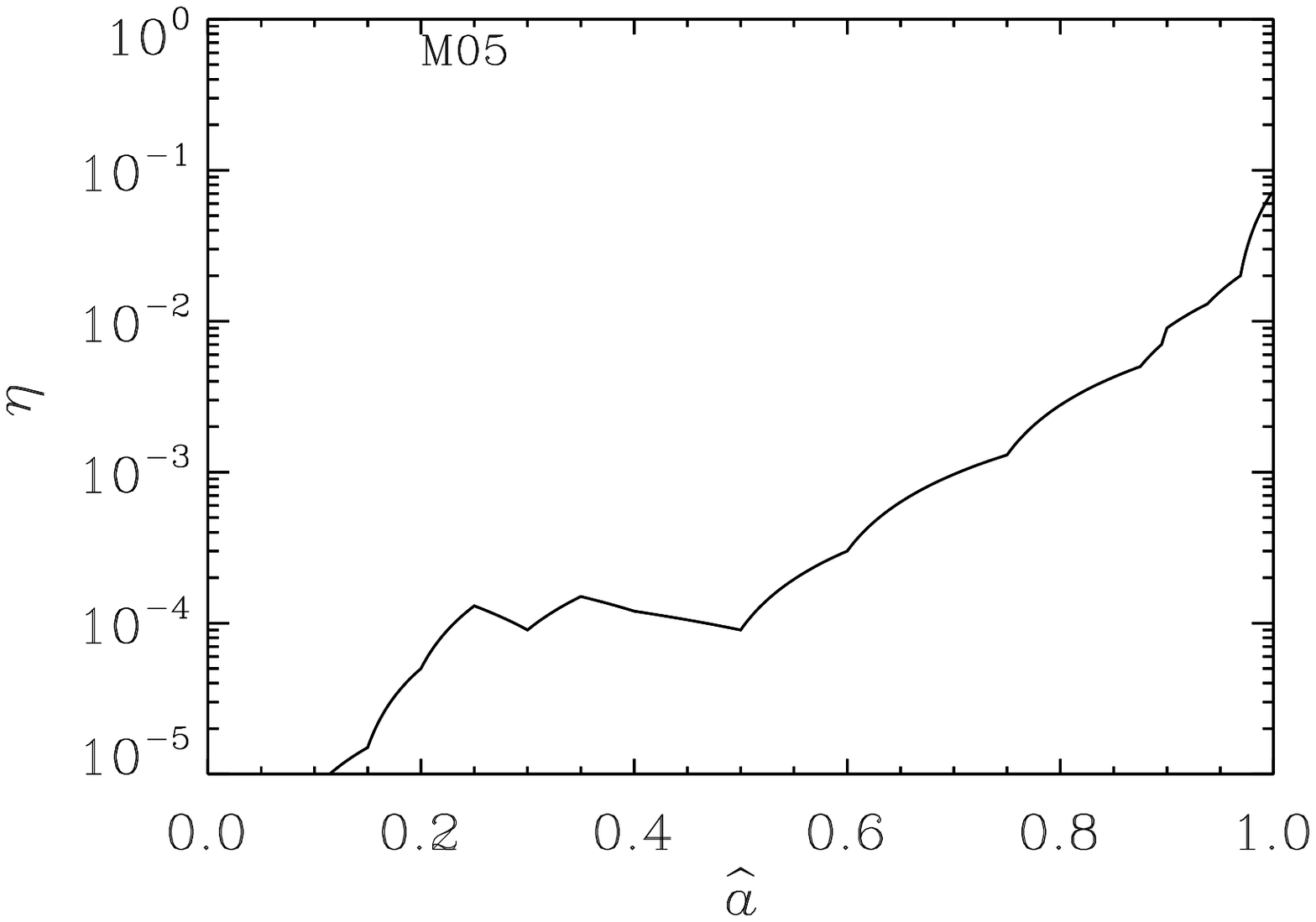, width=8cm, angle=0} }
\caption{\noindent Jet efficiencies $\eta$ as a function of black hole
  spin, $\hat{a}$.  The different efficiencies from the literature are shown in reverse chronological order, so most
  recent first. The acronyms are: T10 for \citet{2010ApJ...711...50T}, T10A is their model $H/R$$=$0.5, T10B the $H/R$$=$1.0 model.  BB09 for \citet{2009MNRAS.397.1302B},  N07 for
  \citet{2007MNRAS.377.1652N}, HK06 for
  \citet{2006ApJ...641..103H} and M05 for \citet{2005ApJ...630L...5M}.
    The curve depicting results of HK06 is
  the analytic approximation found by these authors (their Equation~10), while for M05 we
  have interpolated between  the discrete values given by  their
  simulations.}
\label{fig:eta}
\end{figure*}

\subsection{Jet efficiencies}\label{sec:jet_eff}

We assume the power available for the production of jets can be approximated by:

\begin{equation}
Q_{\rm jet} = \eta \dot{m} c^{2}
\label{eq:q}
\end{equation}

\noindent where $Q_{\rm jet}$ is the jet power, while $\eta$ is the
jet efficiency and we make the assumption that it is only a function
of spin.

Our definition of the jet efficiency is the same as that of \citet{2008MNRAS.388.1011M} and \citet{2008ApJ...676..131S}. The main conceptual difference is that we assign it to black hole spin and use the theoretical results from the literature.  
We note that our method is different to that of \citet{2009ApJ...696L..32D}, since that work
assumes the jet power to depend on black hole spin and mass, rather than spin and accretion rate. 

It is beyond the scope of this work to discuss in detail the physics
of jet formation around spinning black holes. We use recent jet efficiencies
from the literature, and refer the reader to that work for the details
and assumptions (we list these  in the following paragraph  and in Table~\ref{tab:eta}). However, we do 
outline some of the key aspects and limitations  of these efficiencies and how they 
can affect the results of our work. 

The models we will be using are from \citet{2010ApJ...711...50T}
hereafter T10 (we will actually use two separate results: T10A and T10B),
\citet[][BB09]{2009MNRAS.397.1302B}, \citet[][N07]{2007MNRAS.377.1652N},
\citet[][HK06]{2006ApJ...641..103H}  and \citet[][M05]{2005ApJ...630L...5M}. In figures and tables they will be listed in reverse
chronological order. Their dependence on the spin \sp\, is shown in Figure~\ref{fig:eta}.

\begin{table}
  \begin{center}
    \begin{tabular}{cccc}
      \hline
      \hline
Shortname & Reference &  ${h / r}$   \\
\hline
T10A & \citet{2010ApJ...711...50T} &  0.17$^{a}$ \\
T10B & \citet{2010ApJ...711...50T} &   0.33$^{a}$ \\
BB09 & \citet{2009MNRAS.397.1302B} &   0.35-0.45 \\
N07 & \citet{2007MNRAS.377.1652N} &  0.9$^{b}$ \\
HK06 &  \citet{2006ApJ...641..103H} &  0.18-0.4$^{c}$ \\
M05 & \citet{2005ApJ...630L...5M} &   0.2 \\
    \hline
      \hline
    \end{tabular}
  \end{center}
\caption{\noindent Summary of the jet models used here. The first and
  second columns state the shortname we will use to refer to the
  efficiency and the bibliographic reference, respectively. The last
  column states the range of accretion flow aspect ratios in the models or simulations.  
  $^{a}$The jet efficiencies for T10 were estimated by A.~ Tchekhovskoy  to be $\eta$$=P\gamma v^{r}/[2c\pi(h/r)]$, where $P$ is the jet power shown in Figure~6 of T10 (A.~ Tchekhovskoy, 2010, priv. comm.). This estimate is obtained by assuming that at a radius of $GM/c^{2}$, the magnetic pressure  equals the ram pressure in the accretion disc, and assuming the radial infall velocity of the accreted material in the jet-producing region to be $\gamma v^{r}$$\approx$$-c$. The thickness quoted by T10 includes the corona and jet and corresponds to 3$\times$ the thickness of the accretion disc. Hence the $H/R$$=$0.5 model of T10, labelled here as T10A,  is assigned a scale height $h/r$$=$0.17, while the $H/R$$=$1.0 model of T10 is assigned $h/r$$=$0.33 and labelled T10B.
   $^{b}$Assuming a viscosity parameter $\alpha$$=$0.3.
   $^{c}$These aspect ratios are taken from \citet{2003ApJ...599.1238D}. }
    \label{tab:eta}
\end{table}

Assuming an efficiency which is solely a function of spin is
undoubtedly a simplification which sweeps many issues under the
carpet. One such issue is the possible dependence of accretion rate
with spin, which we assume is small enough to have no significant
effect (note that the efficiencies from simulations,  M05 and HK06, will have
this effect included). Additionally, variations in the magnetic field geometry  can lead to similar accretion flows producing very different jets \citep[M05][]{2008ApJ...678.1180B}, another effect that we are not able to take into account here, other than by considering a variety of different jet efficiencies.

Another effect that has been neglected is the effect the thickness
of the accretion disc will have on the jet power, which has been
discussed in detail in the literature \citep
[e.g.][T10]{1997MNRAS.292..887G,1999ApJ...512..100L,2001ApJ...548L...9M}.

On the one hand, it has been argued that thicker discs can sustain
stronger poloidal magnetic fields and hence lead to more powerful jets
\citep [e.g.][]{1999ApJ...512..100L,2001ApJ...548L...9M}. On the other
hand M05 and T10 argue that thick discs cover a large fraction of the
area of the black hole, so that the energy outflowing from this area
will go to the accretion disc and not into the jet. For this
reason, M05 makes a difference between ``total'' and ``jet'' powers. T10 assume the magnetic flux through the black hole is  constant, not higher with larger disc thickness,  so that the occultation effect leads to lower jet powers for thicker discs (T10B) compared to thinner
ones (T10A). The fact that some of the energy can go 
into the disc rather than into powering the jet is also mentioned in HK06, although 
it is not taken into account into the efficiency we use (their Equation~10).

 Punsly (2007, 2010) discusses a component of the jet driven by the disc in the ergosphere. This component  is only present in the BB09 and HK06 efficiencies.  For HK06, however, we are using their analytic approximation which does not always reproduce exactly the results from their simulations.  In particular their simulations have maximum spins of \sp$=$0.99, while we will be using the approximation up to \sp$=$0.998, a spin for which no simulations have been carried out.

In order to calculate efficiencies, M05 and HK06 used the accretion rates from
their simulations. 
 T10 present dimensionless jet powers only, and to convert these to efficiencies one must assume an accretion rate.  
This accretion rate was estimated by assuming that at a radius of $GM/c^{2}$ the magnetic pressure in the  approximately equals the ram pressure of the accretion disc, and assuming the radial infall velocity to be $\approx$$c$ in this region (A.~ Tchekhovskoy , 2010, priv. comm., see Table~\ref{tab:eta}).  The M05 simulations have  a magnetic pressure that is typically (but not always) lower than the ram pressure. For this reason the T10A efficiency is higher than M05, despite having a similar disk thickness, $h/r$$\sim$0.2. The T10B efficiency, on the other hand, is very similar to the M05 one as can be seen in Figure~\ref{fig:eta}.

Neither M05, HK06 or T10 include cooling in the accretion disc, while N07 and BB09 assumed an ADAF accretion flow to estimate the accretion rate, with negligible cooling. Hence, all of these jet efficiencies are estimated for accretion flows with no significant radiative cooling.  One concern is therefore whether the jet powers from these models will be appropriate
for AGN with high accretion rates.

Here we justify our reasons for using Equation~\ref{eq:q}, in order to minimise the effect of these uncertainties. 

{\it 1. Keeping the jet power an explicit function of both jet efficiency and of accretion rate.}
 This will partly account for the fact that different spins will lead to different accretion rates, by keeping the total jet power explicitly dependent on both. It also attempts to reflect the fact that the magnetic fields powering the jets originate in the accretion disc and are not independent of it \citep[e.g.][ M05, HK06]{2005ApJ...620..878D}. Given that the accretion rate depends on the black hole mass via an Eddington ratio $\dot{m}$$\propto$$\lambda$\mbh, there is a natural scaling where black holes with higher masses can achieve higher jet powers. However, the jet power is not only dependent on \mbh, it also depends on the amount of accretion the black hole is undergoing.

{\it 2. Using an efficiency derived for accretion flows with negligible cooling.} While this appears as a major shortcoming for black holes with high Eddington rates, we believe it is still a useful approximation. The higher accretion rates are already considered in the $\dot{m}$ term. The worry is then if the jet efficiency is appropriate for an accretion flow with a high $\lambda$. As mentioned earlier, the aspect ratio is believed to be an important factor in determining the jet power. The jet efficiencies used here are derived from accretion flows with typical aspect ratios 0.17$\lesssim$$h/r$$\lesssim$0.9, therefore covering a wide range of disc thickness. In this paper we will concern ourselves mainly with two types of AGN: those at very low Eddington ratios (ADAFs) and those at very high ones (quasars).  Equation~\ref{eq:q} is appropriate for ADAFs, but an important  fact is that the accretion flows in quasars are radiation-pressure dominated, not gas-pressure dominated, and will have typical aspect ratios $h/r$$\sim$$\lambda$ \citep[NT73,][]{1988ApJ...332..646A}. For quasars, $\lambda$$\sim$0.25 \citep{2004MNRAS.352.1390M} so that the aspect ratio $h/r$$\sim$0.25. Hence the thickness of the two types of flow are similar. This is the reason we make the assumption that the jet efficiencies $\eta$ shown in Figure~\ref{fig:eta} 
can be used for both quasars and ADAFs.

It is immediately clear
that $\eta$ can differ significantly from model to model, which will
affect quantitatively our results. However, given that in all of these
cases $\eta$ increases with increasing spin \sp, our results will be
qualitatively the same for different assumed jet efficiencies.

We stress that all the jet efficiencies have assumptions and limitations. We believe that  showing the results for six different jet efficiencies from the literature is the most informative way of presenting our results. Now that our assumptions have been clearly stated, we proceed to compare the results to observations.

\section{The radio loudness of quasars}\label{sec:loud}

A necessary test of our parametrisation is whether it can produce the 
range in radio loudness observed amongst quasars. 

Quasars are SMBHs with black holes of \mbh$\grtsim10^{8}$~\msol\,
accreting close to the Eddington limit
\citep[e.g.][]{2004MNRAS.352.1390M}.  For a given optical luminosity,
a proxy for bolometric luminosity in unobscured objects, quasars are
found to have a vast range of radio luminosities. Most of the quasars
are found at the `radio-quiet' end, where the ratio of radio to
optical luminosity density is $\lesssim$10. However, a small fraction
show radio luminosities up to 1000 times larger, and it has been
suggested that spin is the hidden variable dominating this scatter
\citep[e.g. ][]{1992ersf.meet..368S,1995ApJ...438...62W,2007ApJ...658..815S}.

\begin{figure*} 
\hbox{ \psfig{file=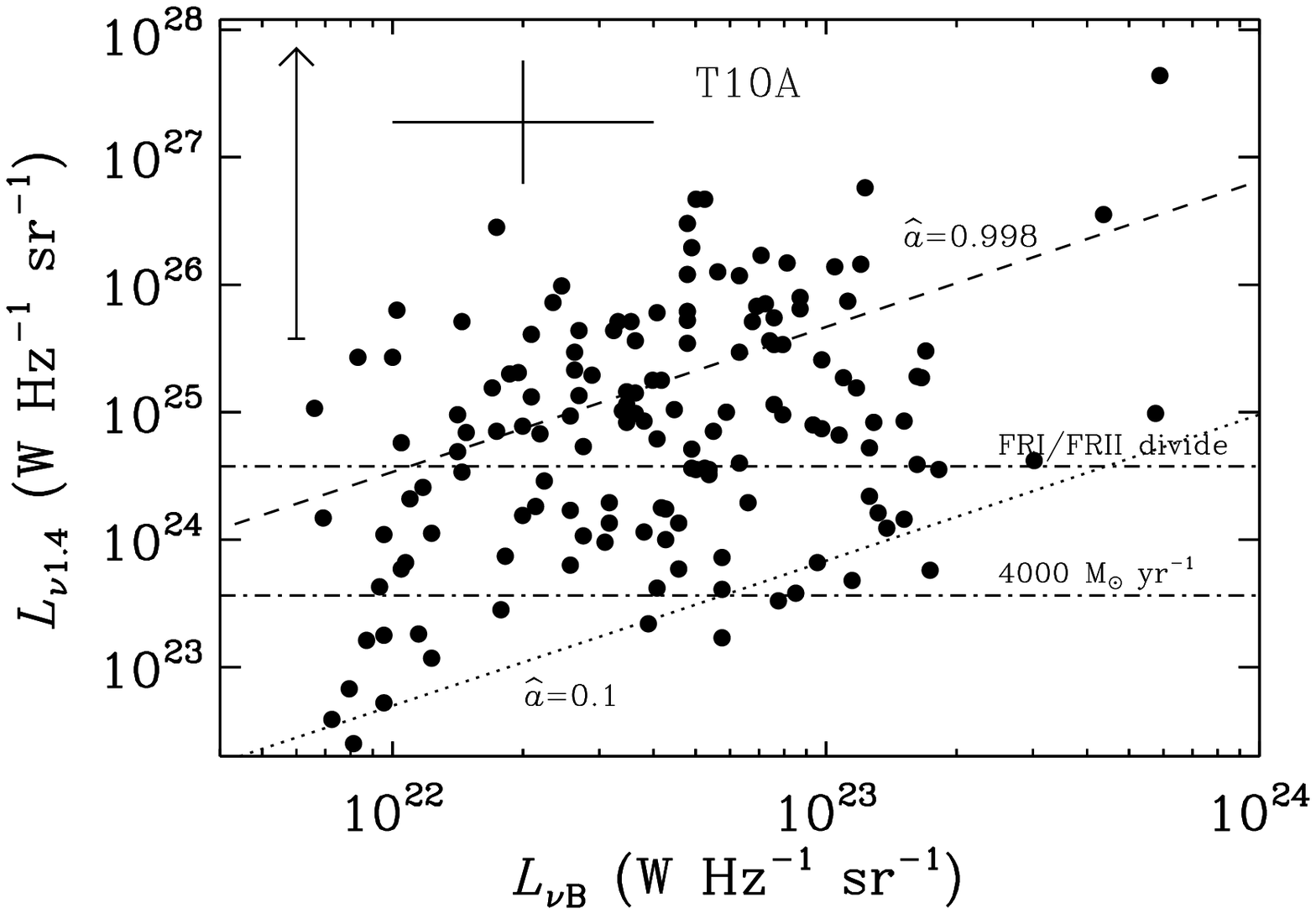, width=8cm, angle=0} \psfig{file=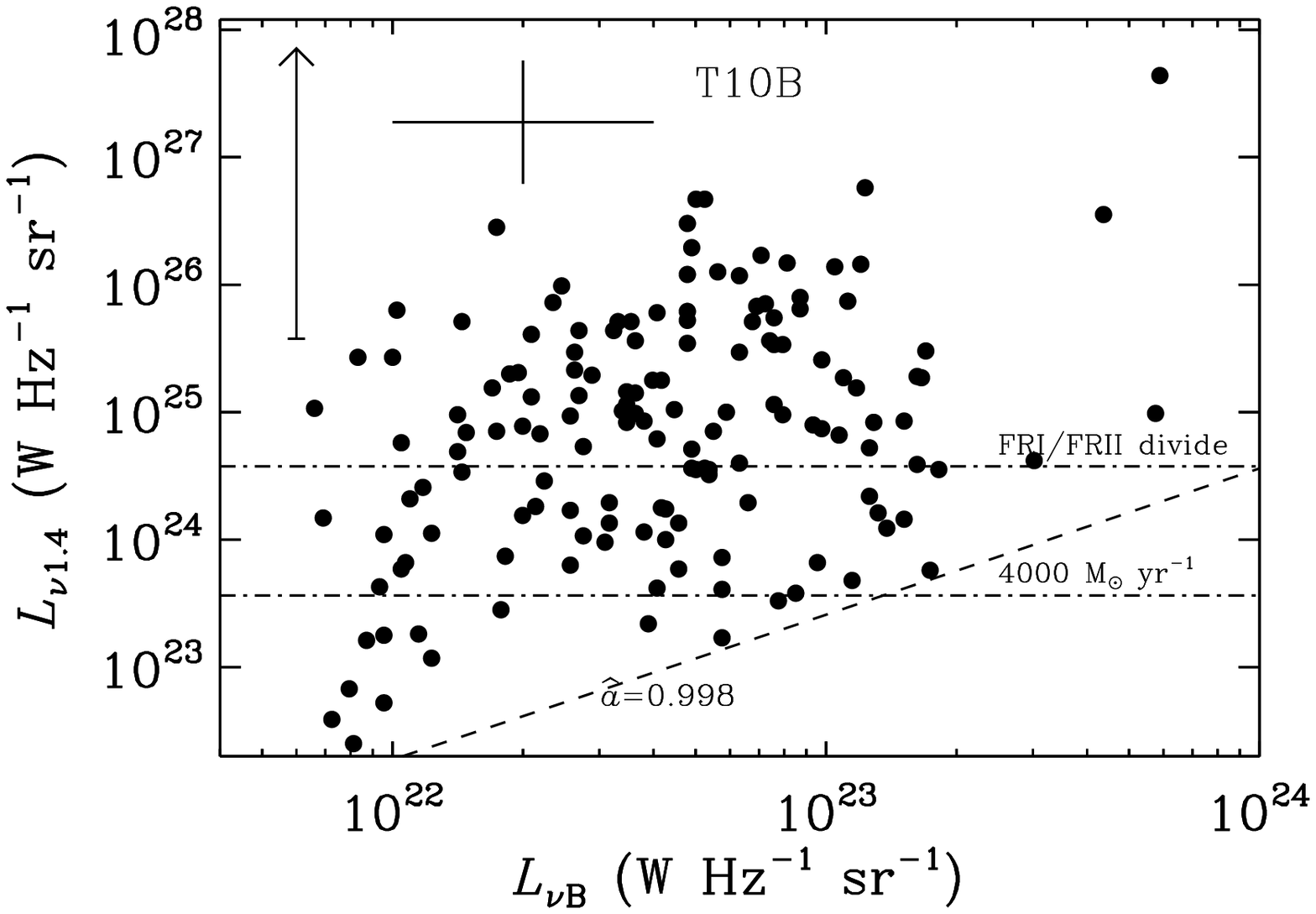, width=8cm, angle=0} }
\hbox{ \psfig{file=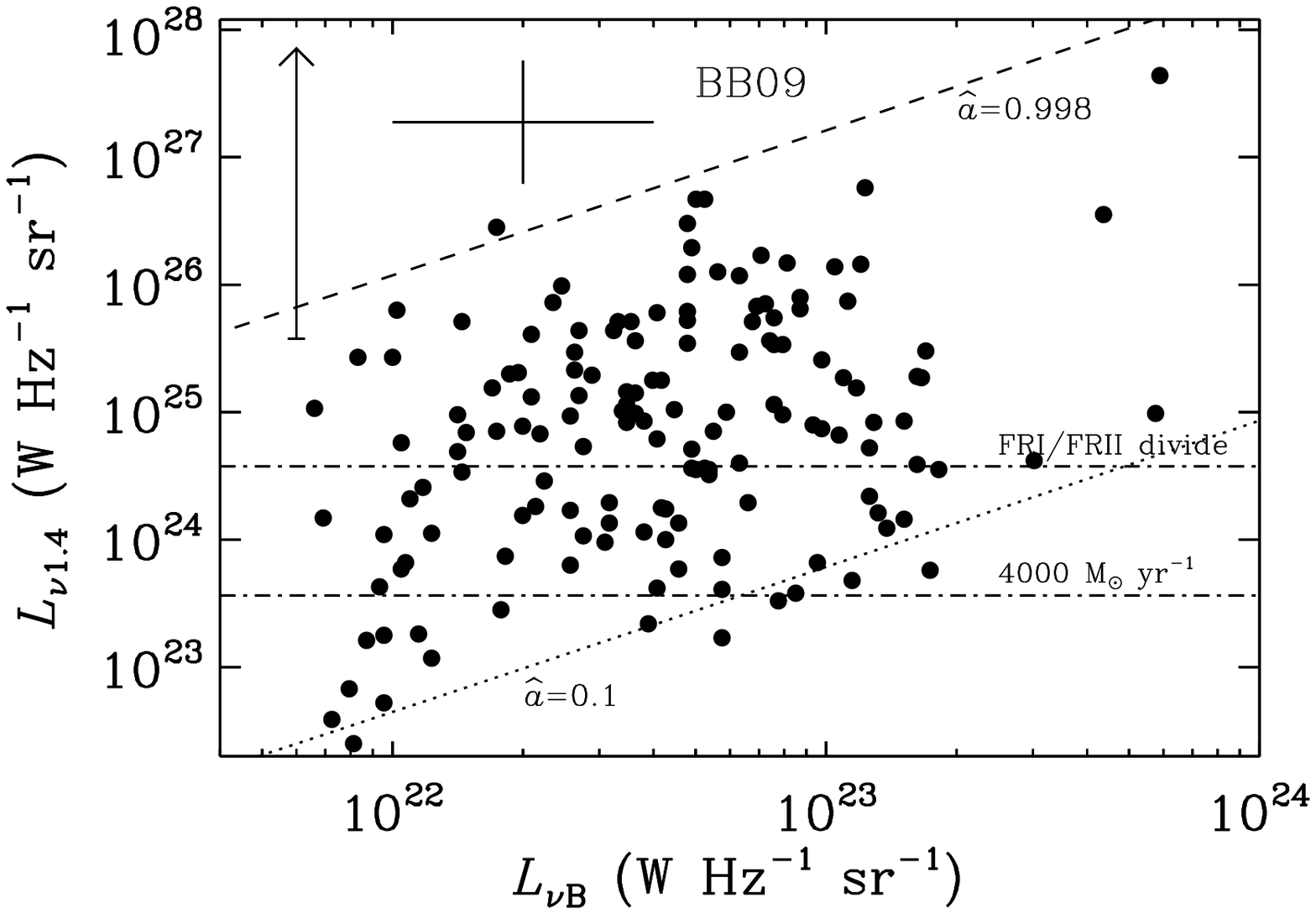, width=8cm, angle=0} \psfig{file=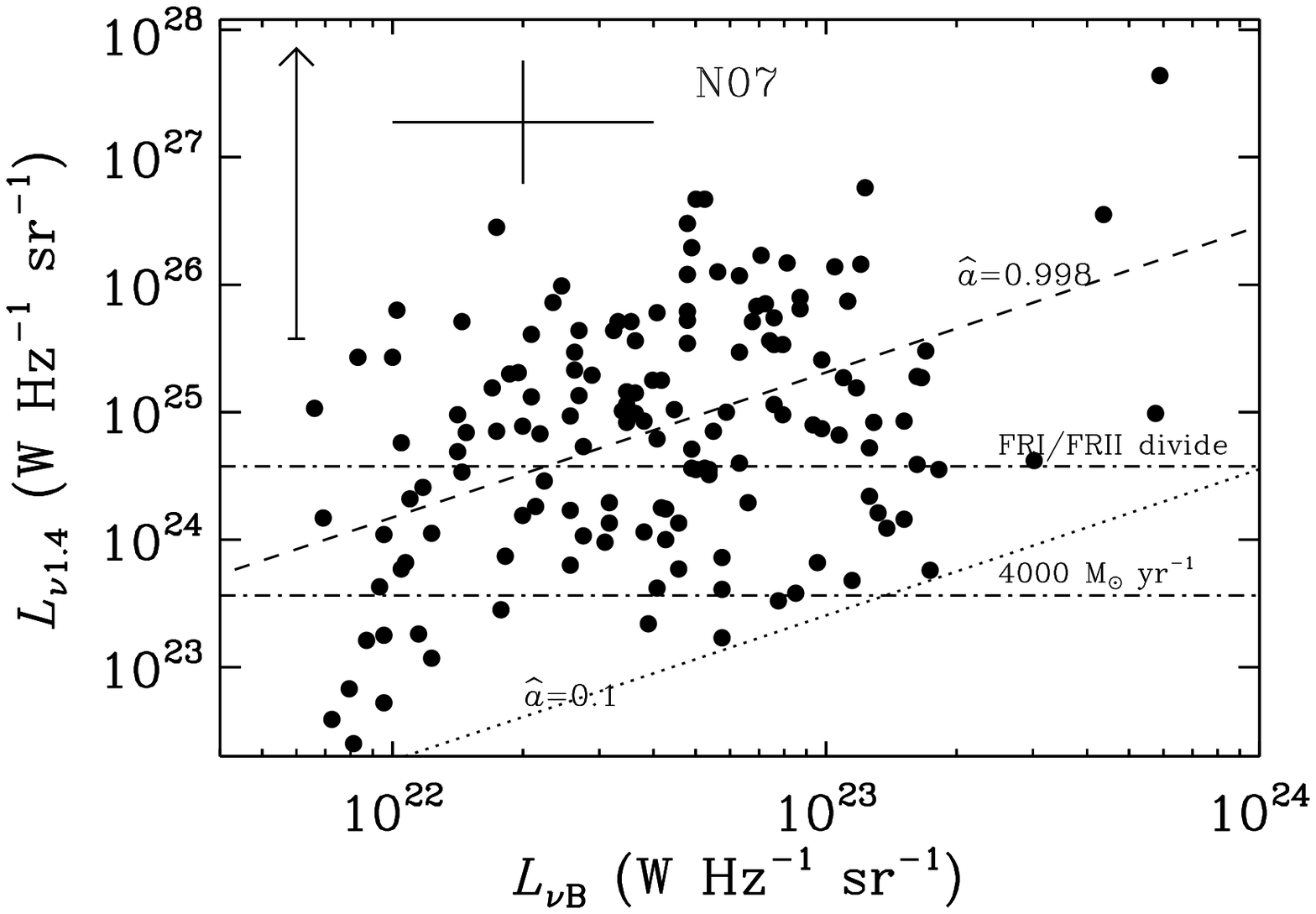, width=8cm, angle=0} }
\hbox{ \psfig{file=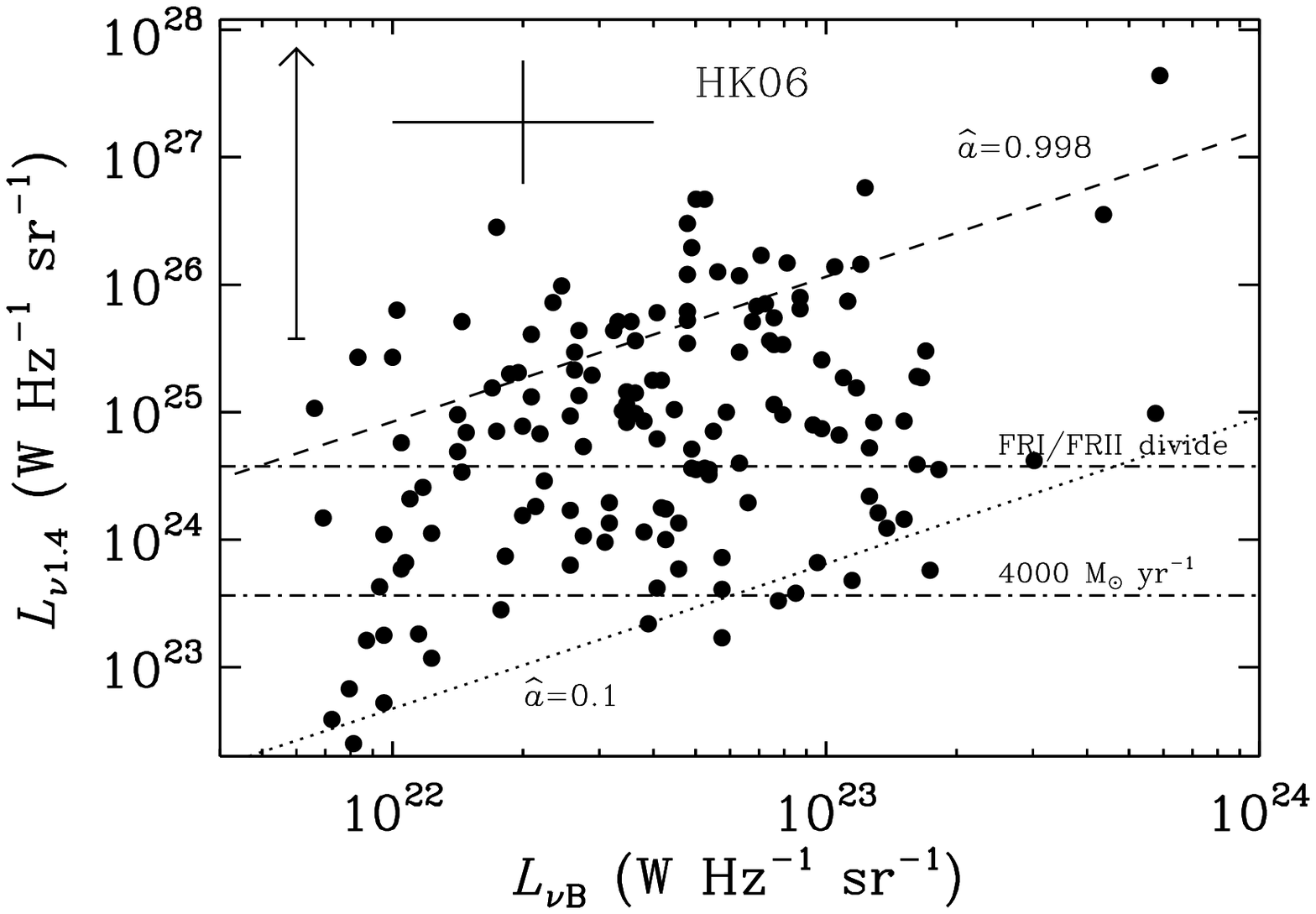, width=8cm, angle=0} \psfig{file=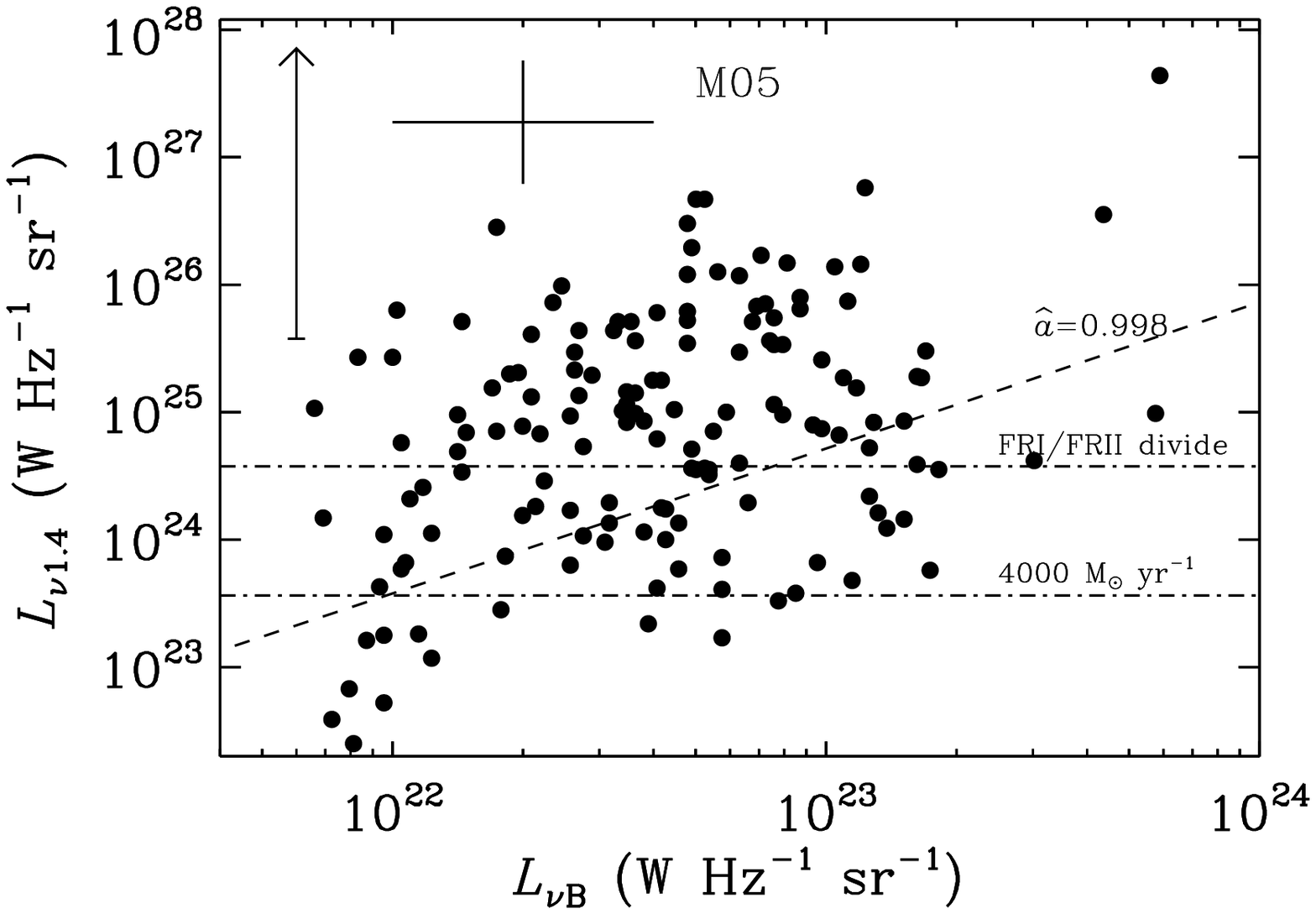, width=8cm, angle=0} }
\caption{\noindent Radio vs optical luminosity densities for a sample
  of 0.35$\leq z \leq$2.1 quasars. These were selected by matching an
  optically-selected sample of quasars to a survey at 1.4~GHz
  (Cirasuolo et al. 2003). Only sources with large enough radio
  luminosities to make the radio flux density cut make it into the
  sample, so that it is not complete below $\sim10^{24}$~\whzsr.  The
  accretion rates are estimated from Equation~\ref{eq:accn} for the
  range $3\times10^{6} \leq$$\lambda$\mbh$\leq 3\times10^{9}$~\msol.  The
  dotted and dashed lines show the resulting tracks for spins
  \sp$=$0.0 and 0.998 respectively, using the jet efficiencies $\eta$
  plotted in Figure~\ref{fig:eta}.  The cross shows the 95\%
  confidence interval for the assumed bolometric corrections
  (horizontal line) and the effect of varying the radio spectral
  indices in the range $0.25 \leq \alpha \leq 1.25$ (vertical line).  The arrow shows
  the effect on the tracks of changing $f$ from 20 to 1. The lower
  dash-dotted horizontal line represents the radio luminosity from a
  starburst forming massive stars at a rate of 1000~\msolyr\,
  \citep[for stars more massive than $\geq$5~\msol\, only,
  ][]{1992ARA&A..30..575C}.  This corresponds to a total
  star-formation rate of $\sim$4000~\msolyr\, assuming a Salpeter
  initial mass function. The upper dash-dotted line shows the FRI/FRII
  divide \citep {1974MNRAS.167P..31F}. }
\label{fig:cira} 

\end{figure*}

\subsection{Estimating observable quantities}\label{sec:obs}

We use Equations~\ref{eq:lbol} and \ref{eq:q} to predict the observed
properties associated with quasars, and test these against the sample
of quasars studied by \citet{2003MNRAS.341..993C}.  The accretion
rates for our predictons are given by:

\begin{equation}
\dot{m} c^{2} = { L_{\rm E}\lambda m_{\bullet} \over \epsilon}
\label{eq:accn}
\end{equation}  
 
\noindent where $L_{\rm E}$ is the Eddington-rate luminosity,
$1.3\times10^{31}$~W per solar mass, $m_{\bullet}$ is the black hole
mass, and the bolometric luminosity $L_{\rm bol}$ is given by $L_{\rm
  bol}=L_{\rm E}\lambda m_{\bullet} $. Typical values for quasars are
$m_{\bullet} \grtsim 10^{8}$~\msol, $\lambda\sim0.3$
\citep[e.g.][]{2004MNRAS.352.1390M}, so that $L_{\rm bol}$$\sim4\times10^{38}$~W
and $\dot{m}\sim$0.7 \msolyr.
 
To convert from bolometric luminosity, $L_{\rm bol}$, to B-band
monochromatic luminosity, $ L_{\nu \rm B}$, we use:

\begin{equation}
L_{\rm bol} = C_{\nu \rm B}\nu_{\rm B} L_{\nu \rm B}
\label{eq:bolcorr}
\end{equation}

\noindent where, in the B band, $\nu_{\rm B}$ is the frequency and
$C_{\nu \rm B}$ is the bolometric correction that converts from the
monochromatic luminosity to bolometric luminosity. Here we use the bolometric corrections from \citet{2007ApJ...654..731H}.

To convert $Q_{\rm jet}$ to $L_{\nu 151}$ we use the conversion given
by \citet{1993MNRAS.263..425M} for radio-quiet quasars and
\citet{1999MNRAS.309.1017W} for radio-loud quasars:

\begin{equation}
\left( { Q_{\rm jet} \over {\rm W}}\right)= 3\times10^{38}
f^{3\over2}\left( {L_{\nu 151} \over 10^{28} ~{\rm W Hz^{-1}
    sr^{-1}}}\right)^{6\over 7}
\label{eq:will}
\end{equation}

\noindent The term $f$ represents the combination of several
uncertainty terms when estimating $Q_{\rm jet}$ from $L_{\nu 151}$,
which includes for example the filling factor in the lobes, departure
from minimum energy,  and  the fraction
of energy in non-radiating particles.

The plausible range of values for $f$ has been estimated to be $1 \leq
f \leq 20$ by Willott et al. (1999, for details see Section~4 of their
work). An independent method of estimating jet power from the radio
luminosity has been developed by estimating the energy stored in X-ray
cavities in the intra-cluster medium
\citep{2004ApJ...607..800B,2010ApJ...720.1066C}.
\citet{2010ApJ...720.1066C} find that $f$$\sim$20 is required for both
methods to agree.

For the remainder of this paper, we will
therefore use a value of $f$$=$20, which means the jet powers
estimated using the relations of Willott et al. (1999) and
\citet{2010ApJ...720.1066C} agree within their uncertainties.  However, we note that the 
studies of \citet{2004ApJ...607..800B} and \citet{2010ApJ...720.1066C} consisted mostly
of sources of moderate radio luminosity. Sources with higher radio luminosities might have typically lower values of $f$, close to 1, as we discuss below.

 The value of  
$f$$=$20 assumes a large contribution from non-radiating particles ($\sim$200 times more energy in non-radiating particles than in radiating ones, see Section~4 of Willott et al. 1999). 
 A large fraction of  non-radiating particles can be inferred observationally if the pressure provided by radiating particles inside the lobes (derived from minimum energy arguments) is much lower than the external pressure acting on the lobes due to the intra-cluster gas. In this case the extra pressure must be provided by particles that are not emitting synchrotron radiation. There exist cases where the pressure from radiating particles is comparable to the external pressure, so that the fraction of non-radiating particles is small.  In those cases
$f$$\sim$1 is more appropriate \citep[see][and Figure~2 of Cavagnolo et al. 2010]{2008MNRAS.386.1709C}.  From studies of  the cavities in the intra-cluster medium, it seems possible that 
the sources that are found to have low jet powers for a given radio luminosity have 
a lower content of non-radiating particles, due to a lower entrainment of external material by the jets. \citep{2008MNRAS.386.1709C,2010ApJ...720.1066C}. 

 Fewer detailed studies exist of the more powerful radio sources, since these are rare in the nearby Universe. However, some studies suggest the fraction of non-radiating particles is small in these cases \citep{2004MNRAS.353..879C}. This is possibly due to the fact that the most powerful radio sources, \citet{1974MNRAS.167P..31F} class II objects, have strong bow shocks that push back the intergalactic medium so that  entrainment is minimal. In these objects,  a value of $f$$\sim$1 might be more appropriate in this case \citep{2004MNRAS.353..879C}.  The arrows shown in Figure~\ref{fig:cira} show how the curves would move if $f$ were  changed from $f$$=$20 to $f$$=$1.  

 Source-to-source variations are likely important, as reflected by the scatter between jet power and radio  luminosity \citep{2004ApJ...607..800B,2010ApJ...720.1066C}. However, there might be systematic variations with luminosity, with the high-radio luminosity sources having a lower fraction of non-radiating particles, and hence requiring a lower value of $f$.  In such cases, assuming $f$$=$20 will systematically underestimate the radio luminosity that can be produced for a given jet power. For a given jet power, $L_{\nu 151} \propto f^{-{7\over4}}$, meaning that changing $f$ from 20 to 1 would increase the radio luminosity by practically 2 dex (see the arrow on Figure~\ref{fig:cira}).

 Since the comparison data is at 1.4~GHz, we convert from 151~MHz
 assuming a single power law $L_{\nu} \propto \nu^{-\alpha}$ assuming
 a spectral index of $\alpha=0.75$.

\subsection{Comparison of the models to the data}\label{sec:comparison}

Figure~\ref{fig:cira} shows the loci of radio luminosity density
versus optical luminosity density from our assumptions in
Section~\ref{sec:assum} and Equations~\ref{eq:accn}, \ref{eq:bolcorr}
and \ref{eq:will}. Each panel assumes the jet efficiencies of
Figure~\ref{fig:eta}, in the same order. We have plotted two tracks
for each jet efficiency: one for \sp$=$0.1 (dotted line) and another one for
\sp$=$0.998 (dashed line). The cutoff  maximum spin is somewhat 
arbitrary, but our choice is motivated by the theoretical 
maximum value predicted  by \citet{1974ApJ...191..507T}.

The black points are the data for the quasars studied by Cirasuolo et
al. (2003), who matched two samples of optically selected quasars
\citep[the 2dF QSO redshift survey and the large bright quasar
  survey][]{2001MNRAS.322L..29C,1995AJ....109.1498H} to the FIRST
radio survey \citep{1995ApJ...450..559B}. The sensitivity of the FIRST
survey is low enough that several decades of radio loudness can be
explored, yet the sample is probably incomplete below radio luminosity
densities of $L_{\nu 1.4}\sim 10^{24}$~\whzsr\, (as noted by Cirasuolo
et al. 2003).

The data show three main features: a high-radio luminosity envelope, a
low-radio luminosity envelope, and as a result of the difference
between these two, a scatter of $\sim$3 dex. The high-luminosity
envelope is not a selection effect but the lower envelope might well
be due to incompleteness below $L_{\nu 1.4}\sim 10^{24}$~\whzsr.

Given a maximum achievable jet efficiency, $\eta$(\sp$\sim$1),
Equations~\ref{eq:q} and \ref{eq:lbol} do predict an upper envelope
showing a correlation between radio and optical luminosity, $Q_{\rm
  jet} \propto \dot{m}$. The exact location of the envelope, however,
depends on the model assumed for $\eta$, and different models
reproduce the data in Figure~\ref{fig:cira} with varying degrees of
success.

The BB09 model and HK06 produce upper envelopes in reasonable
agreement with the data (although we stress the caveats mentioned in Section~\ref{sec:jet_eff}). The agreement is not perfect and some sources
appear $\sim$1 dex higher than the  curves. However, given the
uncertainties in converting from bolometric to monochromatic
luminosity, from jet power to radio luminosity density and the
uncertainties in the actual radio spectral indices, we do not consider
this discrepancy a worry.

The T10A and N07,  efficiencies produce upper envelopes that are
significantly lower. This is because the maximum efficiency these
two models reach is $\eta$$\sim$0.2-0.3, while the most powerful
radio-loud quasars and radio galaxies require jet efficiencies $\grtsim$1
\citep{2007MNRAS.374L..10P,2011ApJ...727...39M,2010arXiv1010.0691F}. Therefore, any
model or simulation that does not reach such high values will not be
able to reproduce the upper envelope \citep[see also][]{2010arXiv1012.1910P}. This is the case for the T10B  and M05 models, which produce even lower tracks.

However, we remind the reader that in Section~\ref{sec:obs}, we have assumed
a value of $f$$=$20 when converting from jet power to radio luminosity density. This
implies a large component of non-radiating particles.  The arrows shown in Figure~\ref{fig:cira} show how the curves would move if $f$ were  changed from $f$$=$20 to $f$$=$1. As discussed in Section~\ref{sec:obs},  it is possible that the most powerful radio sources systematically require lower values of $f$.  For $f$$=$1, the high-spin tracks for  T10A, N07 and HK06 will easily produce high enough radio luminosities, the BB09 will predict far too much radio luminosity for \sp$\sim$1, and the T10B and M05 high spin tracks will be high enough to match the upper envelope.

The upper envelope can also explain the result of
\citet{1991Natur.349..138R}, who found a linear correlation between
jet power and narrow-line luminosity ($L_{\rm NL}$, an isotropic proxy
for $L_{\rm bol}$) for radio-loud AGN (their work assumed
$f$$\approx$2 to convert from radio luminosity to jet power).  If we
assume $L_{\rm bol}\sim 500-1000\times L_{\rm NL}$,  our the curves
with maximum efficiencies $\eta$$\sim$0.1-1.0  
 can also explain the data of
\citet{1991Natur.349..138R} as an approximate upper envelope. \citet{1998MNRAS.294..494S} found an
upper envelope for the radio luminosity density versus optical
magnitude of steep-spectrum quasars, and argued it was not due to any
selection effects or relativistic beaming.  A similar envelope is also
found between infrared luminosity density (another approximately
isotropic estimator of $L_{\rm bol}$) and low-frequency radio
luminosity density for radio-selected AGNs, an envelope that can again
be reasonably reproduced by our curves  \citep{2010arXiv1010.0691F}.

The minimum jet efficiency, $\eta$(\sp$=$0), from HK06, N07 and BB09
predict lower envelopes, which appear in good agreement with the data.
However, as Cirasuolo et al. (2003) point out, the sample of quasars
might be incomplete below $L_{\nu 1.4}\sim 10^{24}$~\whzsr, so that
the agreement between the lower envelope and the data might be
fortuitous. Again, the uncertainties when converting $L_{\rm bol}$ and
$Q_{\rm jet}$ into $L_{\nu \rm B}$ and $L_{\nu 151} $ are large (see
Figure~\ref{fig:cira}), so that quasars with slightly lower
radio-to-optical luminosity density ratios are not strictly in
disagreement with our curves.

The fact that in the the M05, T10A and T10B models $\eta$$\rightarrow$0 as \sp$\rightarrow$0, means that they do not
have a lower envelope, but this is not in disagreement with current
data, as explained above. The existence of a possible lower envelope
requires further studies using deep radio observations of powerful
quasars.

Discriminating whether this lower envelope is a real physical effect
or just a selection effect, is further complicated by the fact that
very powerful starbursts, with massive star-formation rates
$\sim$1000~\msolyr\, (for stars $\geq$5~\msol, which would correspond
to a total star-formation rate of $\sim$4000~\msolyr\, for a Salpeter initial mass function), can
show radio luminosity densities as high as
$\sim3\times10^{23}$~\whzsr\, without the presence of an AGN \citep
[this is shown by the bottom horizontal line in Figure~\ref{fig:cira},
  following][]{1992ARA&A..30..575C}.

\subsection{Implications for AGN}

Also overplotted on Figure~\ref{fig:cira} is the divide between
Fanaroff-Riley class I and class II objects \citep[][hereafter FRI and
  FRII sources]{1974MNRAS.167P..31F}. FRI and FRII sources have
powerful extended jets, and the bulk of the quasars in the figure lie
around or above the FRI/FRII divide. This suggests that extended jets
are common around quasars, and indeed the common assumption that
quasars were never associated with FRI-type jets has been proven false
\citep[see][]{2001ApJ...562L...5B,2007MNRAS.381.1093H}. 

\citet{1993MNRAS.263..425M}
also found that many of the brightest optically-selected quasars at
$0.3 \leq z \leq 0.5$ were FRII sources, and the high jet powers inferred by \citet{2004MNRAS.349.1419C} for
powerful nearby FRIs ($\geq$$10^{36}$~W)  seem difficult to reach
without high accretion rates.

Turning this around, even with a high spin, no AGN with a low
radiative luminosity is expected to show FRII luminosities (and
presumably by extension no FRII jet structures either). For example,
the FRI/FRII division is at a 151~MHz radio luminosity of
$\sim2\times10^{25}$~\whzsr, which corresponds to
\qjet$\sim1.3\times10^{38}$~W (assuming $f$$=$20). Even with jet efficiencies
 as high
high as $\eta$$\sim$0.5, this still
requires an accretion rate of 0.05~\msolyr, which can only be achieved
by SMBHs with $\lambda$\mbh$\grtsim3\times10^{7}$~\msol. This is the
typical value for optically-selected quasars
\citep{2004MNRAS.352.1390M}, and corresponds to moderate to high
Eddington rates for black hole masses $\sim$$10^{7}-10^{9}$~\msol, and
only corresponds to modest Eddington rates for the rare black holes
with masses $\grtsim$$3\times10^{9}$~\msol. This explains why FRII
sources are mostly associated with HEG spectra, and only sometimes
with LEGs spectra, while FRIs which can be associated with both HEG or
LEG spectra
\citep[e.g.][]{1996AJ....112....9L,2001ApJ...562L...5B,2006MNRAS.370.1893H}.
Section~\ref{sec:res} describes this quantitatively.

An upper limit on the observed radio luminosity density, $L_{\nu
  151} $, can be estimated by assuming $\eta_{\rm max}$$\sim$1, $\epsilon_{\rm max}$$\sim$0.3  as well as
$\lambda_{\rm max}\sim1$ and \mbh$_{\rm ,max}$$\sim3\times10^{9}$~\msol. 
This yields maximum values of $Q_{\rm
  jet,max}\sim10^{41}$~W, and $L_{\nu 151,\rm max} \sim5\times10^{28}$~\whzsr\, or
$L_{\nu 1.4,\rm max} \sim1\times10^{27}$~\whzsr. No radio galaxy observed so
far has a radio luminosity significantly above this value \citep[e.g.][] {1991Natur.349..138R,1999MNRAS.309.1017W}.

We now proceed to test whether under reasonable assumptions the spin paradigm can explain the different contributions to the local radio LF from AGN with high and low accretion rates (HEGs and LEGs, respectively).

\section{Modelling the local radio luminosity function}\label{sec:fit_rlf}

We proceed to fit the local radio LF of LEGs and HEGs to find the
distribution of spins that best describes each subpopulation. The
best-fitting parametrisation allows us to interpret the
physical conditions that lead to the spin distribution, it can be used
to derive an estimate of the spin history of SMBHs.

Above $L_{\nu 1.4}\grtsim 10^{23}$~\whzsr, all the sources
contributing to the radio LF have \mbh$\geq10^{8}$~\msol
\citep[e.g.][]{2004MNRAS.351..347M,2009ApJ...696...24S}. We will
therefore limit our discussion to SMBHs with \mbh$\geq10^{8}$~\msol.  The radio luminosity is
determined from the jet power \qjet\, (see below), which from
Equation~\ref{eq:q} depends on the jet efficiency, $\eta$ as well as
the accretion rate, $\dot{m}$.

Throughout this section we follow the convention that the sign
$\phi$(L) refers to a luminosity function while the sign $P$(x) refers
to a normalised probability distribution function (PDF):

\begin{equation}
 \int P(x){\rm d}x =1
\end{equation}

\noindent Luminosity functions are PDFs that are not normalised to
unity, but rather to a space density. Hence we will manipulate LFs in
an analogous way to PDFs (note that in practice all the integrals are
done using base-10 logarithms).

  Assuming that the jet efficiency is essentially a monotonic function of spin,
  the PDF for jet efficiencies is related to the PDF for spins by:

\begin{equation}
P(\eta){\rm d}\eta =  P(\hat{a}){\rm d}\hat{a}
\end{equation}

The space density of sources with a given radio luminosity density
$L_{\nu}$ is related to the space density of sources with a given jet
power $Q$, via:

\begin{equation}
\phi (L_{\nu}){\rm d }L_{\nu}  = \phi (Q){\rm d }Q .
\end{equation}

From Equation~\ref{eq:q}, the distribution of jet powers will depend
both on the distribution of accretion rates and spins. The PDF of a
given jet power $Q$ can be thought of as the probability of $Q$ and an
accretion rate $\dot{m}$ and an efficiency $\eta$, marginalised over
the `nuisance' parameters $\dot{m}$ and $\eta$.

\begin{eqnarray}
 P(L_{\nu})  = {{\rm d}Q_{\rm jet}\over {\rm d}L_{\nu}} P(Q_{\rm jet}) \\
= { {\rm d}Q_{\rm jet}\over {\rm d}L_{\nu}} \int \int P(Q_{\rm jet} ,\dot{m}, \eta) {\rm d}\dot{m}{\rm d \eta}  
 \end{eqnarray}
 
 Using Bayes' theorem, we can write this in terms of the PDF of $Q$ given  $\dot{m}$ and $\eta$ and the PDFs of $\dot{m}$  and $\eta$. 
 
 \begin{eqnarray}
 P(Q_{\rm jet} ,\dot{m}, \eta) 
 =  P(Q_{\rm jet} |\dot{m}, \eta)P(\eta, \dot{m})
 \end{eqnarray} 
 
 For a given accretion   and jet efficiency, only one value of the jet power is possible, so that  the probability of   \qjet\, given  $\dot{m}$  and  $\eta$ is a delta-function:
 
  \begin{equation}
P(Q_{\rm jet} | \dot{m}, \eta)= \delta(Q_{\rm jet} - \eta  \dot{m} c^{2}), 
 \end{equation} 
 
\noindent and hence

  \begin{eqnarray}
\int \int P(Q_{\rm jet} |\dot{m}, \eta)P(\eta, \dot{m}){\rm d}\dot{m}{\rm d}\eta   \\ \nonumber
= \int \int \delta(Q_{\rm jet} -\eta  \dot{m} c^{2}) P(\eta, \dot{m}){\rm d}\dot{m}{\rm d}\eta  \\ \nonumber
 =  \int P(\eta, \dot{m}'=Q_{\rm jet} -\eta  c^{2}) {\rm d}\eta 
  \end{eqnarray} 
 
For a particular class of objects, we make the approximation that the
spin is independent from the instantaneous accretion rate, so that:

\begin{equation}
P(\eta, \dot{m}')  = P(\eta)P(\dot{m}'). 
\label{eq:spin_indep_mdot}
\end{equation}

The space density of sources with a given radio luminosity is
therefore given by:

 \begin{eqnarray}
 \phi(L_{\nu})  = {{\rm d}Q_{\rm jet}\over {\rm d}L_{\nu}}  \int P(\eta)\phi(\dot{m}') {\rm d}\eta  \nonumber \\
 = {{\rm d}Q_{\rm jet}\over {\rm d}L_{\nu}}  \int P(\hat{a})\phi(\dot{m}') {\rm d}\hat{a}
\end{eqnarray}

We model the LEGs and HEGs separately, and we begin with the HEGs.

\subsection{High-accretion rate radio galaxies}\label{sec:hi_acc}

The HEGs are modelled by SMBHs with high accretion rates, which we
refer to as QSOs for simplicity. The space density of SMBHs with a
given accretion rate can be estimated from the X-ray LF, since the
X-ray luminosity can be used to estimate the bolometric luminosity via

\begin{equation}
L_{\rm bol}=C_{\rm X} L_{\rm X}
\end{equation}

\noindent where $C_{\rm X} $ is the bolometric correction at X-ray
energies. Hence, the accretion rate $\dot{m}$ can be estimated as:

\begin{equation}
\dot{m} = { C_{\rm X} L_{\rm X}  \over \epsilon c^{2} }.
\end{equation}

The space density of HEGs with a given radio luminosity is therefore:

\begin{eqnarray}
\phi_{\rm QSO}(L_{\nu})
=  {{\rm d}Q_{\rm jet}\over {\rm d}L_{\nu}}\int P_{\rm QSO}(\hat{a})\phi( L_{\rm X}')  {{\rm d} L_{\rm X}' \over {\rm d}\dot{m}'}  {\rm d}\hat{a} 
\end{eqnarray}

To avoid including SMBHs with \mbh$<10^{8}$~\msol, we impose a minimum
 bolometric luminosity of  $L_{\rm bol}\geq3\times10^{38}$~W. This can only be achieved by SMBHs with \mbh$\geq10^{8}$~\msol\, at 25\% of the Eddington luminosity \citep[the characteristic value found by][]{2004MNRAS.352.1390M}.
 
Assuming typical
bolometric corrections, this corresponds to an X-ray luminosity of $9\times10^{36}$~W. We
have also limited the upper X-ray luminosity to $\leq
1\times10^{39}$~W since no sources brighter than this have been
observed \citep[e.g.][]{2008ApJ...679..118S}.

 The X-ray LF of  \citet{2008ApJ...679..118S} is selected in the hard band (2-8 keV), meaning it is sensitive to unabsorbed and moderately absorbed sources (with column densities \nh$\leq$$10^{26}$ m$^{-2}$ and $10^{26}$ m$^{-2}$ $\leq$\nh$\leq10^{28}$ m$^{-2}$, respectively). However, sources with column densities $\geq$1.5$\times10^{28}$ m$^{-2}$ are optically-thick to Compton scattering (`Compton-thick') and will be missed by the X-ray LF. 

In the X-ray luminosity range of interest, $9\times10^{36}$~W $\leq$\lx$\leq
1\times10^{39}$~W, approximately half of the hard X-ray sources are unabsorbed and the other half are absorbed but Compton thick. The fraction of Compton-thick AGN is estimated to be similar to the fraction of Compton-thin absorbed AGN, meaning the space density of  quasars given by the hard X-ray LF has to be corrected by a factor $\sim$1.5 \citep[see e.g.][]{1999ApJ...522..157R,2007A&A...463...79G,2010MNRAS.tmp.1757G}. Although there might be a slight evolution in the fraction of absorbed AGN \citep[e.g.][]{2006ApJ...652L..79T,2008A&A...490..905H}, this correction is likely still correct at high redshift \citep[e.g.][]{2007MNRAS.379L...6M,2007A&A...463...79G,2009ApJ...693..447F}. We therefore multiply the space density of the X-ray LF by a factor of 1.5, to estimate the space density of all QSOs (whether hidden or not).

\subsection{Low-accretion rate radio galaxies}\label{sec:low_acc}

We model the LEG component of radio LF as SMBHs with low accretion
rates, which we call ADAFs.  This requires some extra assumptions,
since no luminosity function is available to estimate their accretion
rates. However, the local mass function of supermassive black holes,
hereafter BHMF, provides the distribution of masses, which can be
related to the distribution of accretion rates via a distribution of
Eddington rates:

\begin{eqnarray}
P(\dot{m}')  = \int P(\dot{m}', m_{\bullet}){\rm d}m_{\bullet}\nonumber \\
 = \int  P(\dot{m}' | m_{\bullet}) P(m_{\bullet}){\rm d}m_{\bullet}\nonumber \\
= \int P(\lambda ) {{\rm d} \lambda \over {\rm d} \dot{m}'} P(m_{\bullet}){\rm d}m_{\bullet} \end{eqnarray}

We have expressed the conditional probability of an accretion rate
given a black hole mass, $P(\dot{m}' | m_{\bullet})$, as a probability
of an Eddington rate $P(\lambda )$.

The radio LF due to ADAFs  can therefore be modelled as:

\begin{eqnarray}
\phi_{\rm ADAF}(L_{\nu}) = \nonumber \\
{{\rm d}Q_{\rm jet}\over {\rm d}L_{\nu}} \int P_{\rm ADAF}(\hat{a})\int P(\lambda )  \phi(m_{\bullet}) {{\rm d} \lambda \over {\rm d} \dot{m}'} {\rm d}m_{\bullet} {\rm d}\hat{a}
\end{eqnarray}

Given our ignorance about the distribution of Eddington ratios, we
choose a prior PDF which is flat in logarithmic space, known as a
Jeffreys prior:

\begin{equation}
P(\lambda) \propto {1 \over (\lambda_{\rm max}-\lambda_{\rm min})  }
\end{equation}

 We assume the ADAFs are described by SMBHs accreting at Eddington
 ratios in the range $10^{-8} \leq$$\lambda$$\leq 10^{-2}$, hence
 $\lambda_{\rm max}$$=$$10^{-2}$ and $\lambda_{\rm min}$$=$$10^{-8}$.

\subsection{Model selection for the spin distribution}\label{sec:modelsel}

We now proceed to use the data from the observed local radio LF of
LEGs and HEGs to constrain the spin distribution. Three models are
considered.

Model A  consists of a power law PDF for the spin:

\begin{equation} 
P_{\rm A}(\hat{a}) = N_{\rm A} \hat{a}^{-p}
\end{equation}

\noindent The power law index $p$ is the only free parameter in this
model, and it is given a flat prior PDF in the range $-4 \leq p \leq
4$.  $N_{\rm A}$ is simply a normalisation constant.

The next model, B,  is a single gaussian distribution:

\begin{equation} 
P_{\rm B}(\hat{a}) = N_{\rm B} e^{-{1\over 2}\left( \mu_{B} -\hat{a} \over \sigma_{\hat{a}} \right)^{2}} .
\end{equation}

\noindent We fix $\sigma_{\hat{a}}$$=$0.05 and $N_{\rm B}$ is the
normalisation term, so this model has also only one free parameter,
$\mu_{B}$, to which we assign a flat prior in the range $0 \leq
\mu_{B} \leq 1$.

The value of $\sigma_{\hat{a}}$$=$0.05 is assumed since it can provide
a reasonable approximation to the width of the spin distributions
expected to arise from accretion onto SMBHs as well as from major
mergers (see Section~\ref{sec:disc}). Note that if $\mu_{B}$ is close
to one edge, the PDFs will be truncated, so we calculate the
normalisation term numerically, rather than using the analytic value
of $N_{\rm B}= 1/\sigma_{\hat{a}}\sqrt{2\pi}$.

Model C consists of two gaussians distribution with variable relative
height:

\begin{equation} 
P_{\rm C}(\hat{a}) = N_{\rm C}\left[  H_{\rm C1} e^{-{1\over 2}\left( \mu_{C1} -\hat{a} \over \sigma_{\hat{a}} \right)^{2}} + e^{-{1\over 2}\left( \mu_{C2} -\hat{a} \over \sigma_{\hat{a}} \right)^{2}} \right] .
\end{equation}

\noindent There are three free parameters: the two mean spins
$\mu_{C1}$, $\mu_{C2}$ as well as the relative height of the first
gaussian $H_{C1}$. The standard deviation of both gaussians is kept
constant at the value $\sigma_{\hat{a}}$$=$0.05, and $N_{\rm C}$ is a
normalisation term. All three parameters, $\mu_{C1}$ and $\mu_{C2}$
are assigned flat priors in the range 0-1, while due to our ignorance
of scale, $H_{C1}$ is assigned a flat prior in log$_{10}$, in the
range -3 to 0.

For each model, we calculate the `likelihood' $P({\rm data | \{{\it
    x_{\rm 1}, ... x_{\rm n}} \}, model})$ as a function of the
$n$-parameters, $\{{\it x_{\rm 1}, ... x_{\rm n}} \}$:

\begin{eqnarray} 
P _{\rm ADAF}({\rm data | \{{\it x_{\rm 1}, ... x_{\rm n}} \}, model})  = \nonumber \\ 
\prod_{i} P _{\rm ADAF}({\rm data_{i} | \{{\it x_{\rm 1}, ... x_{\rm n}} \}, model}) =  \nonumber \\ 
{1 \over \sigma_{\phi_{\rm LEG}} \sqrt{2\pi}} e^{-{1\over 2}\Sigma_{i}\left( \phi_{i \rm LEG} - \phi_{\rm  ADAF} \over \sigma_{\phi_{\rm LEG}} \right)^{2} } 
\end{eqnarray} 

\noindent and 
\begin{eqnarray} 
P _{\rm QSO}({\rm data | \{{\it x_{\rm 1}, ... x_{\rm n}} \}, model})  = \nonumber \\ 
\prod_{i} P _{\rm QSO}({\rm data_{i} | \{{\it x_{\rm 1}, ... x_{\rm n}} \}, model}) =  \nonumber \\ 
{1 \over \sigma_{\phi_{\rm HEG}} \sqrt{2\pi}} e^{-{1\over 2}\Sigma_{i}\left( \phi_{i \rm HEG} - \phi_{\rm  QSO} \over \sigma_{\phi_{\rm HEG}} \right)^{2} } . 
\end{eqnarray}

\noindent where $\phi_{i \rm LEG}$ and $\phi_{i \rm HEG}$ are the data
points from the observed radio LFs for LEGs and HEGs respectively,
while $\phi_{\rm ADAF}$ and $\phi_{\rm QSO}$ are the modelled radio
LFs of high- and low-accretion rate SMBHs.  These are independent, so
that the combined likelihood is given by:

\begin{eqnarray}
P _{\rm total}({\rm data | \{{\it x_{\rm 1}, ... x_{\rm n}} \}, model})  = \nonumber \\
P _{\rm QSO}({\rm data | \{{\it x_{\rm 1}, ... x_{\rm n}} \}, model}) P _{\rm ADAF}({\rm data | \{{\it x_{\rm 1}, ... x_{\rm n}} \}, model}) 
\end{eqnarray}

The integral of the likelihood over the parameter space $\{{\it x_{\rm
    1}, ... x_{\rm n}} \}$ yields the `evidence', $P({\rm model |
  data})$:

\begin{eqnarray}
P({\rm data | model}) = \nonumber \\
 \int  P(\{{\it x_{\rm 1}, ... x_{\rm n}} \}) P _{\rm total} ({\rm model | \{{\it x_{\rm 1}, ... x_{\rm n}} \}, data}){\rm d}^{\rm n}x,
\label{eq:evidence}
\end{eqnarray}

\noindent and Bayes' theorem allows us to infer the probability of the model given the data: 

\begin{equation}
P({\rm model | data})  = { P({\rm data | model})P({\rm model}) \over P({\rm data})}.
\end{equation}

\noindent Where $P({\rm model})$ is the prior for a model. When
comparing two models, the denominator can be eliminated by taking the
`odds ratio', ${ P({\rm model1 | data}) / P({\rm model2 | data})}$.

\begin{equation}
{ P({\rm model1 | data}) \over P({\rm model2 | data})}= { P({\rm data | model1})P({\rm model1})  \over P({\rm data | model2})P({\rm model2}) }
\label{eq:odds}
\end{equation}

Hence, using the odds ratio, we can compare models two at a time, and
keep the model which fares best agains the others. In
Equation~\ref{eq:evidence}, the integral over the parameter space, and
the priors $P(\{{\it x_{\rm 1}, ... x_{\rm n}} \})$ penalise the use
of extra parameters as well as the exploration of parameter space with
a low likelihood. Hence, models with different number of parameters
can be compared in a fair way.

Finally, we  use the different jet efficiencies $\eta(\hat{a})$ described in
Section~\ref{sec:assum} and shown in Figure~\ref{fig:eta}.
Using these jet efficiencies and our parametrisation we are able to fit the local
radio LF, and Section~\ref{sec:res} presents the results.

\begin{table}
  \begin{center}
    \begin{tabular}{ccc}
      \hline
      \hline
  $\eta$ &     Models  & log$_{\rm 10}$[Odds Ratio] \\
\hline
T10A  &    B / A   & 257\\
T10A   &    C / B  & 100\\
T10A  &   C / A  &  357\\
\hline
  T10B &   B / A   & $>$247 \\
  T10B &   C / B  &  34 \\
T10B   &  C / A  & $>$281\\
\hline
BB09   &    B / A   & $>$177 \\
BB09   &    C / B  & 31 \\
BB09   &   C / A  & $>$209\\
\hline
N07    &   B / A   & $>$261 \\
 N07   &   C / B  & 51 \\
 N07   &  C / A  & $>$311 \\

\hline
HK06   &    B / A   & $>$161 \\
HK06   &     C / B  &  3\\
HK06   &    C / A  & $>$163 \\
\hline
M05    &   B / A   & $>$264\\
 M05   &   C / B  &   54\\
M05    &  C / A  & $>$319 \\
      \hline
      \hline
    \end{tabular}
  \end{center}
\caption{\noindent Summary of results of the model selection. Section~\ref{sec:modelsel} describes the models and the model selection method. The
  first column indicates the two models being compared, with the most
  likely model on the left. The second column shows the base-10
  logarithm of the odds ratio (given by Equation~\ref{eq:odds}, see
  Section~\ref{sec:modelsel} for more details).  The right-censored results marked with a $>$ occur due to the fact that the evidence of model~A is $<$$10^{-324}$, too small for the program.} 
    \label{tab:mod_comp}
\end{table}

\begin{table}
  \begin{center}
    \begin{tabular}{ccccc}
      \hline
      \hline
$\eta$ & Population & $\mu_{C1}$ & $\mu_{C2}$ &  $H_{C1}$ \\
\hline
T10A &   ADAF-LEGs & 0.99  & 0.29  & 0.316 \\
T10A &  QSO-HEGs &  0.70 & 0.01  & 0.079 \\
\hline
T10B &   ADAF-LEGs & 0.99  & 0.99  & - \\
T10B &  QSO-HEGs &  0.99 & 0.00  & 0.398 \\
\hline
BB09 &   ADAF-LEGs & 0.89  & 0.08  & 0.631 \\
BB09 &  QSO-HEGs &  0.00 & 0.00  &  - \\
\hline
N07 &   ADAF-LEGs&  0.99 & 0.16  & 0.501 \\
N07 &  QSO-HEGs&  0.99 &  0.00 &  0.040 \\
\hline
HK06 &   ADAF-LEGs &  0.99 & 0.06  & 0.794 \\
HK06 &  QSO-HEGs &  0.00 & 0.00 & - \\
\hline
M05 &   ADAF-LEGs&  0.09  & 0.99  & 0.100 \\
M05 &  QSO-HEGs  & 0.99 & 0.09  & 0.398 \\
     \hline
      \hline
    \end{tabular}
  \end{center}
\caption{\noindent Summary of best-fitting parameters for model C, chosen ahead of the other two models. The quasars (QSOs) have been modelled using the X-ray LF as described in Section~\ref{sec:hi_acc} and the parameters are those that best fit the observed radio LF of HEGs. The ADAFs are modelled using the local BHMF, as described in Section~\ref{sec:low_acc} and have been compared to the observed radio LF of LEGs. }
    \label{tab:best_param}
\end{table}

\begin{table*}
  \begin{center}
    \begin{tabular}{ccccccc}
      \hline
      \hline
      Parameters for model A  \\
$\eta$ & $p_{\rm~ADAF}$ & $p_{\rm~QSO}$ & $\langle \hat{a} \rangle_{ADAF}$  & $\langle \hat{a} \rangle_{QSO}$  & $\langle \hat{a} \rangle_{SMBH~z=0}$ & $\langle \hat{a} \rangle_{SMBH~z=1}$ \\
\hline
T10A &  0.5 & 1.3 & 0.34 & 0.05 & 0.32 & 0.20\\
T10B &  -4.0 & 0.7 & 0.83 & 0.26 & 0.42 & 0.27 \\
BB09 & 0.3 & 4.0 & 0.41 & 0.00 & 0.25 & 0.04 \\
N07 &  -0.6 & 4.0 & 0.61 & 0.00 & 0.52 & 0.16 \\
HK06  & -3.2 & 4.0 & 0.81 & 0.00 & 0.59 & 0.12 \\
M05 & -4.0 & 0.5 & 0.83 & 0.34 & 0.71 & 0.42 \\
     \hline
  \hline
      Parameters for model B \\
    $\eta$ &  $\mu_{\rm B~ADAF}$ & $\mu_{\rm B~QSO}$   & $\langle \hat{a} \rangle_{SMBH~z=0}$ & $\langle \hat{a} \rangle_{SMBH~z=1}$ \\

\hline
T10A & 0.48 & 0.13 & 0.32 & 0.16 \\
T10B & 0.04 & 0.49 & 0.49 & 0.49 \\
BB09 & 0.85 & 0.04 &  0.61 & 0.15 \\
N07 & 0.79 & 0.04 &  0.53 & 0.12 \\
HK06 & 0.91 & 0.04 &   0.66 & 0.16 \\
M05 & 0.96 & 0.73 &   0.92 & 0.78 \\
      \hline
      \hline
    \end{tabular}
  \end{center}
\caption{Summary of best-fitting parameters for the rejected models: A and B. The model selection and parameter-fitting procedures are described in Section~\ref{sec:low_acc}.  Models~A and B show the same trend of evolution in \amean\, with $z$ as seen in model~C.}
    \label{tab:rej_models}
\end{table*}

\section{The distribution of spins from  the local radio LF}\label{sec:res}

\subsection{Comparison to the $z$$=$0 radio LF}\label{sec:z0}

\begin{figure*}
%\begin{center}
\hbox{ \psfig{file=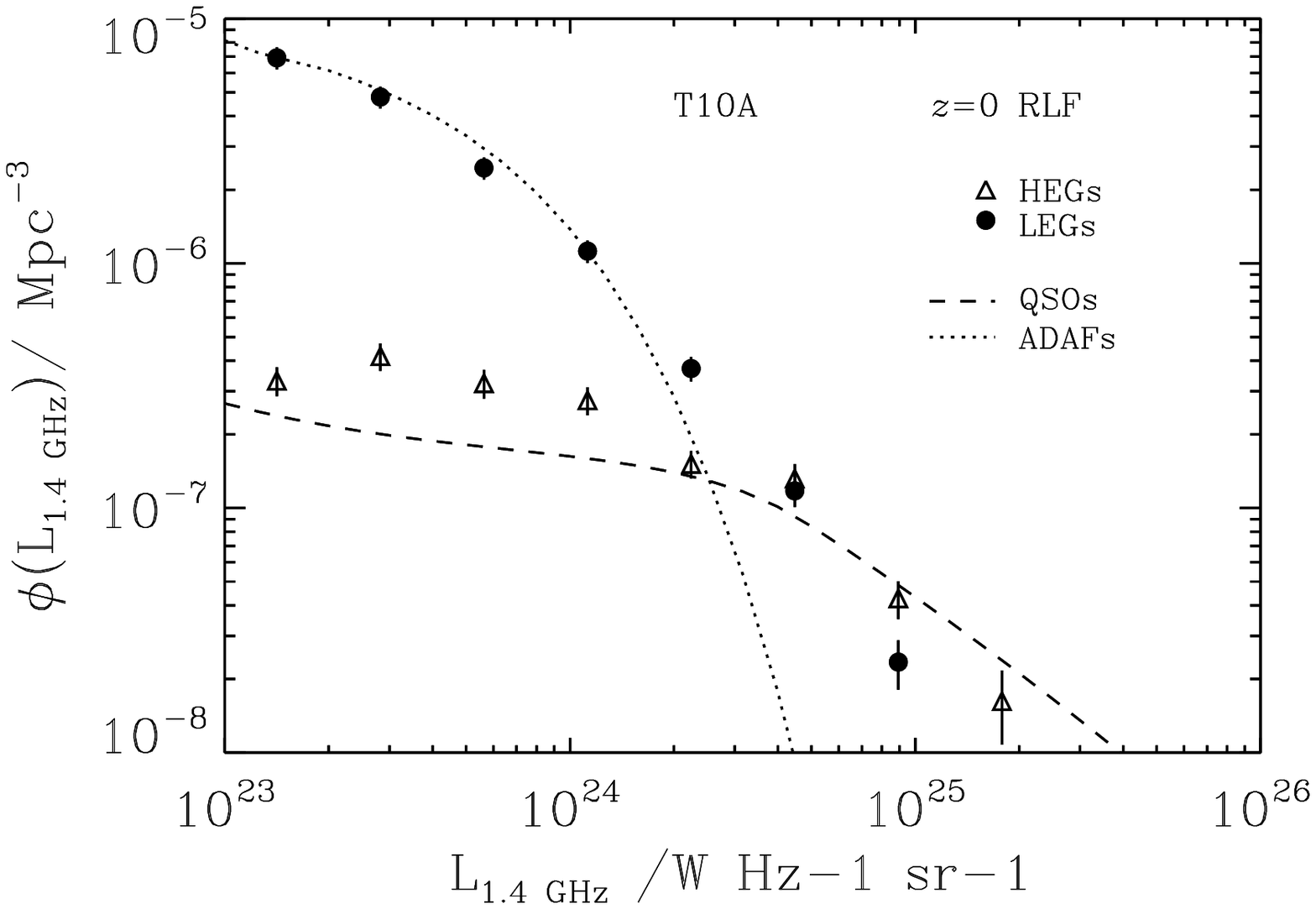, width=8cm, angle=0} \psfig{file=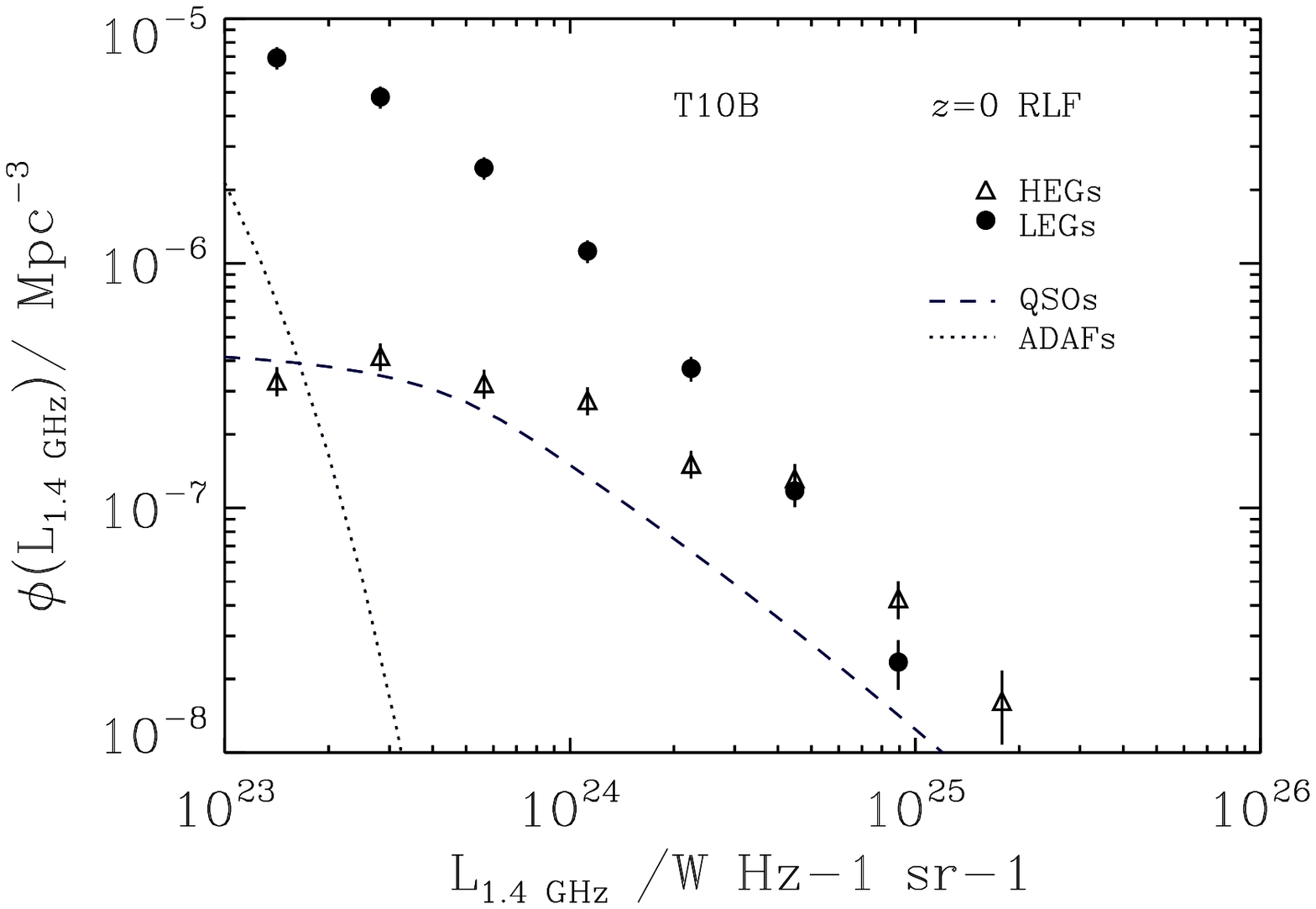, width=8cm, angle=0} }
\hbox{ \psfig{file=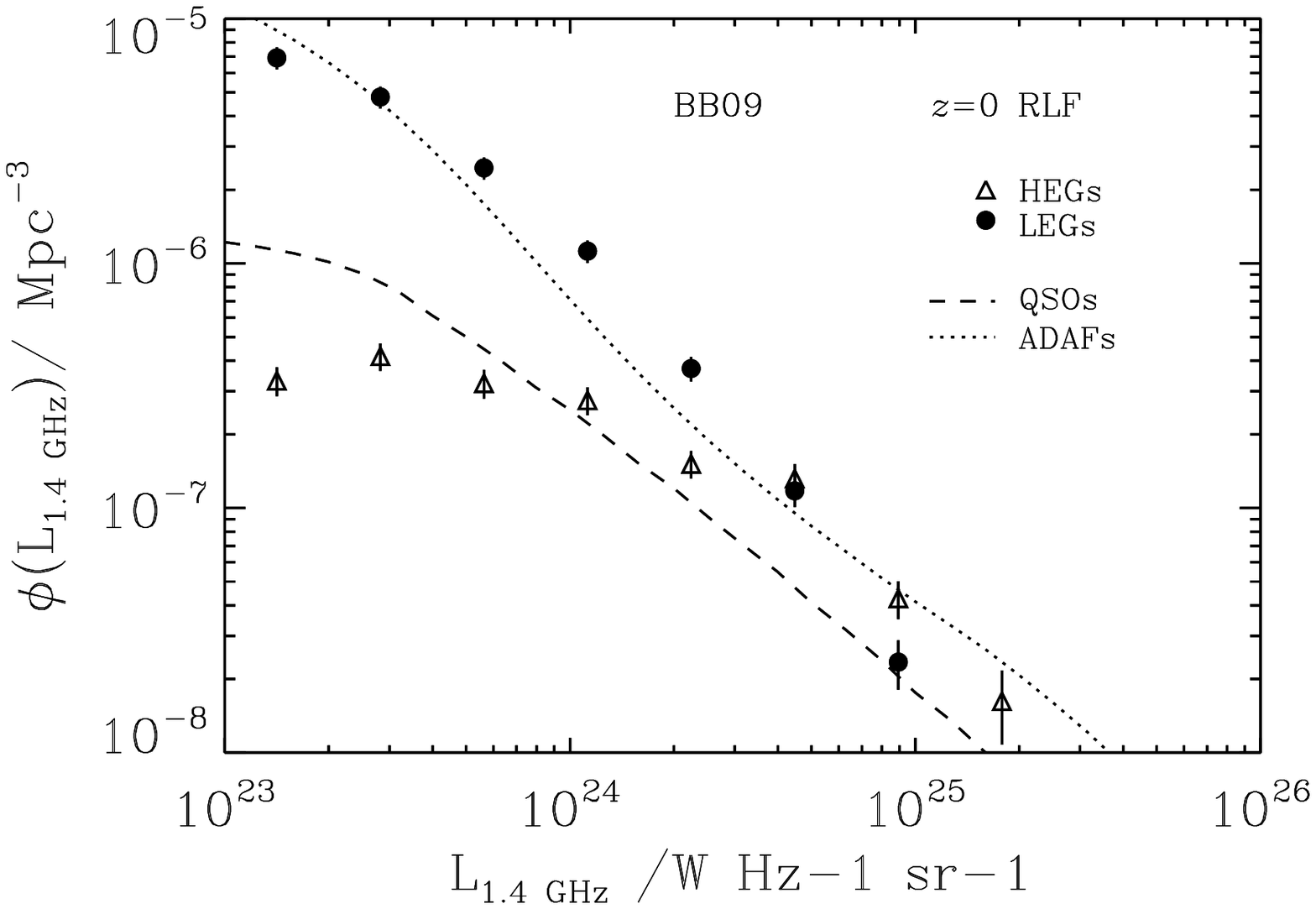, width=8cm, angle=0} \psfig{file=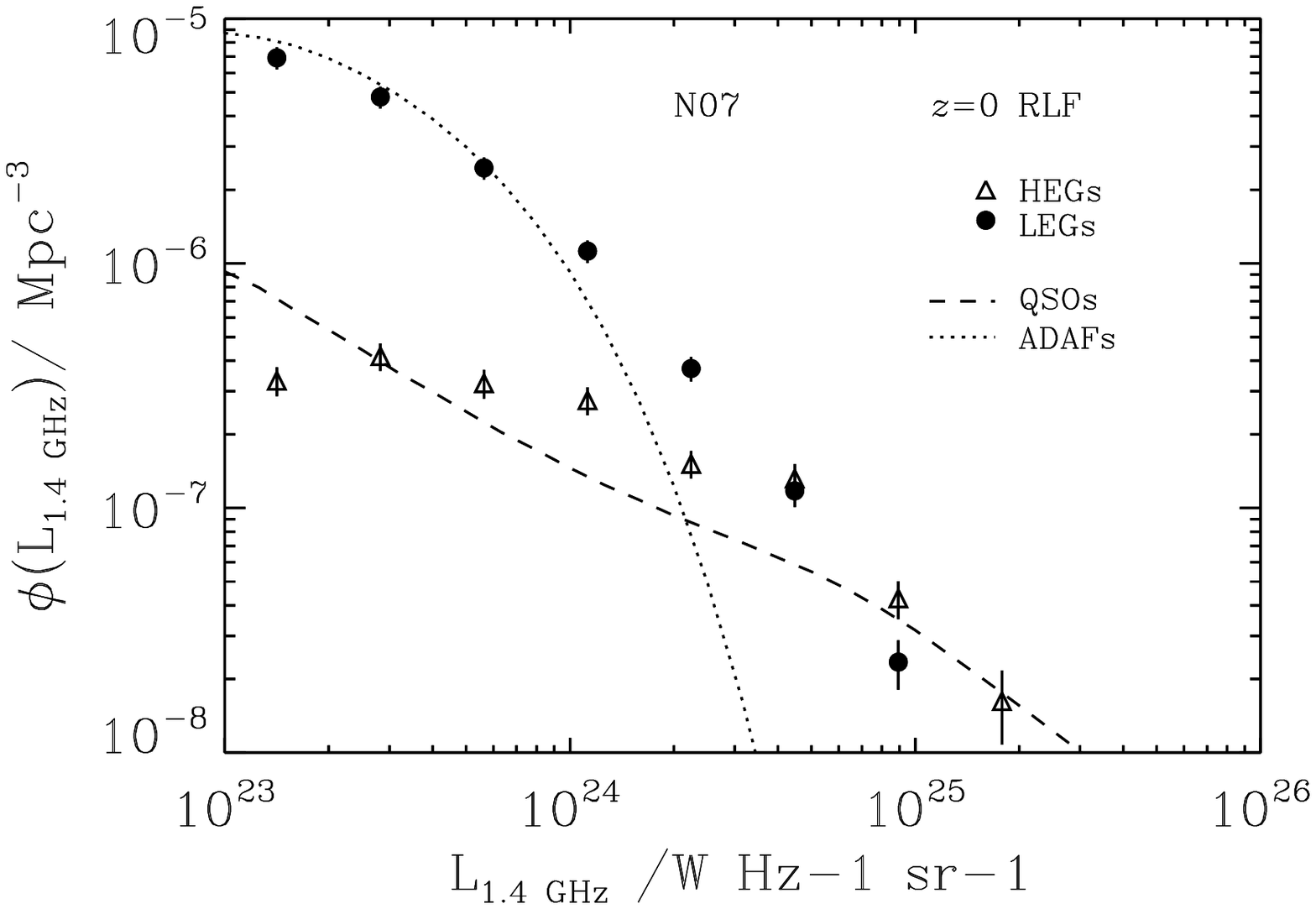, width=8cm, angle=0} }
 \hbox{ \psfig{file=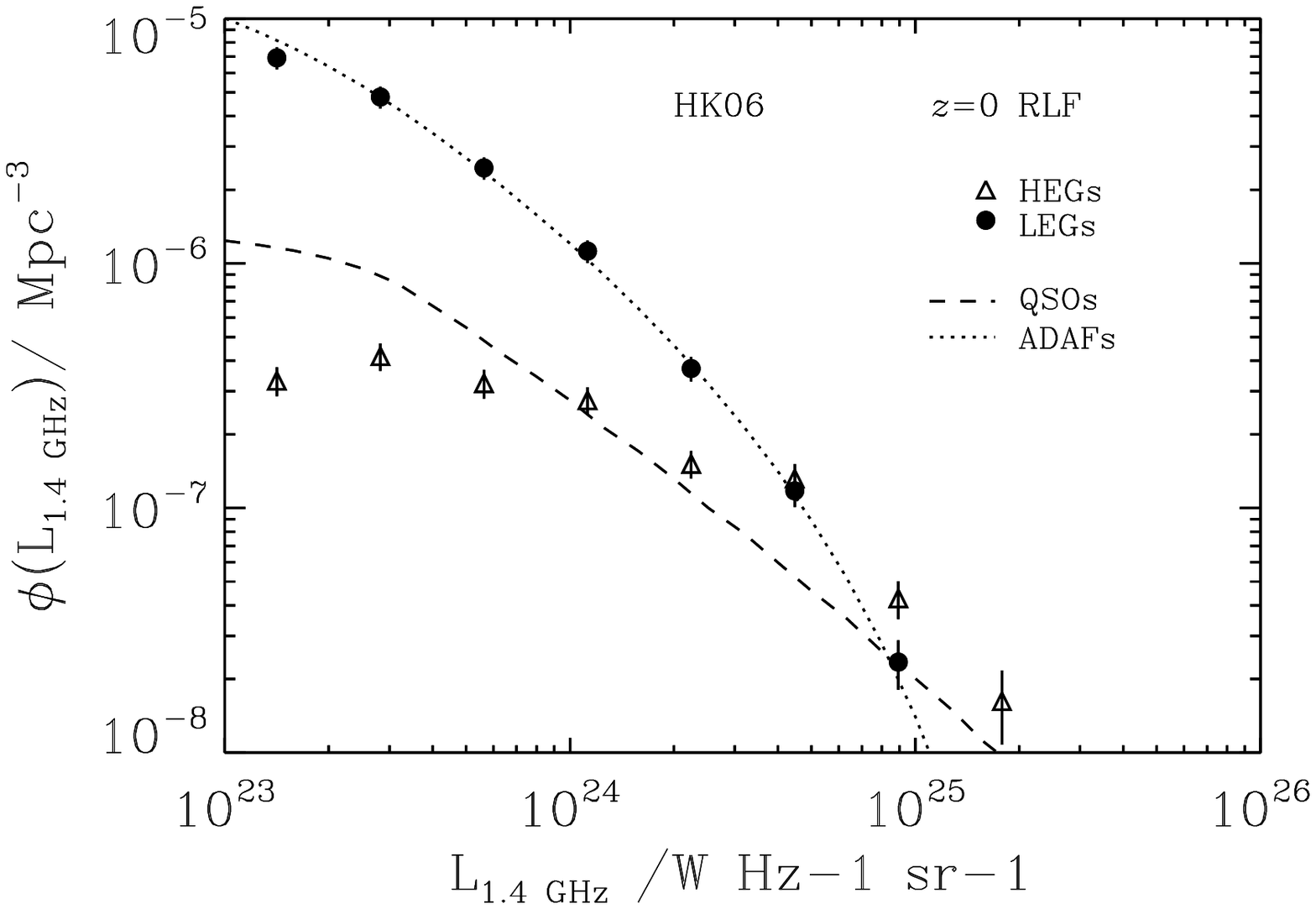, width=8cm, angle=0} \psfig{file=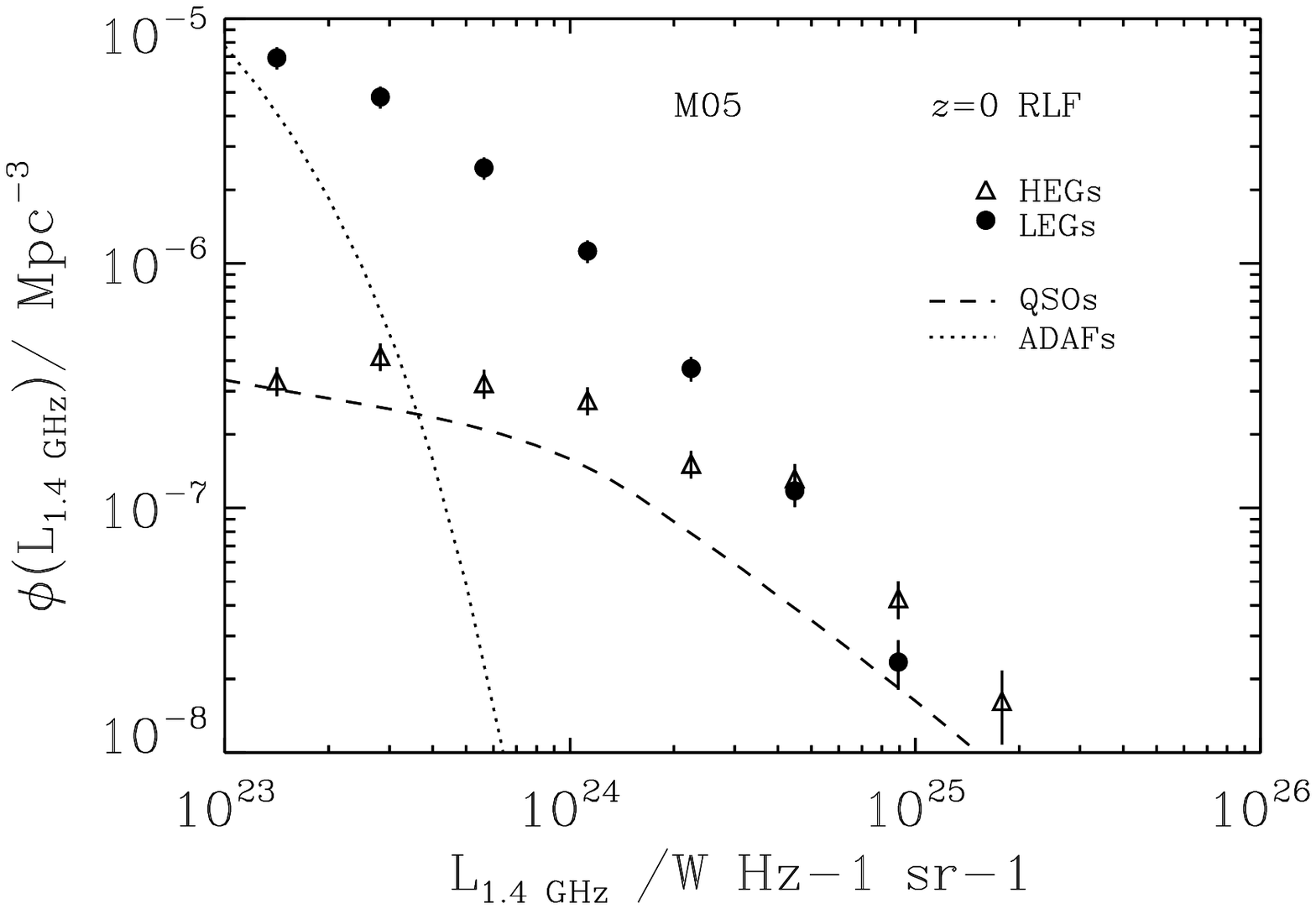, width=8cm, angle=0} }
\caption{\noindent The $z$$=$0 radio luminosity function split into
  high-excitation gas (HEGs, open triangles) and low-excitation gas
  sources (LEGs, black circles, data from P. Best, et~al., in
  prep.). The empty squares indicate the summed radio LF. The
  acronyms summarise the jet efficiencies $\eta$ shown in
  Figure~\ref{fig:eta}. The curves show the best-fitting curves from
  our most appropriate model (model C), chosen using the odds ratio as
  described in Section~\ref{sec:modelsel}.  The dashed line shows the
  radio LF modelled from high-accretion rate SMBHs (labelled QSOs, for
  simplicity), and is compared to the HEGs, while the dotted line
  shows the radio LF modelled from low-accretion rate objects
  (labelled ADAFs) which is compared to the LEGs.  The model has three
  free parameters which describe the spin distribution. These
  parameters are fixed by comparison with the data.  The figure shows
  how the two components of the radio LF (HEGs and LEGs) can be
  explained by a population of high-accretion rate SMBHs and one of
  low-accretion rate SMBHs, different distributions of spins.
  }
\label{fig:rlf_gaus2}
%\end{center}
\end{figure*}

To constrain the best-fitting model and parameters, we use the data kindly provided by P. Best prior to publication \citep[Best et al. 2011, in prep., from the original sample of ][]{2005MNRAS.362....9B}, who have separated the local radio LF into HEGs and LEGs.  These are plotted as
empty triangles and filled circles, respectively, in Figure~\ref{fig:rlf_gaus2}.

Using Equation~\ref{eq:odds}, and assuming equal priors for all
models, $P({\rm model A})$$=$$P({\rm model B})$$=$$P({\rm modelC})$,
we select the best model, which we find to be model C (see
Table~\ref{tab:mod_comp}).  We note that model~C has been chosen
despite the Ockham factors that penalise the exploration of extra
parameters.

Once we have chosen a model, the best-fitting parameters are chosen by
maximising the likelihood, which is equivalent to the traditional
$\chi^{2}$ minimisation. We find the best-fitting parameters for SMBHs
with high and low accretion rates (`QSOs' and `ADAFs' respectively)
separately, and Table~\ref{tab:best_param} summarises the results.  Table~\ref{tab:rej_models} summarises the best-fitting parameters for the rejected models.

Figure~\ref{fig:rlf_gaus2} shows that the best-fitting parameters can
accurately reproduce the local radio LF of LEGs and HEGs in the region
$L_{\nu 1.4}\grtsim 3\times10^{23}$~\whzsr. The region
1-3$\times10^{23}$~\whzsr, is not so well fit, but a few points are
worth mentioning. Although the fits look better if we only fit the
region $\grtsim 3\times10^{23}$~\whzsr, the best-fitting parameters
are essentially unchanged.

\begin{figure*}
%\begin{center}
\hbox{ \psfig{file=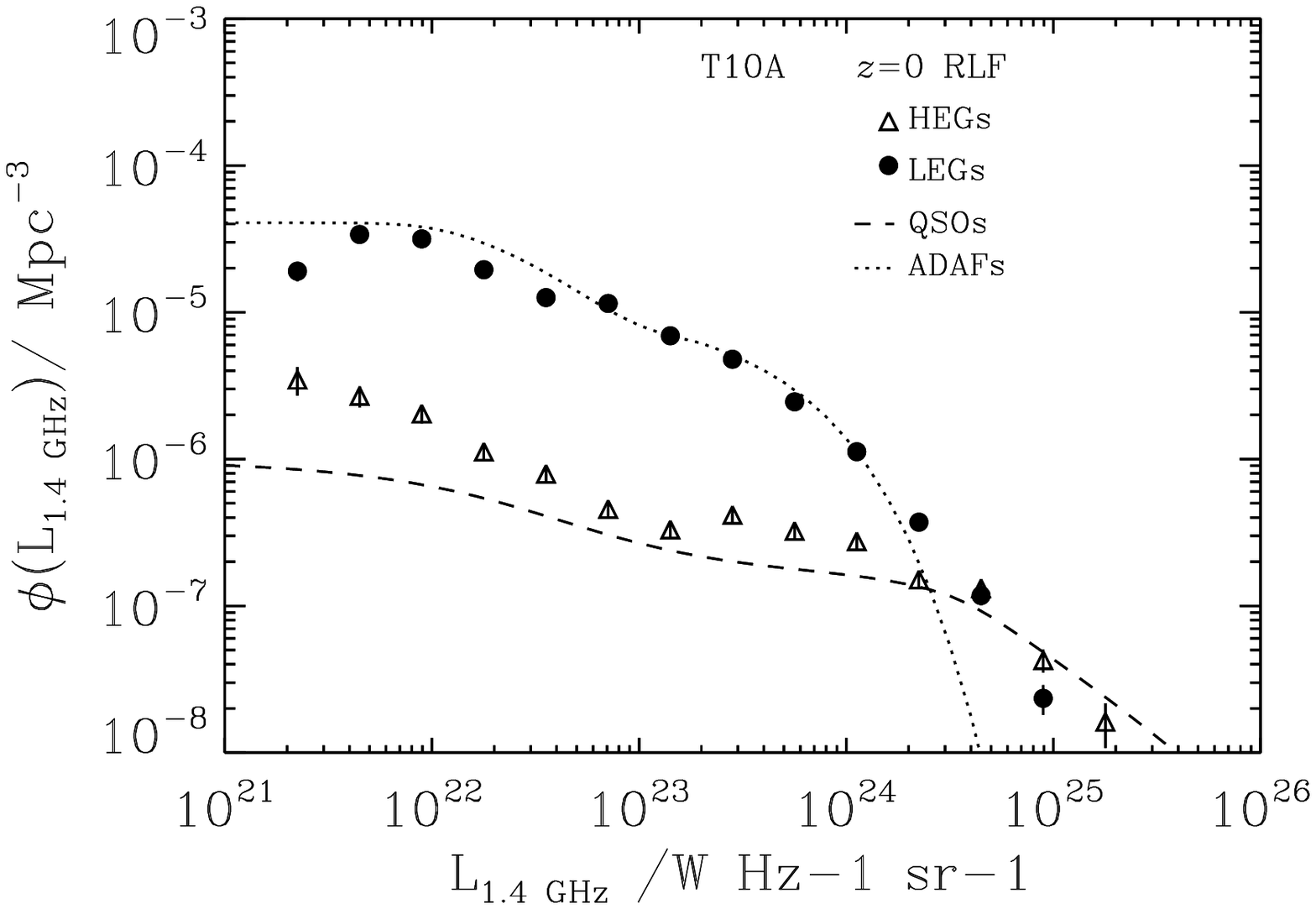, width=8cm, angle=0} \psfig{file=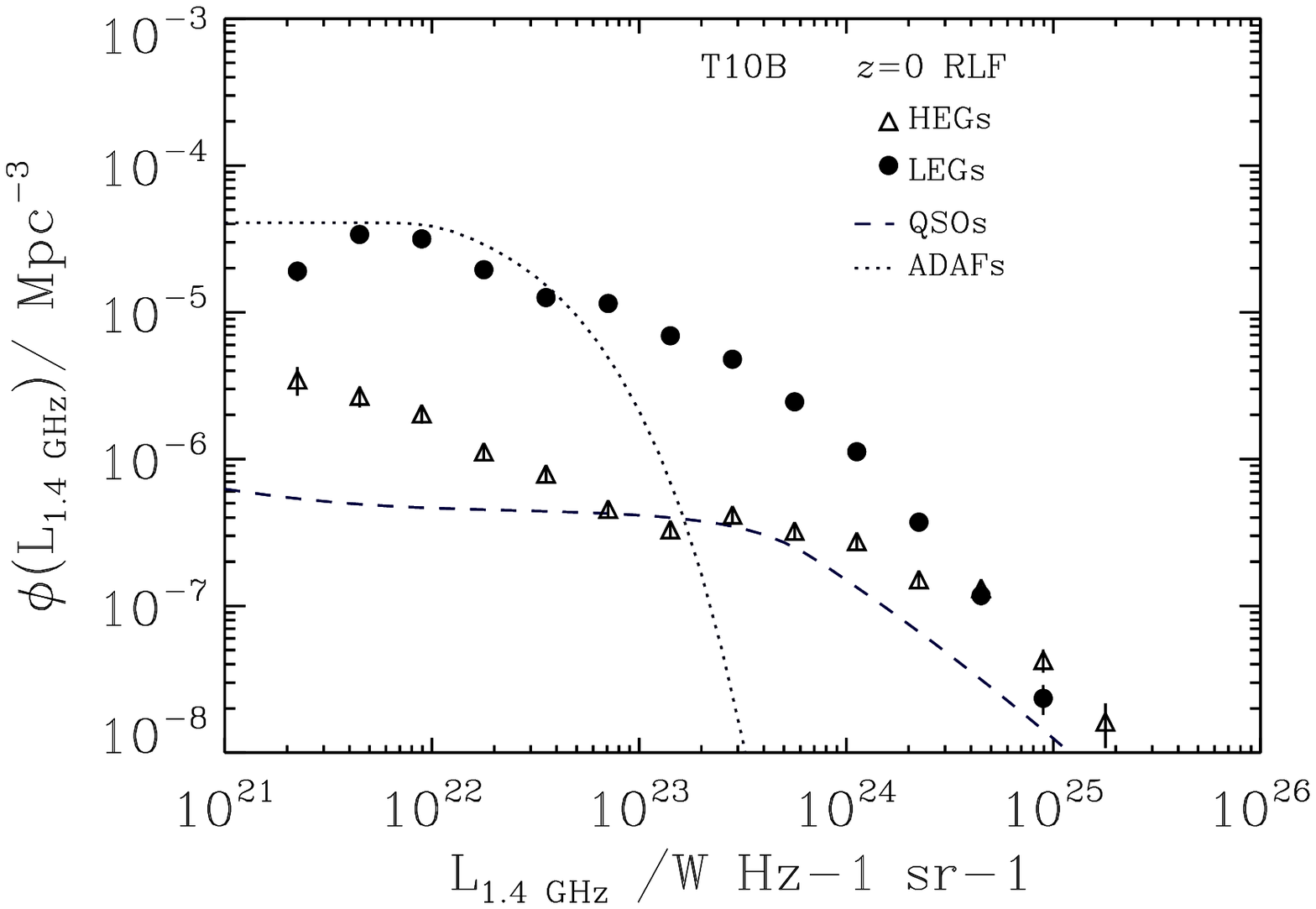, width=8cm, angle=0} }
\hbox{ \psfig{file=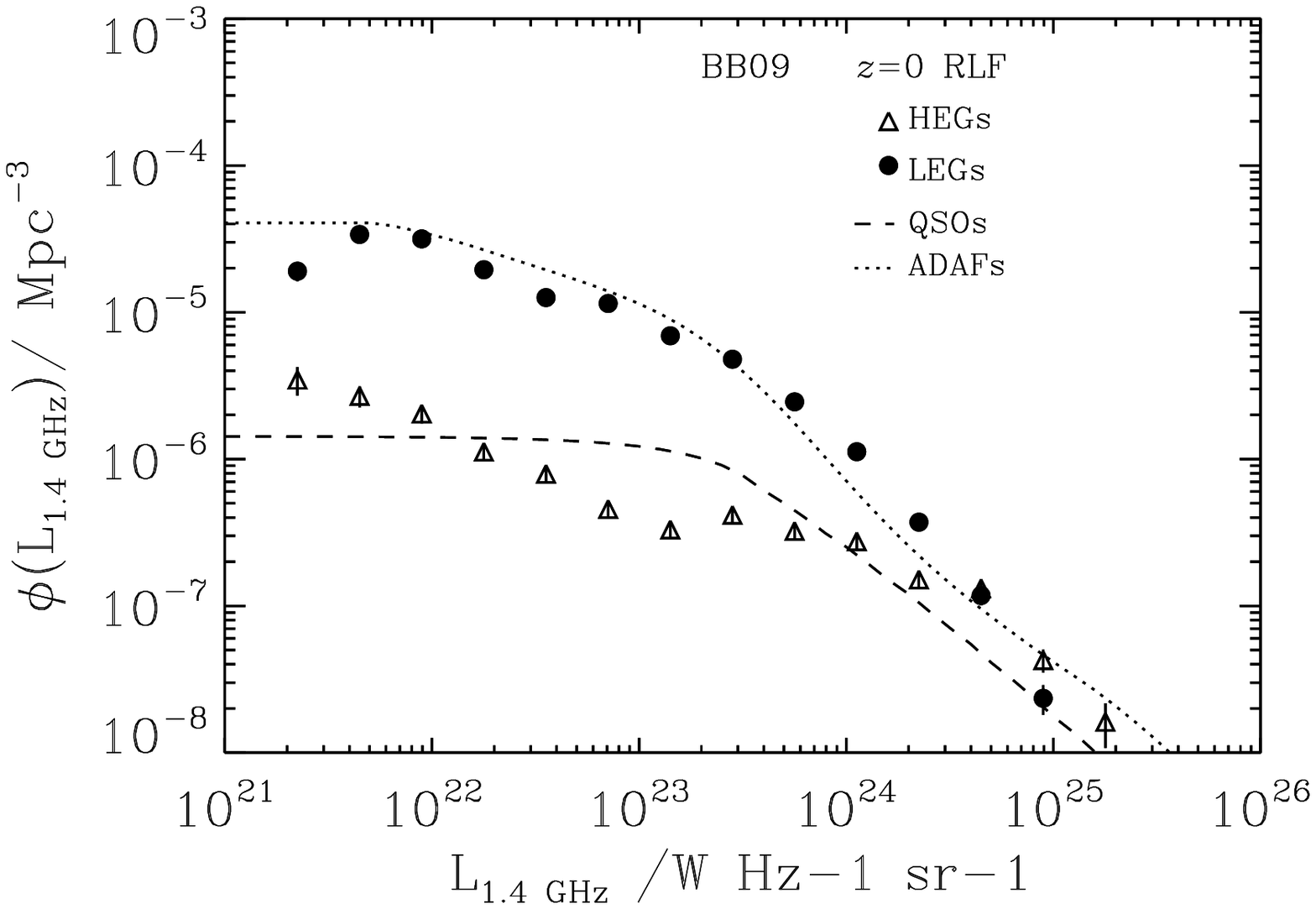, width=8cm, angle=0} \psfig{file=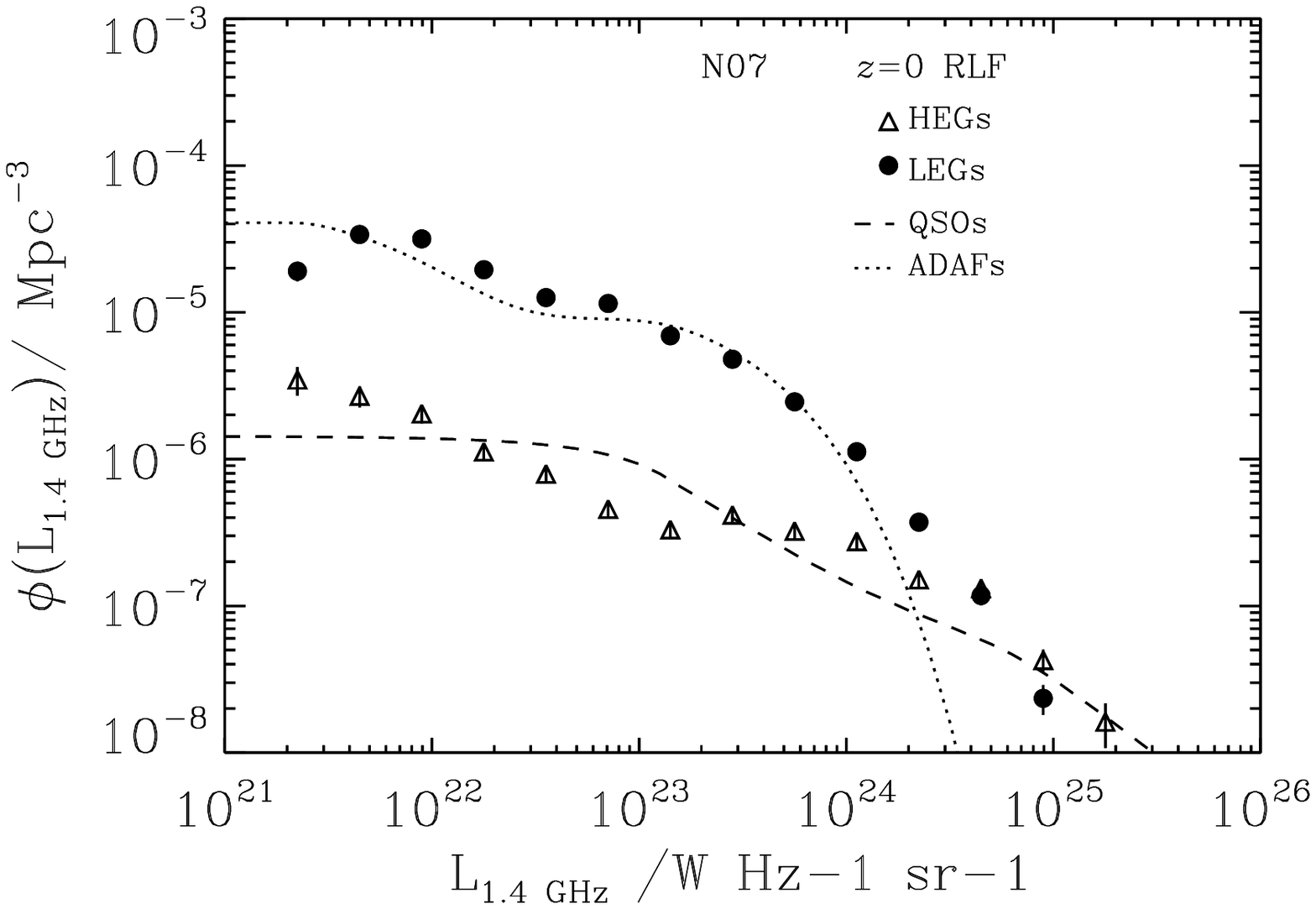, width=8cm, angle=0} }
 \hbox{ \psfig{file=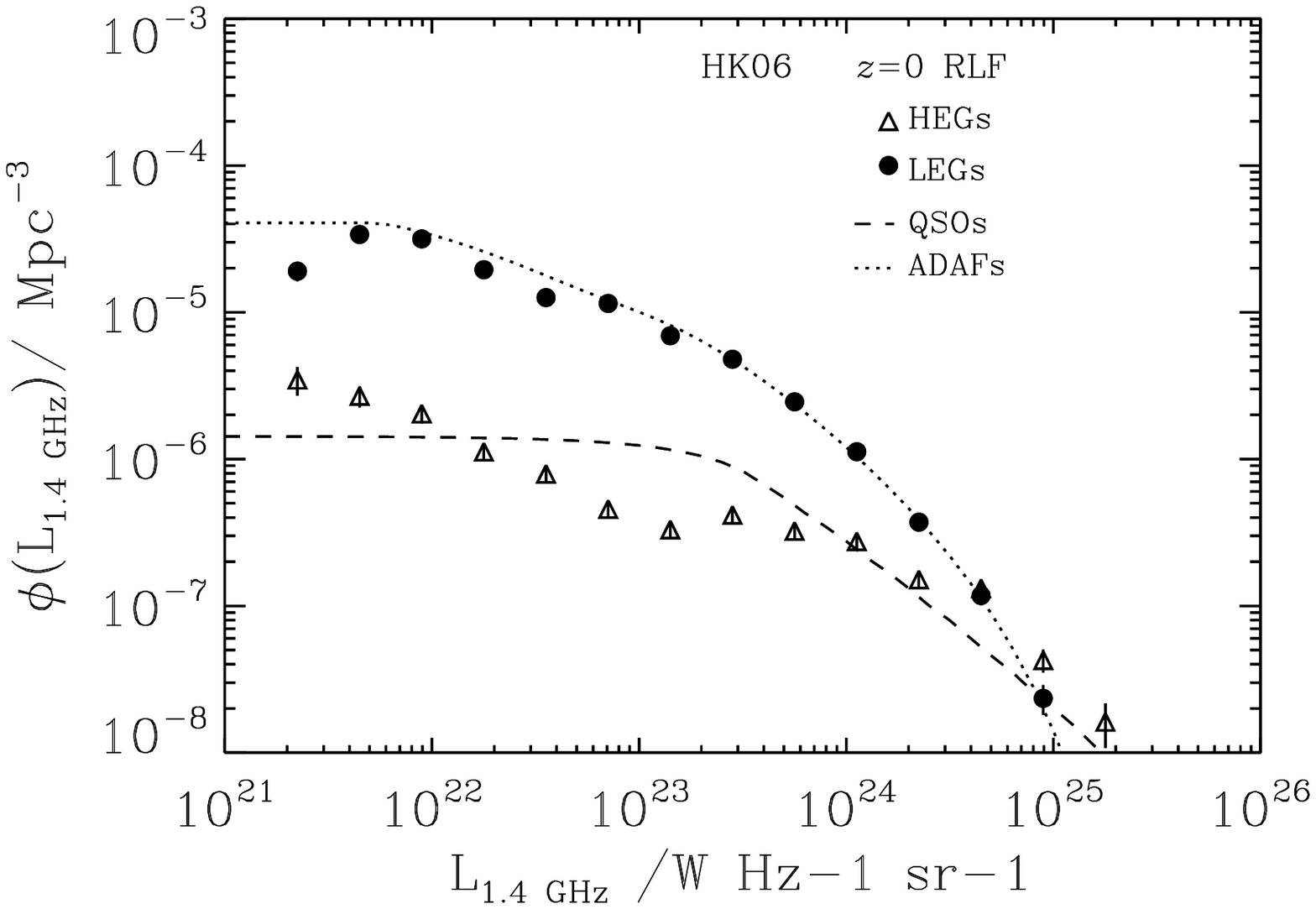, width=8cm, angle=0} \psfig{file=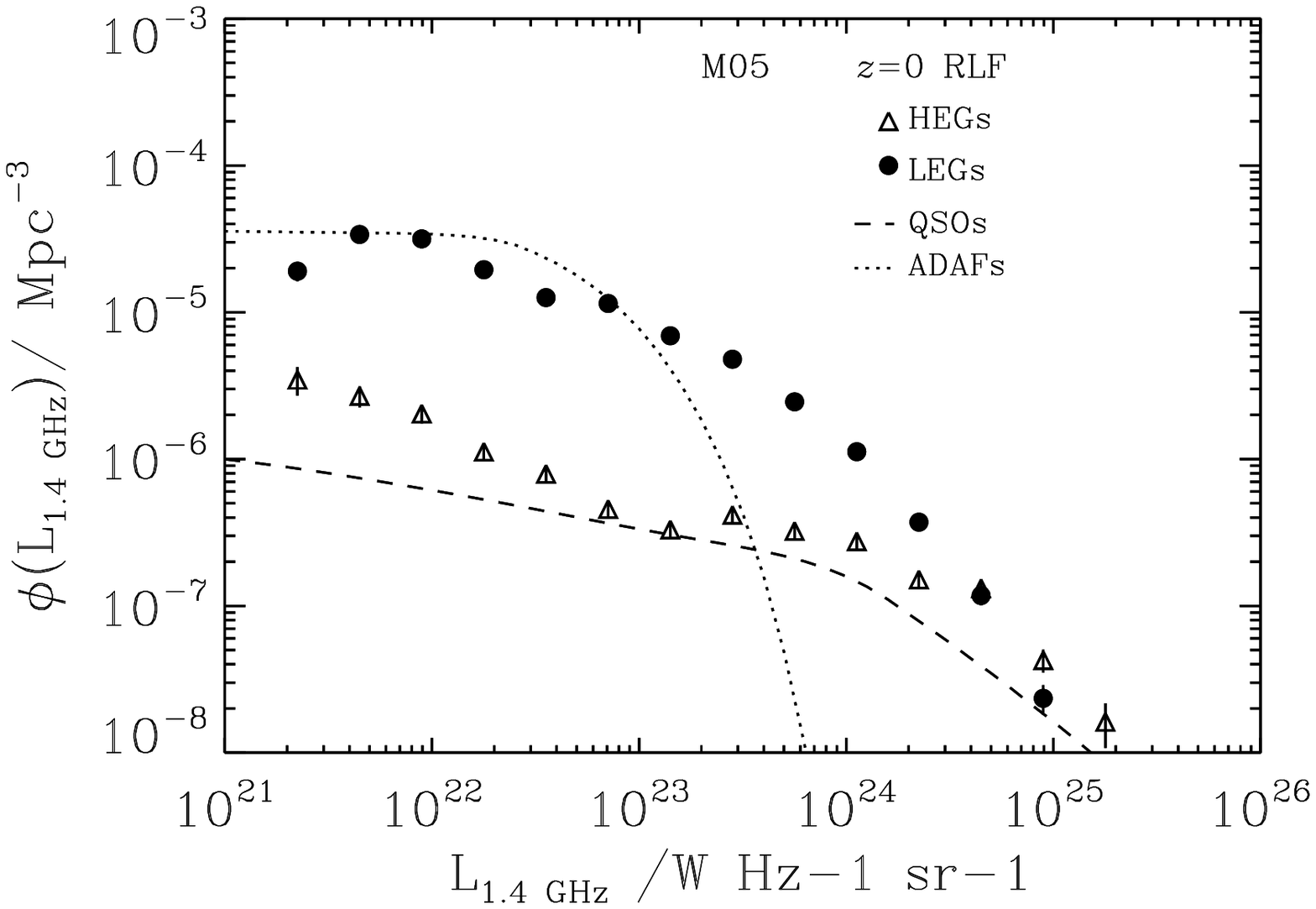, width=8cm, angle=0} }
\caption{\noindent The entire observed $z$$=$0 radio luminosity function down to $10^{21}$~\whzsr\, compared to our models with their best-fitting parameters.  This figure is similar to Figure~\ref{fig:rlf_gaus2}, but it shows a wider range in radio luminosity density. The parameters have only been fit in the region $\geq10^{23}$~\whzsr, so we are now comparing the predictions from our best-fitting models to the observed radio LF. We can see that the ADAF models provide a very good match to the LEGs all the way down to luminosity densities of $10^{21}$~\whzsr. There is no need for a further component to explain the LEGs, so our model predicts that the LEGs down to $10^{21}$~\whzsr\, are all powered by SMBHs with \mbh$\grtsim 10^{8}$~\msol, with a negligible contribution from lower mass SMBHs.  The models do not provide such good fits for the HEGs, indeed an extra component is required to explain the high space density of HEGs at luminosity densities $<10^{23}$~\whzsr. The most likely explanation is that SMBHs with masses below $10^{8}$~\msol\, and moderate accretion rates dominate this region of  the HEG radio LF \citep[these sources are typically Seyfert galaxies and low-luminosity AGN, LLAGN, e.g.][]{2005A&A...435..521N,2006A&A...451...71F}.  }
\label{fig:entire_rlf}
%\end{center}
\end{figure*}

\begin{figure*}
%\begin{center}
\hbox{ \psfig{file=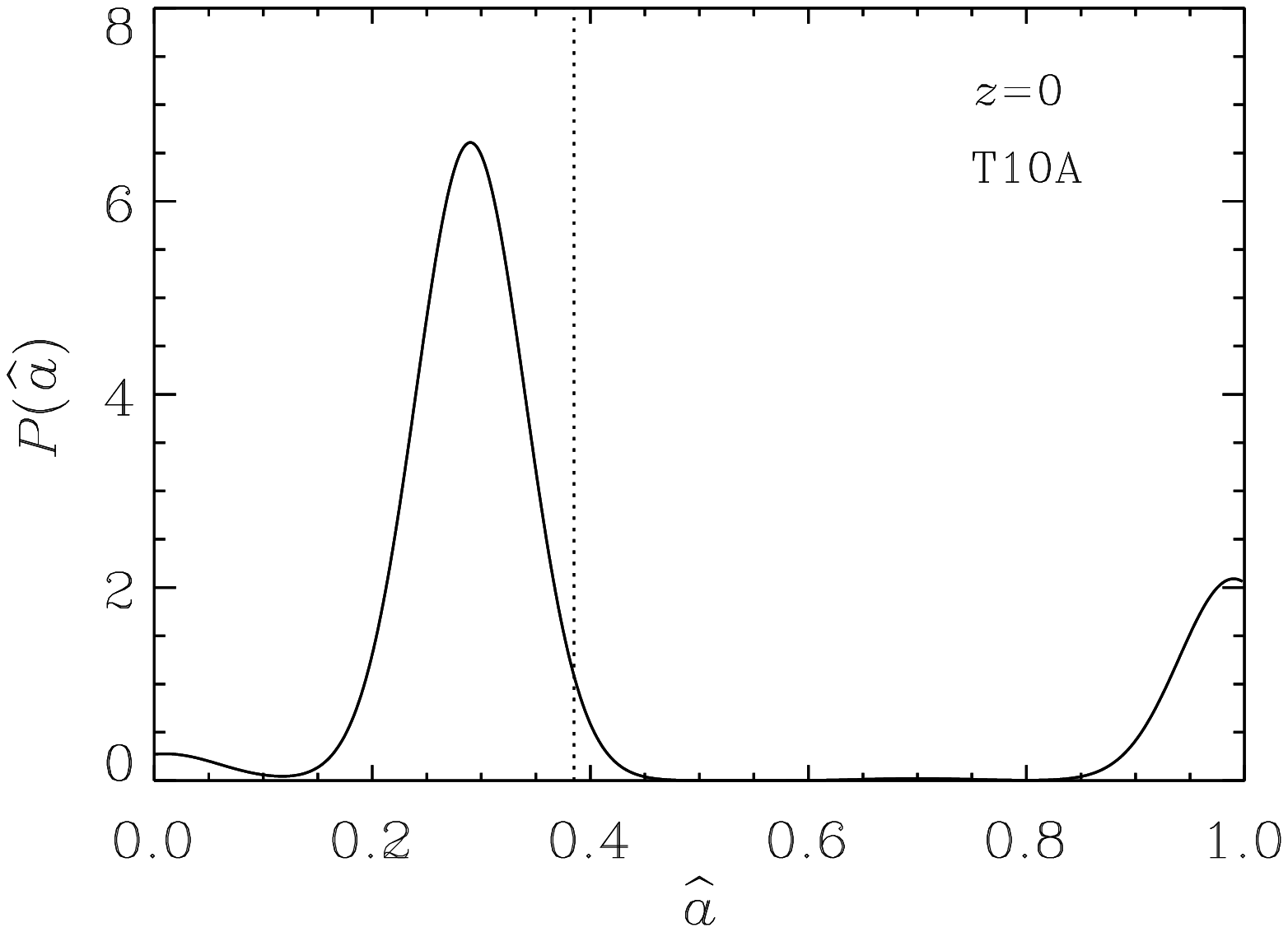, width=8cm, angle=0} \psfig{file=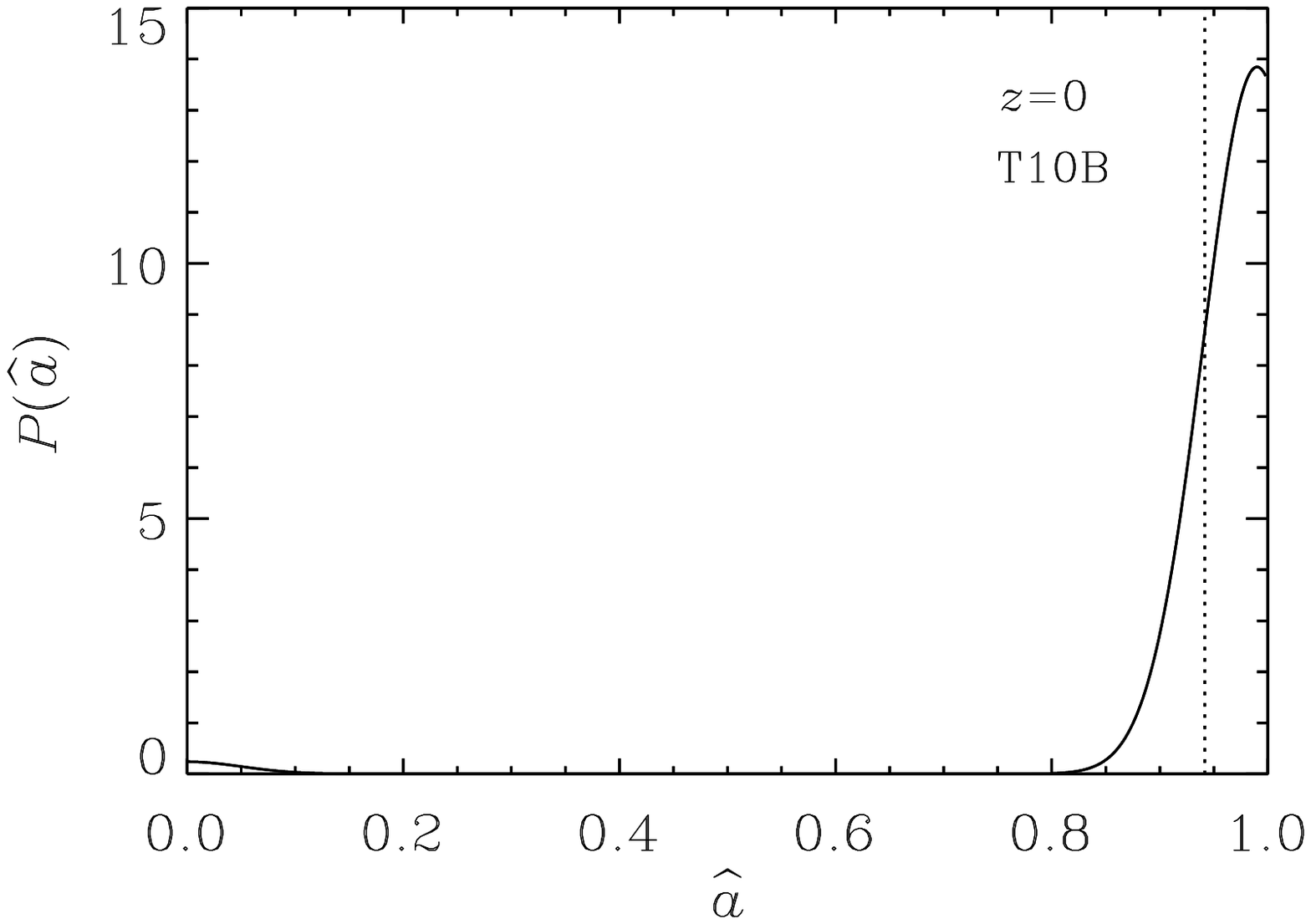, width=8cm, angle=0} }
\hbox{ \psfig{file=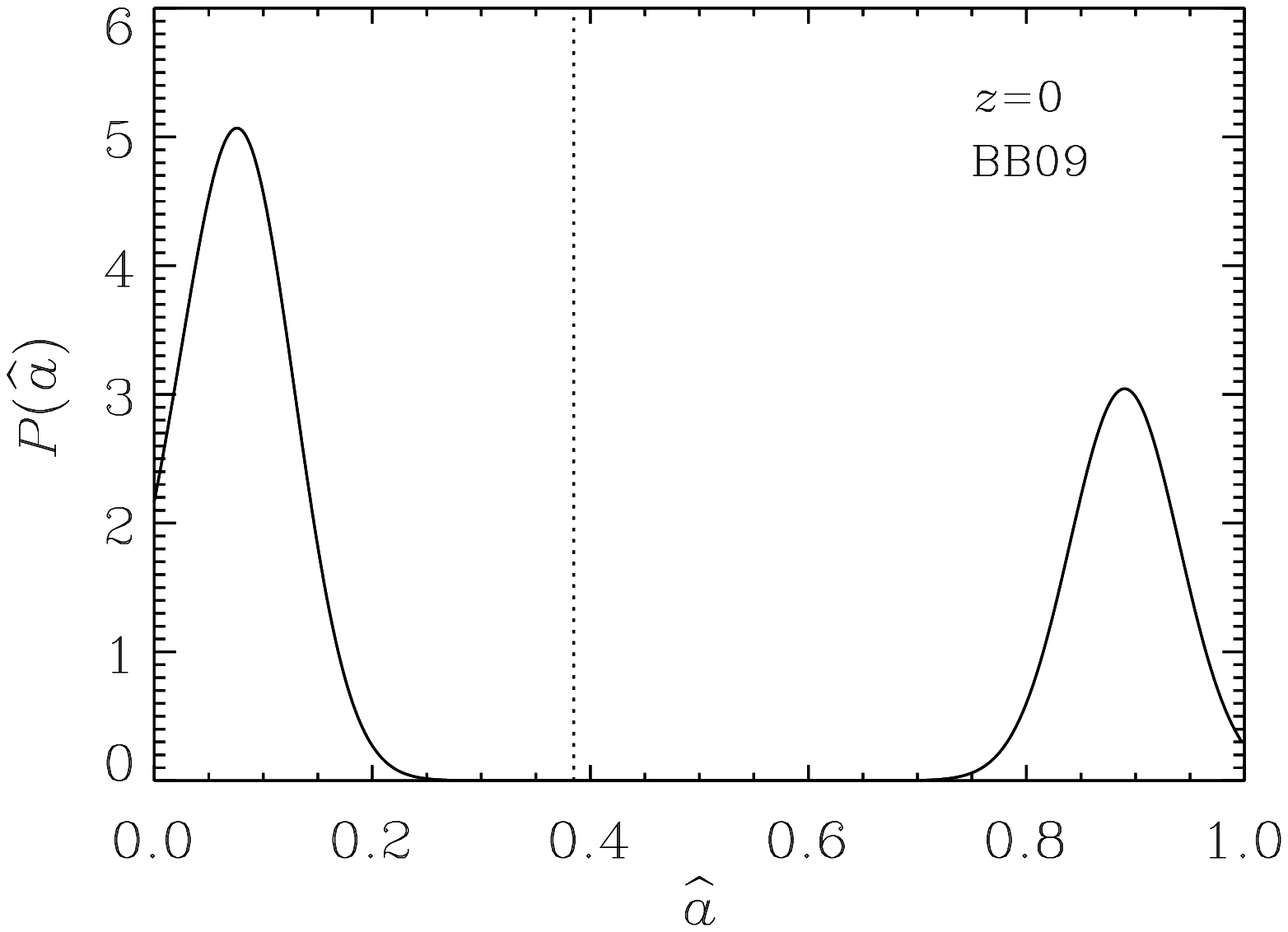, width=8cm, angle=0} \psfig{file=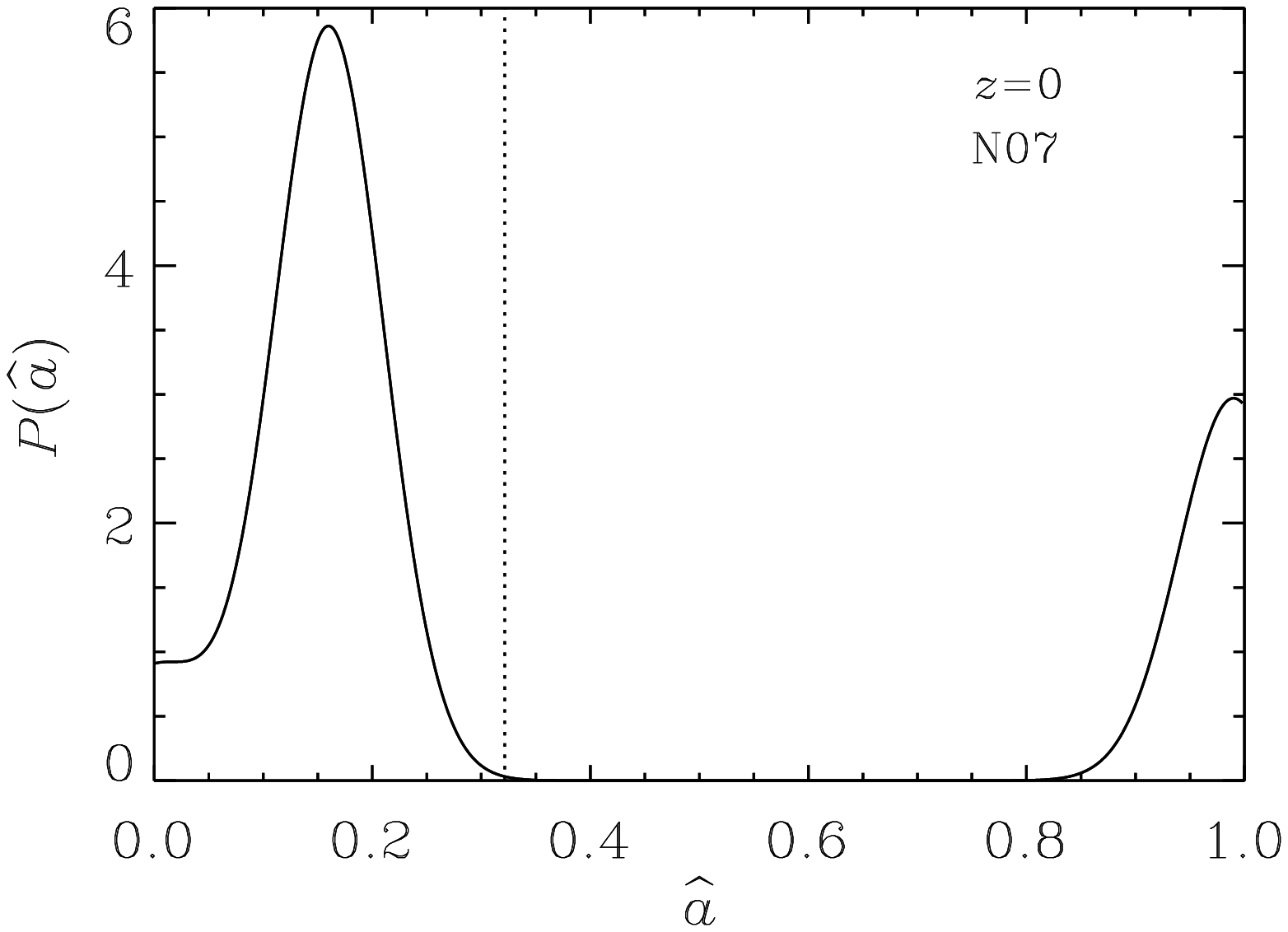, width=8cm, angle=0} }
 \hbox{ \psfig{file=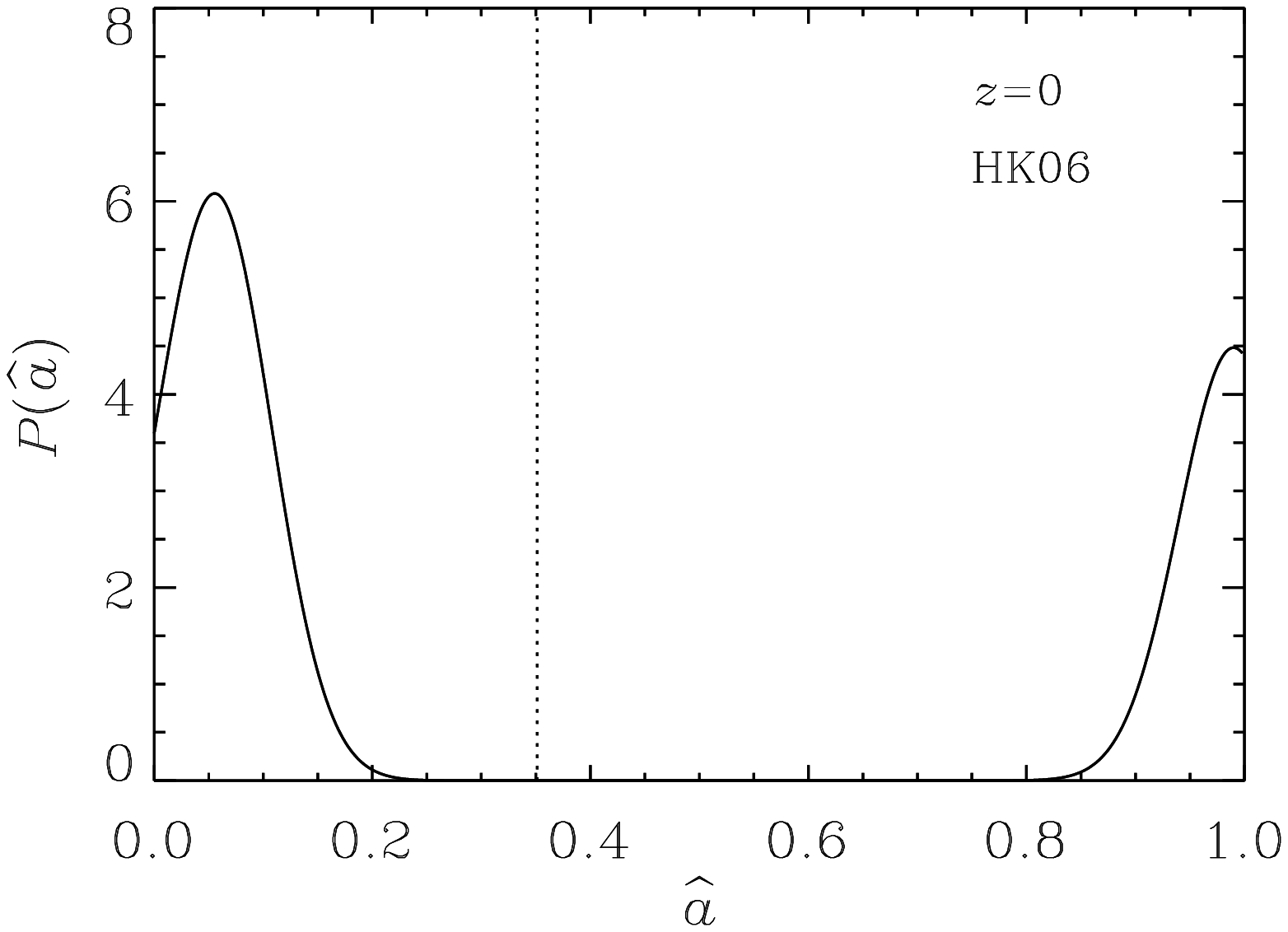, width=8cm, angle=0} \psfig{file=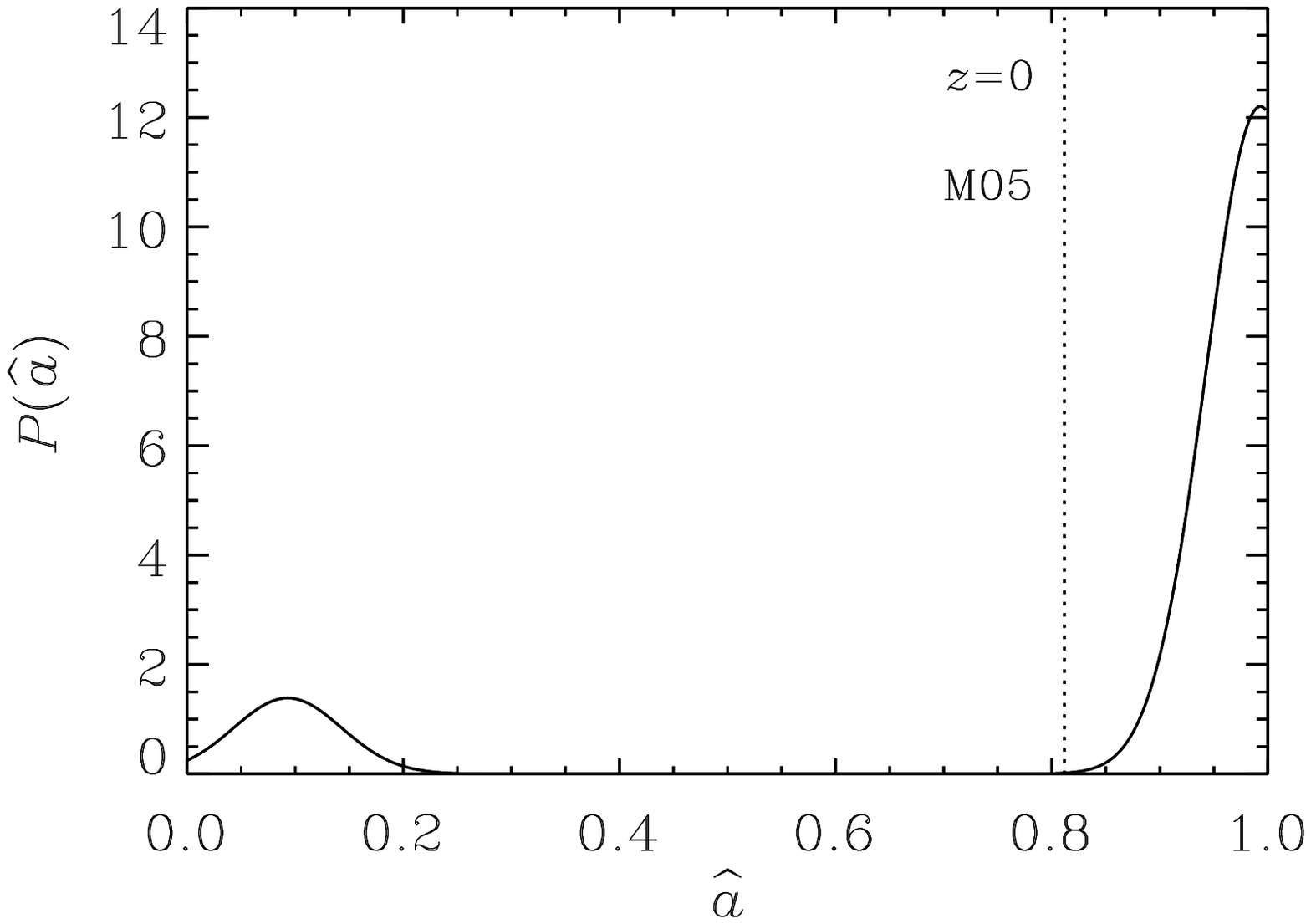, width=8cm, angle=0} }
\caption{\noindent Spin distributions at $z$$=$0 for all SMBHs resulting from fitting the two Gaussian model to the local radio luminosity function. For each of the jet-efficiencies, the distribution shown is the combination of the ADAF-LEG and the QSO-HEG components quoted in Table~\ref{tab:best_param}, weighted by their relative space densities. Since the observed space density of LEGs is higher than that of HEGs, the distributions shown here are dominated by the ADAF-LEG component, with a small contribution from the QSO-HEG component. Note that the M05 distribution has a component peaking at \sp$\sim$0.16, with a very small amplitude. }
\label{fig:spin_dbn_z0}
%\end{center}
\end{figure*}

Some 
models for the efficiencies do a remarkable job of explaining the radio 
LF, particularly the LEGs (e.g. BB09, HK06).  Models that cannot reach
very high values of $\eta$ struggle to fit the LEGs, but can still 
fit the HEGs reasonably well. For example the shape of the QSO 
model function from M05 is almost identical to the HEG LF, although the
normalisation is slightly off. 

We have assumed a value of $f$$=$20, but if we
change this to $f$$=$1 (at all radio luminosities)  the T10B and M05 models  reproduce the LEGs  better, and provide a reasonable fit. In this case the M05 model requires all the spin to lie around \sp$=$0.14 for the HEGs, while the LEGs require a bimodal distribution (with means of $\mu_{C 1}$$=$0.86 and $\mu_{C 2}$$=$0.10). The T10B requires bimodal distributions for both the HEGs (0.33 and 0.00 respectively).  The T10A, BB09, N07 and HK06, on the other hand,  require all the probability density to lie around zero spin (both $\mu_{C 1}$ and $\mu_{C 2}$$=$0.00 for both HEGs and LEGs). Despite the low spin values, all three models overpredict the LEGs and HEGs, and provide a very poor fit in the case of $f$$=$1.

As was discussed in Section~\ref{sec:obs}, the typical value of $f$ might 
change systematically with luminosity. A possible improvement to our modelling would be to vary $f$ with radio luminosity. Currently we do not wish to follow this approach since 
we do not have enough independent constraints, so  allowing $f$ to vary constrained only by the radio LF of HEGs and LEGS would lead to  `fine tuning' of parameters rather than allowing us to make inferences. The value of $f$$=$20 is justified independently and is appropriate for the radio sources in the local Universe (see Section~\ref{sec:obs}).

The QSO LF includes a break (due to the break in the X-ray LF) which looks
 similar to the break observed in the HEGs,
although the normalisation is not exactly in the right place. The shape of the 
break in the X-ray LF is approximately preserved due to the fact that the 
jet efficiencies are generally shallow at the low-spin end (e.g. HK06, N07, BB09). 
The jet efficiencies that are steeper at the low-spin end  have smoother breaks (M05, T10).  

Similarly,  in many models there is a break in the LEG LF. This originates in the break 
of the mass function, but in some models with very steep values of $\eta$ at
high spin the break is smoothed out. This is most noticeable in BB09.  The observed
 break in the LEG radio LF is not as marked as that in the mass function since, due to 
 their different accretion rates and jet efficiencies, SMBHs with a range of masses 
 contribute to that region of the LEG radio LF. 

We note that removing the  Compton-thick correction, modelled as a 50\% increase in the space density of HEGs (Section~\ref{sec:hi_acc}), makes no difference to our results. For all the jet efficiencies, the resulting best-fitting parameters for the HEG-QSOs are practically identical to those when including the Compton-thick sources. The only exception is for BB09: when removing the Compton-thick correction the smaller gaussian has a mean spin of 0.99 and an amplitude 0.01 relative to the gaussian with  mean spin 0.00 (compare with the results in Table~\ref{tab:best_param}).  

Overall, we consider the fits acceptable, given that they
are based on a few simple physically-motivated prescriptions. We 
remind the reader that the uncertainties in converting jet powers 
to radio luminosity densities, discussed in Section~\ref{sec:obs}
apply here equally. 

Most notably, assuming $f$$=$20 requires
the models to achieve $\eta_{\rm max}$$\sim$1 in order to 
explain the LEGs, due to their low accretion rates but high radio
luminosities.  Reducing $f$ to $\sim$1 would mean the low-$\eta$ 
models would fare much better (e.g. M05, T10B and to a certain
extent N07) but then the models with high $\eta_{\rm min}$, such
as HK06 and BB09, would overpredict the radio LF. 

The exact choice of cutoff for the maximum spin also makes a difference. For example BB09 argue for lower values of maximum spin in ADAFs, and indeed if we set an upper cutoff of \sp$\sim$0.95 the BB09 model provides a better  fit to the LEGs than for \sp$=$0.998.

We have only fitted the data above radio luminosity densities of $10^{23}$~\whzsr,
but we can see how our best-fits fare in the lower luminosity region, where
we have not fitted to the data. Figure~\ref{fig:entire_rlf} shows what our models
predict in this region. Indeed, extrapolation of our model radio LFs to lower radio luminosity
densities does not overpredict significantly the observed radio LFs of
HEGs or LEGs. 

The ADAF models provide a very good match to the LEGs in the range  $10^{21}-10^{23}$~\whzsr. There is no need for a further component to explain the LEGs: the prediction from our model is therefore that SMBHs with \mbh$\grtsim 10^{8}$~\msol\, dominate the LEG radio LF down to $10^{21}$~\whzsr\,  so that any contribution from lower mass SMBHs must be small.  

The QSO models do not fit the entire HEG radio LF, but an important positive result is that they do not overpredict the density of lower-radio luminosity HEGs. Indeed, an extra population is probably needed to explain the higher space density of HEGs, and these are most likely SMBHs with lower 
masses and moderate or low accretion rates, for example Seyfert galaxies and low-luminosity AGN \citep[e.g.][]{2005A&A...435..521N,2006A&A...451...71F}.

We can also estimate the distribution for all SMBHs, by combining the spin distributions of ADAFs and QSOs  weighed by their space densities:

\begin{equation}
P_{\rm SMBH}(\hat{a}) = {   w_{\rm QSO}P_{\rm QSO}(\hat{a}) + w_{\rm ADAF}P_{\rm ADAF}(\hat{a})  \over  w_{\rm QSO} + w_{\rm ADAF} }
\label{eq:dbn_smbh}
\end{equation}

\noindent where:

\begin{eqnarray}
 w_{\rm QSO} \equiv \sum \phi_{\rm QSO}(L_{\nu}) \\ 
  w_{\rm ADAF} \equiv \sum \phi_{\rm ADAF}(L_{\nu}) \\
  \label{eq:weights}
\end{eqnarray}

The expectation value for this distribution is given by:

\begin{equation}
\langle \hat{a} \rangle_{\rm SMBHs} = \int \hat{a} P_{\rm SMBH}(\hat{a}) {\rm d}\hat{a},
\label{eq:exp_a}
\end{equation}

The inferred distribution of spins for all SMBHs are shown in Figure~\ref{fig:spin_dbn_z0}, where we have combined the distributions for QSOs and ADAFs, and weighted them by the relative modelled space densities of QSOs and ADAFs respectively. All the distributions are bimodal, although the relative importance of the two components varies greatly between different jet efficiencies (e.g. contrast HK06 and M05).  Note that we weigh by the model distributions $\phi_{\rm QSO}(L_{\nu})$ and $\phi_{\rm ADAF}(L_{\nu})$, so that the expectation value will be most reliable for jet efficiencies that can reproduce the observations better.

For any given jet efficiency, the range of expectation values for the mean spin of all SMBHs is
similar  for  models~A and B as it is for model~C (see Table~\ref{tab:rej_models} and Figure~\ref{fig:spin_dbn_z0}).

In
Section~\ref{sec:disc} we discuss the implications of such a bimodal
distribution, and compare it to theoretical predictions from accretion
and mergers of SMBHs.

Interestingly, regarding the radio-loudness of quasars, our
results suggest that there should
be something resembling a dichotomy in radio-loudness (see Figure~\ref{fig:spin_dbn_z0}). 
There has been
some debate regarding the existence of a dichotomy in radio-loudness
\citep[e.g.][]{2002AJ....124.2364I,2003MNRAS.341..993C}, but we note
here that the dichotomy in spins will translate into a dichotomy in
jet efficiency, which does not necessarily map linearly into a dichotomy
in the ratio of radio-to-optical luminosity density (often used to
define radio-loudness). The uncertainties in converting from jet power
to radio luminosity density, the scatter in bolometric corrections and
selection effects can all blur out such a dichotomy.

\subsection{A prediction for the $z$$=$1 radio LF}\label{sec:z1}

If we assume that the individual spin distributions for ADAFs and QSOs
do not change significantly up to redshift of $z$$=$1, then we can
compare our predictions to the observed radio LF at this redshift.

We can use the observed X-ray LF, which tells us how the space density
of SMBHs with high accretion rates increases with $z$. However, we
need to make an extra assumption about the ADAFs, represented by the
BHMF. If we assume that the space density of SMBHs accreting at very
low rates has not changed significantly between $z$$=$1 and 0, then we
can use the local BHMF, with the same distribution of Eddington ratios
\citep[this is supported observationally by the relatively constant
  evolution of AGN with moderate radio luminosities,
  see][]{2009ApJ...696...24S}.

The larger volumes probed at $z$$=$1, together with the strong evolution
of the radio LF of FRII sources \citep{2001MNRAS.322..536W}, mean that we are able to observe radio sources up to the highest radio luminosities, $\sim$$10^{28}$~\whzsr\, at 1.4~GHz.  Such sources have are typically FRIIs and are  rarely seen in the local radio LF due to the smaller local volume.  

\begin{figure*}
%\begin{center}
\hbox{ \psfig{file=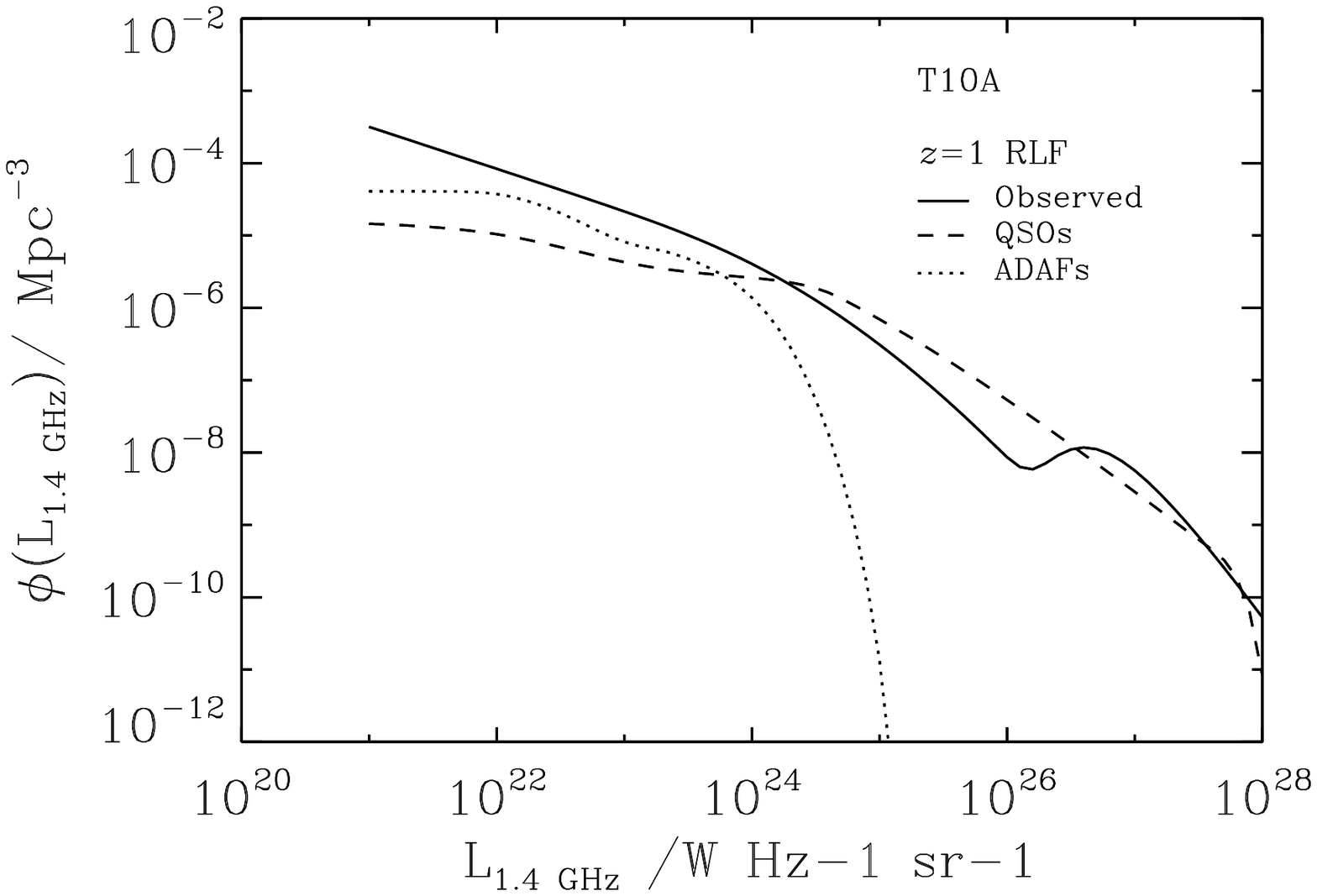, width=8cm, angle=0} \psfig{file=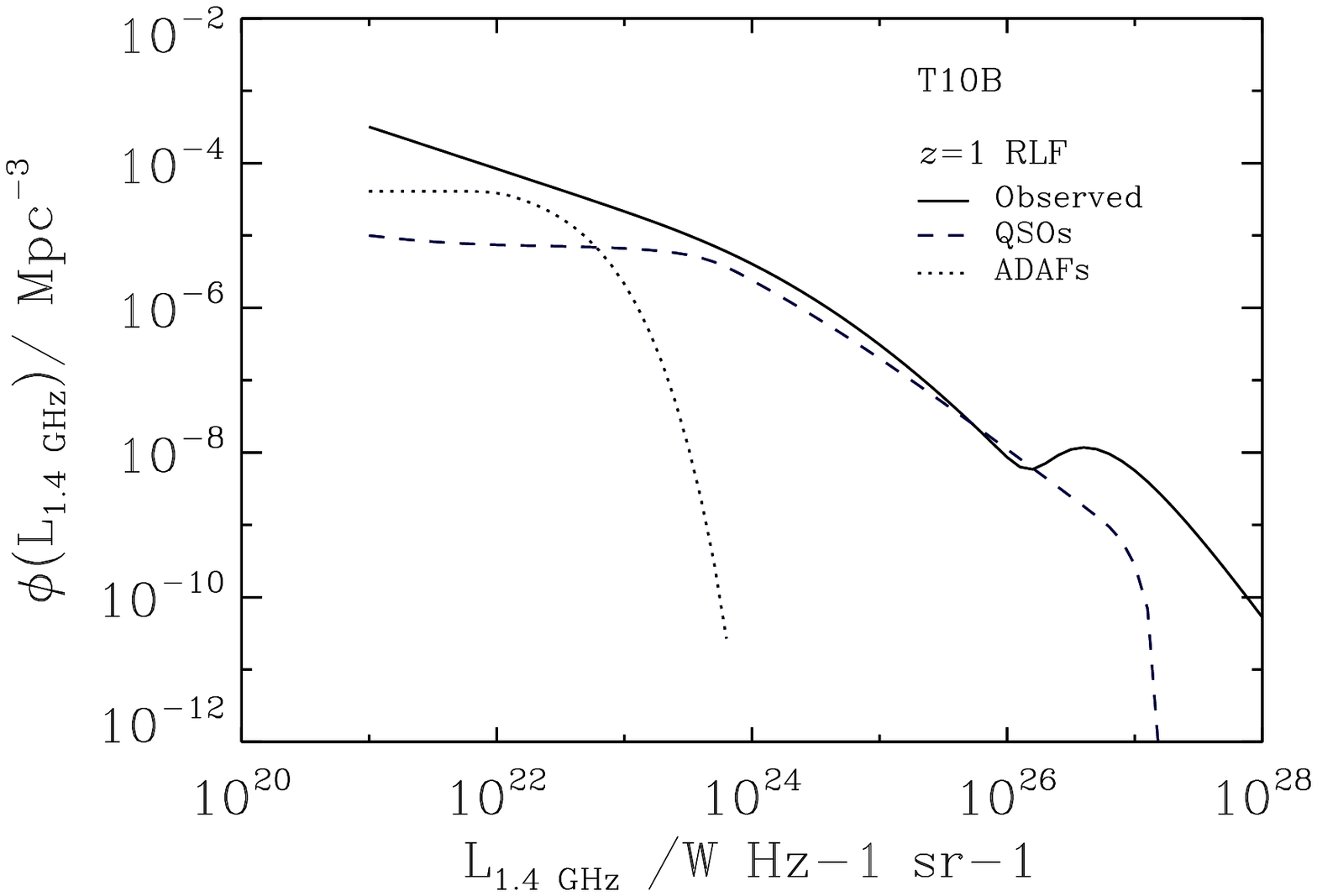, width=8cm, angle=0} }
\hbox{ \psfig{file=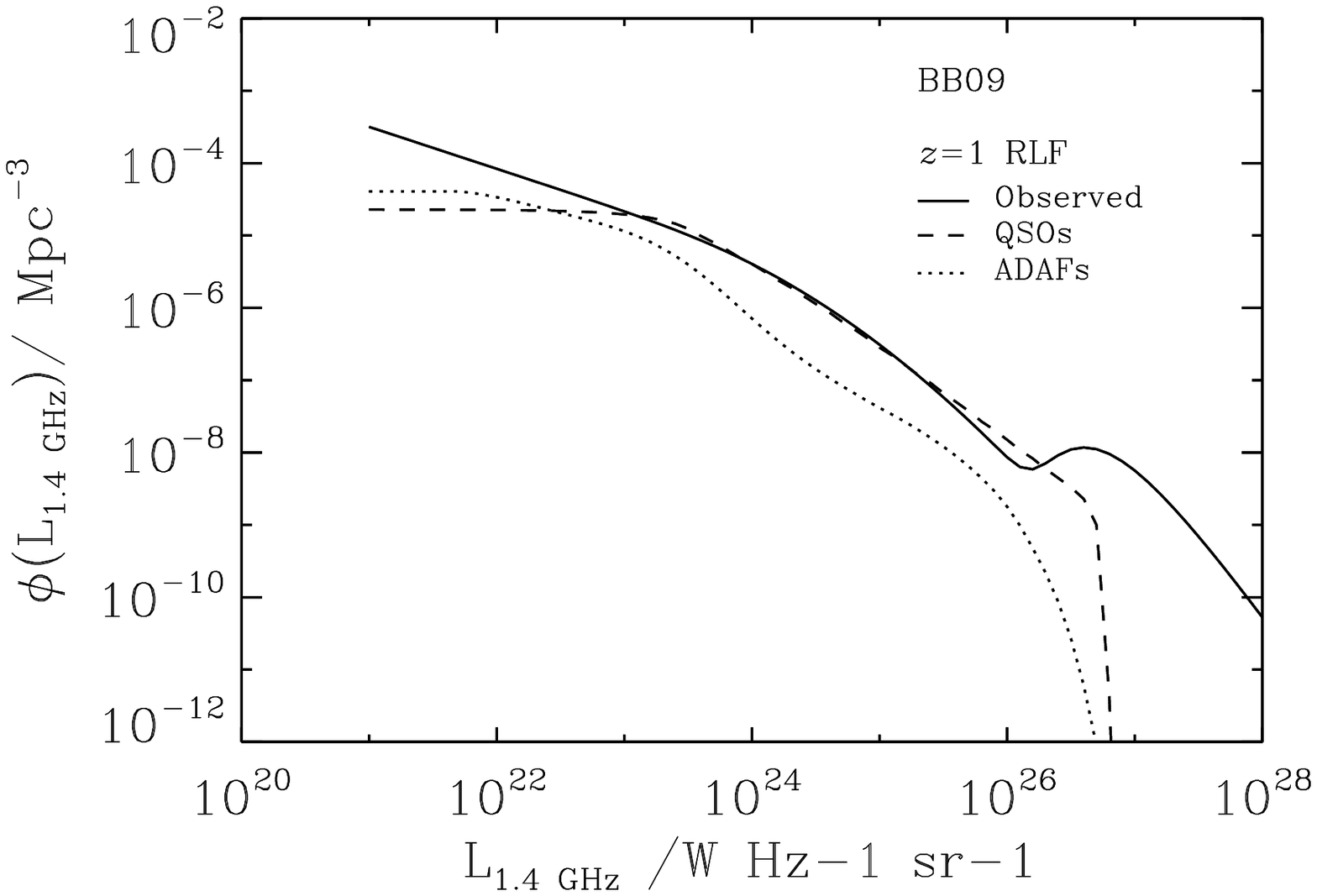, width=8cm, angle=0} \psfig{file=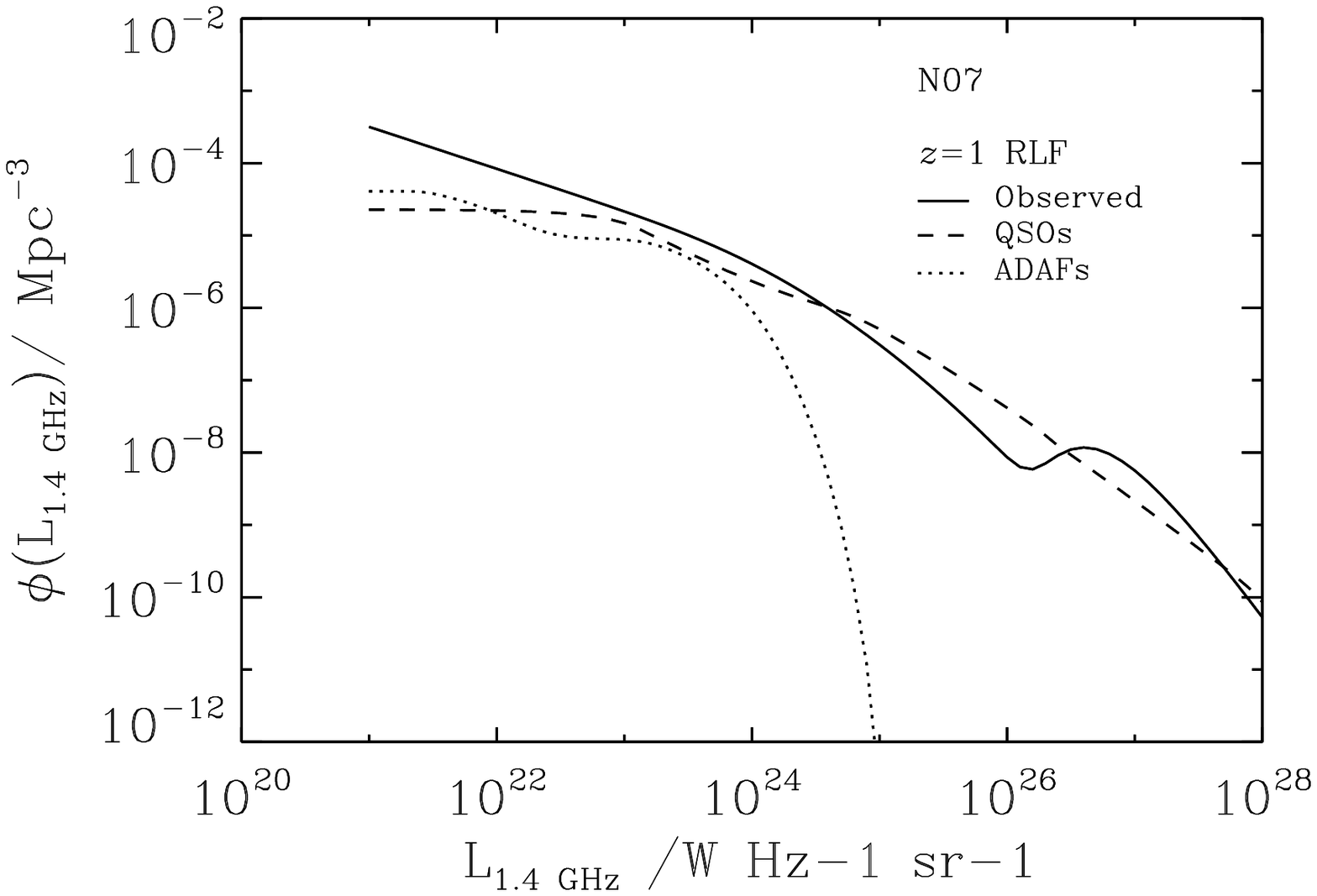, width=8cm, angle=0} }
\hbox{ \psfig{file=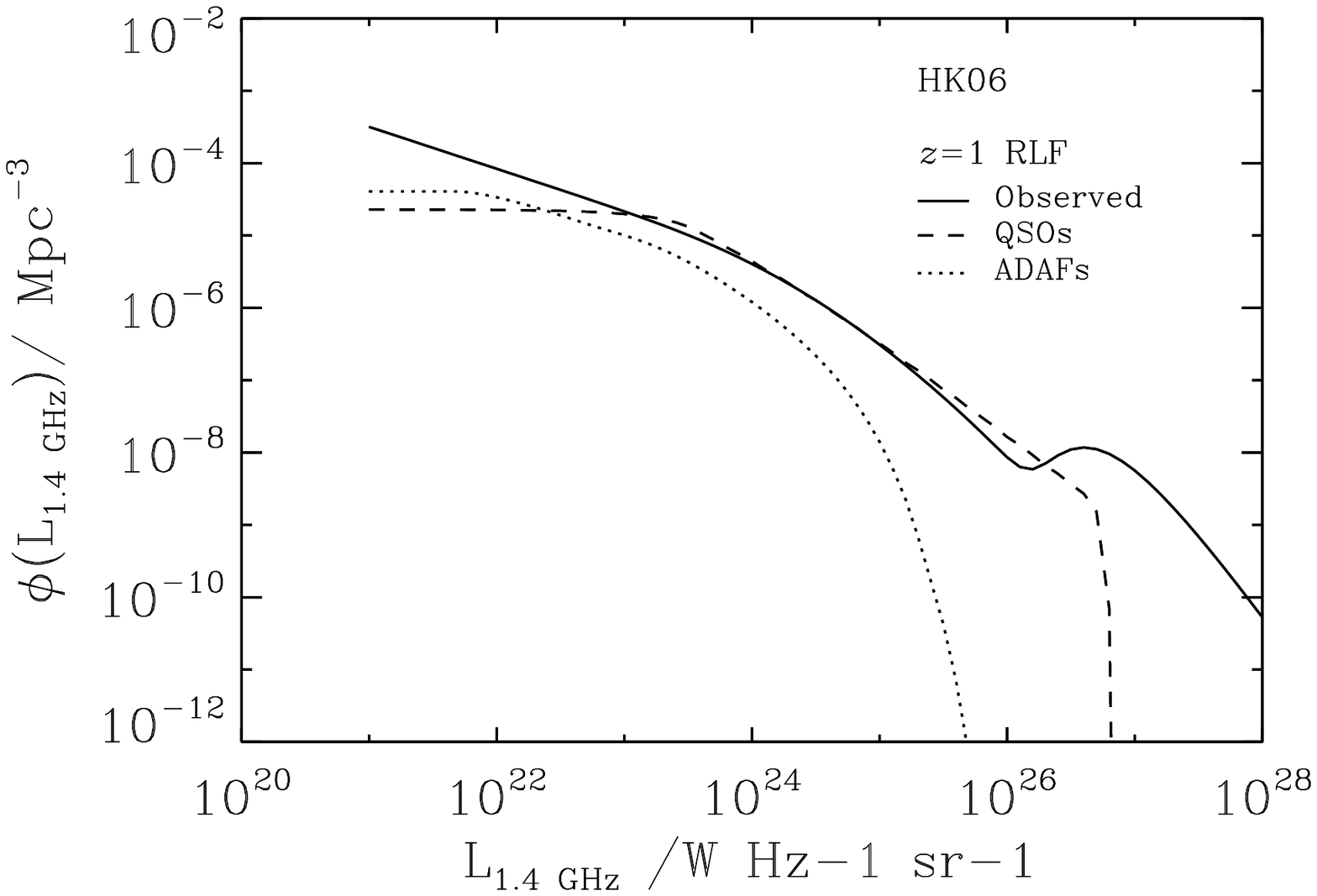, width=8cm, angle=0} \psfig{file=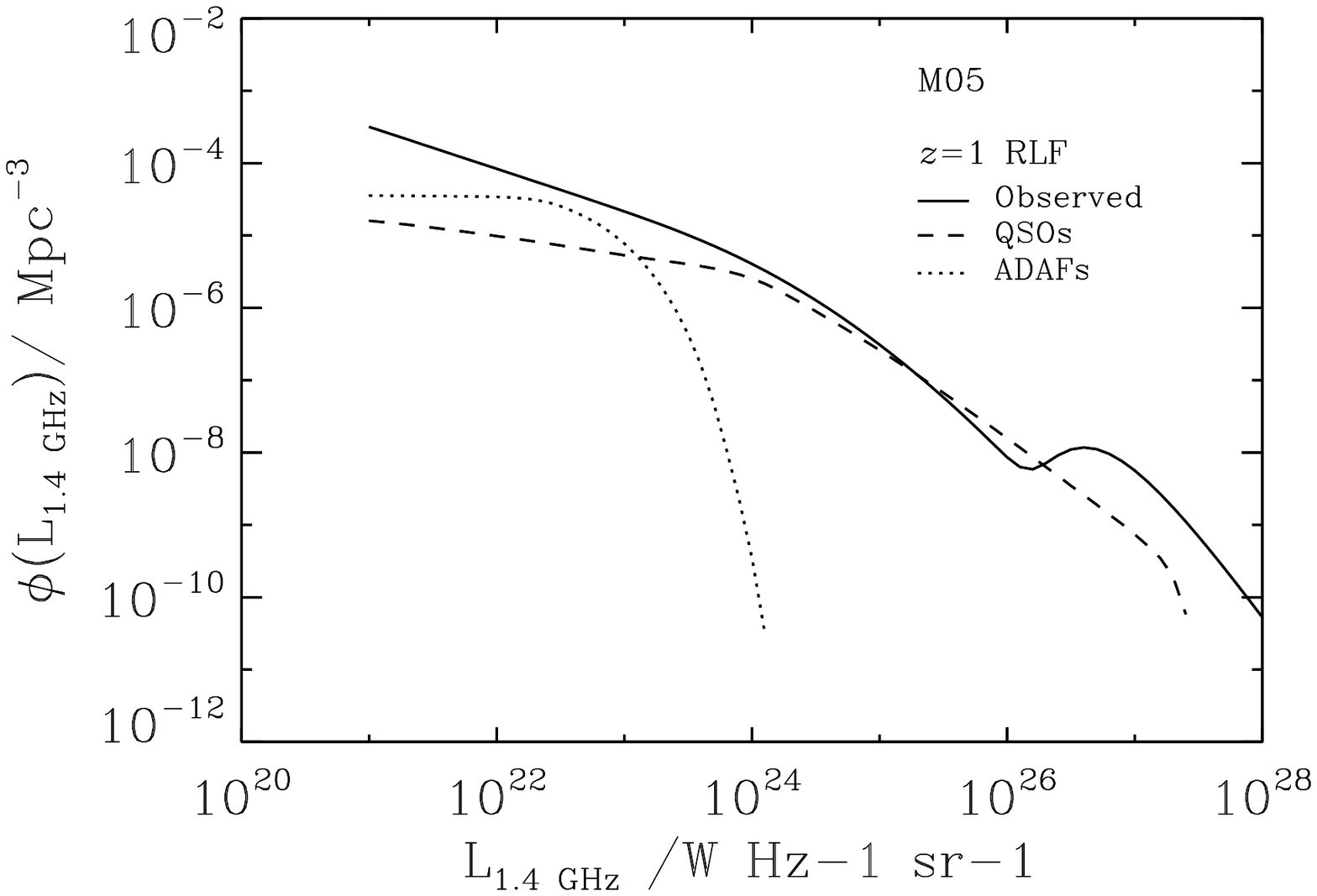, width=8cm, angle=0} }
\caption{\noindent The observed radio LF at $z$$=$1 \citep[solid line, ][]{2001MNRAS.322..536W,2009ApJ...696...24S} compared to the predicted radio LF from our model. We are assuming that the spin distributions for high- and low-accretion rate SMBHs (QSOs and ADAFs) at $z$$=$1 are the same as  those at $z$$=$0. The space density of QSOs is derived using the X-ray LF, at $z$$=$1, while the space density of ADAFs is derived assuming the present day black hole mass function to be already in place at $z$$=$1. The spin distributions of  the QSOs for the BB09 and HK06 models lack a high-spin component (see Figure~\ref{fig:spin_dbn_z0}), hence the sharp drop at high radio luminosities.  The overall agreement is very good, except around $L_{\nu}\sim$$10^{26}$~\whzsr, although we note that this region is not well constrained observationally, and the `wiggle' is  a consequence of the functional form used to describe  the radio LF \citep{2001MNRAS.322..536W}.  Our model predicts that the radio LF at $z$$=$1 is dominated by high-accretion rate SMBHs, which should be observed as sources with HEG spectra. }
\label{fig:pred_z1}
%\end{center}
\end{figure*}

Figure~\ref{fig:pred_z1} shows the resulting radio LFs from the QSOs
and ADAFs, compared to the observed radio LF, for the same range of
radio luminosities that we have observed.  We can see, that unlike at
$z$$=$0, it is the QSOs that dominate the radio LF. The solid line in 
Figure~\ref{fig:pred_z1} shows the observed radio LF at $z$$=$1, combining
the high radio luminosity LF from
 \citet{2001MNRAS.322..536W} and the low radio luminosity LF from \citet{2009ApJ...696...24S}.

Our model
overpredicts the radio LF in some regions, particularly around
$L_{\nu}\sim$$10^{26}$~\whzsr. We note however, that this is an
ill-constrained region of the radio LF: too faint for the samples used
by \citet{2001MNRAS.322..536W}, but too rare in space density for the
area covered by \citet{2009ApJ...696...24S}.  Indeed the apparent
`wiggle' is a consequence of the analytic function fitted to the radio
LF by \citet{2001MNRAS.322..536W}. 

The good agreement over so many
decades of radio luminosity is very encouraging.
Looking at the results  for the different jet efficiencies, 
we can make the following predictions for the $z$$=$1 radio LF:

At radio luminosity densities  $\grtsim$$10^{26}$~\whzsr, the composition of the $z$$=$1 radio LF should comprise $\sim$100\% HEGs. This prediction is in very good agreement with observations (Fernandes et al. 2011, in prep.). In the range $10^{23}-10^{26}$~\whzsr, at the higher luminosities we expect the HEGs to dominate, with LEGs making up only $\sim$10\% of the total population. We then expect the fraction of LEGs to increase rapidly with decreasing luminosity and eventually dominate. The dominance of the  HEG population, unlike  in the local radio LF,  is due to the strong increase  between $z$$=$0 and 1 of the comoving accretion rate onto SMBHs, which is reflected in the strong evolution in the X-ray LF \citep[e.g.][]{2008ApJ...679..118S}.

\begin{figure*}
%\begin{center}
\hbox{ \psfig{file=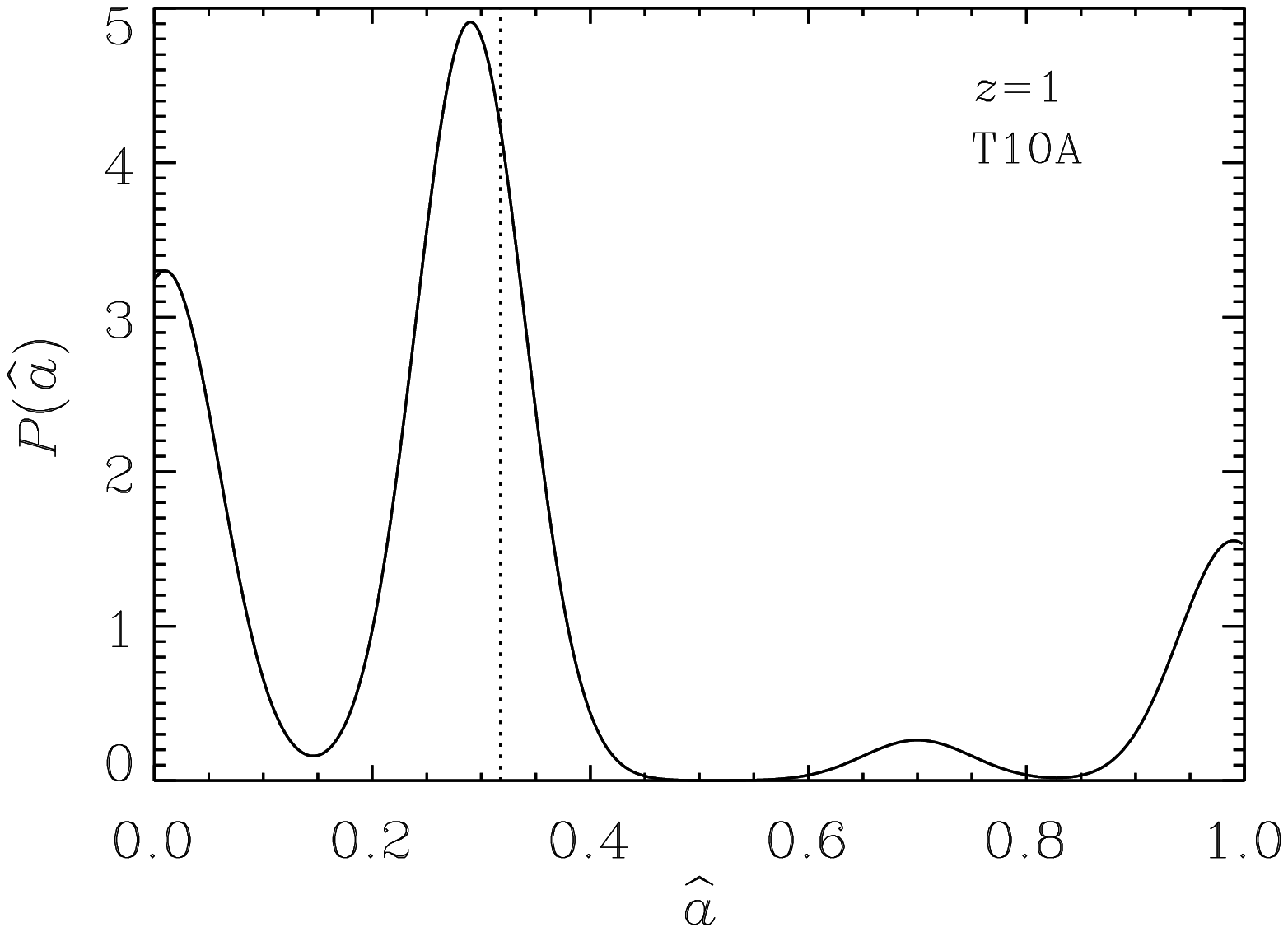, width=8cm, angle=0} \psfig{file=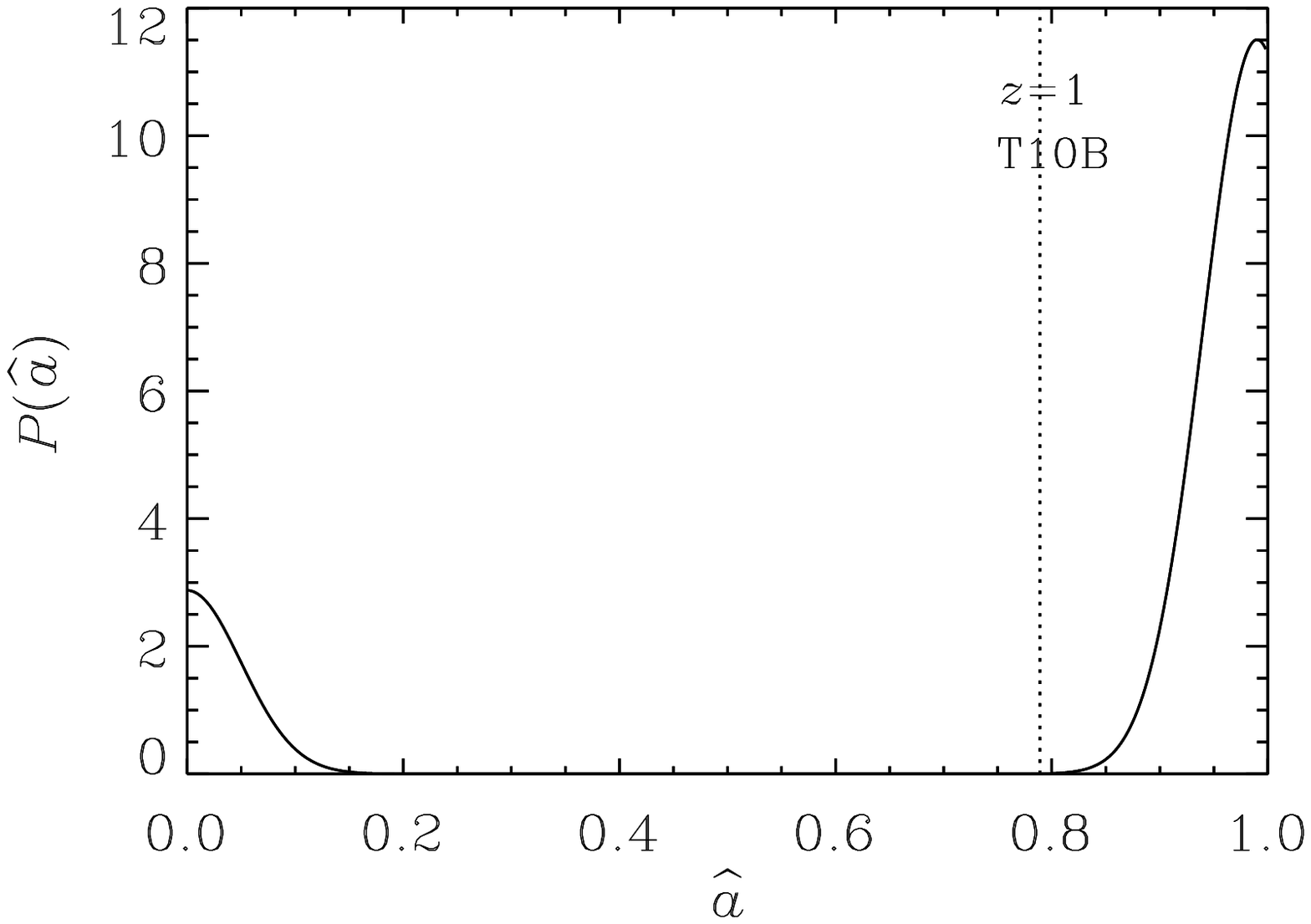, width=8cm, angle=0} }
\hbox{ \psfig{file=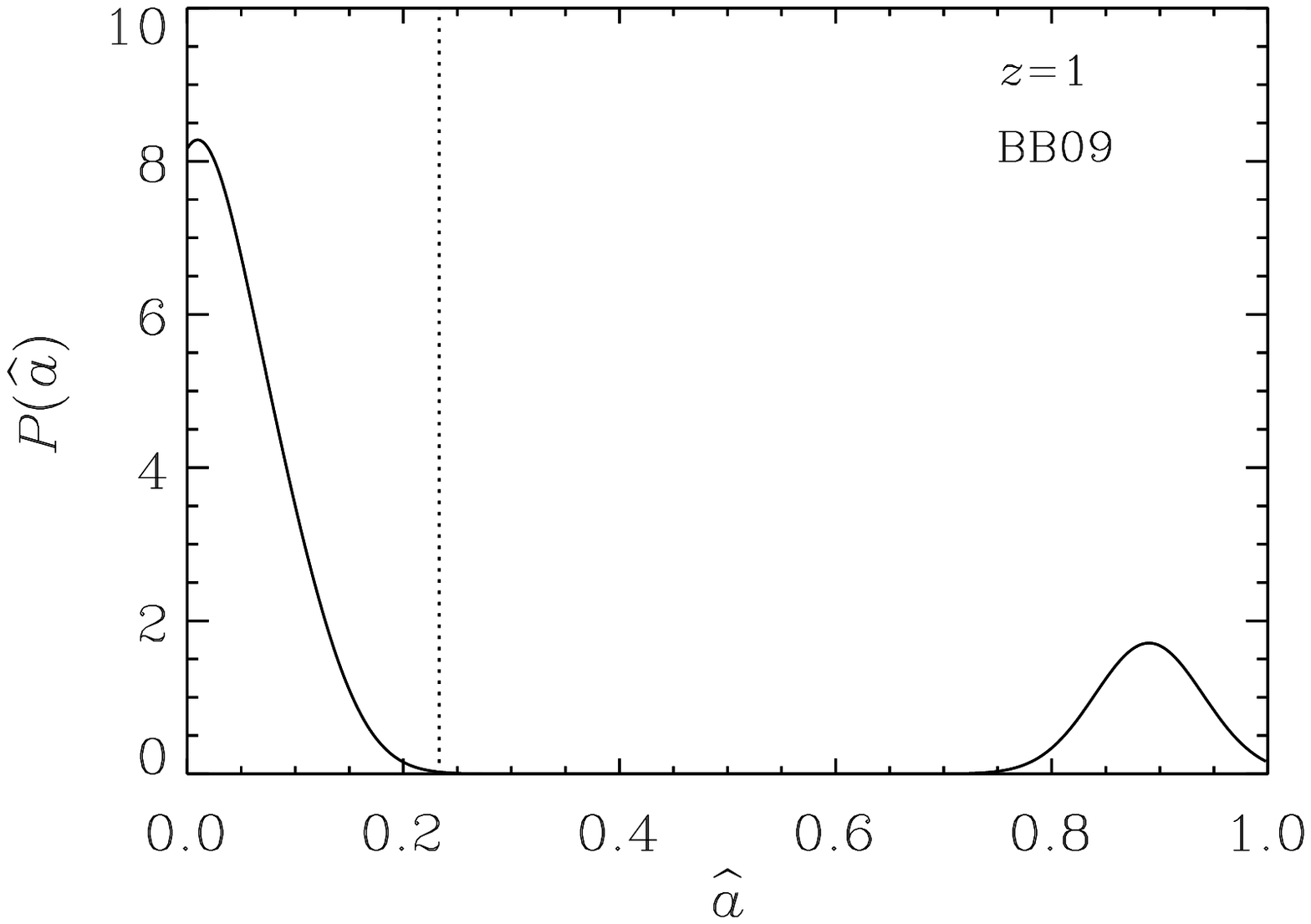, width=8cm, angle=0} \psfig{file=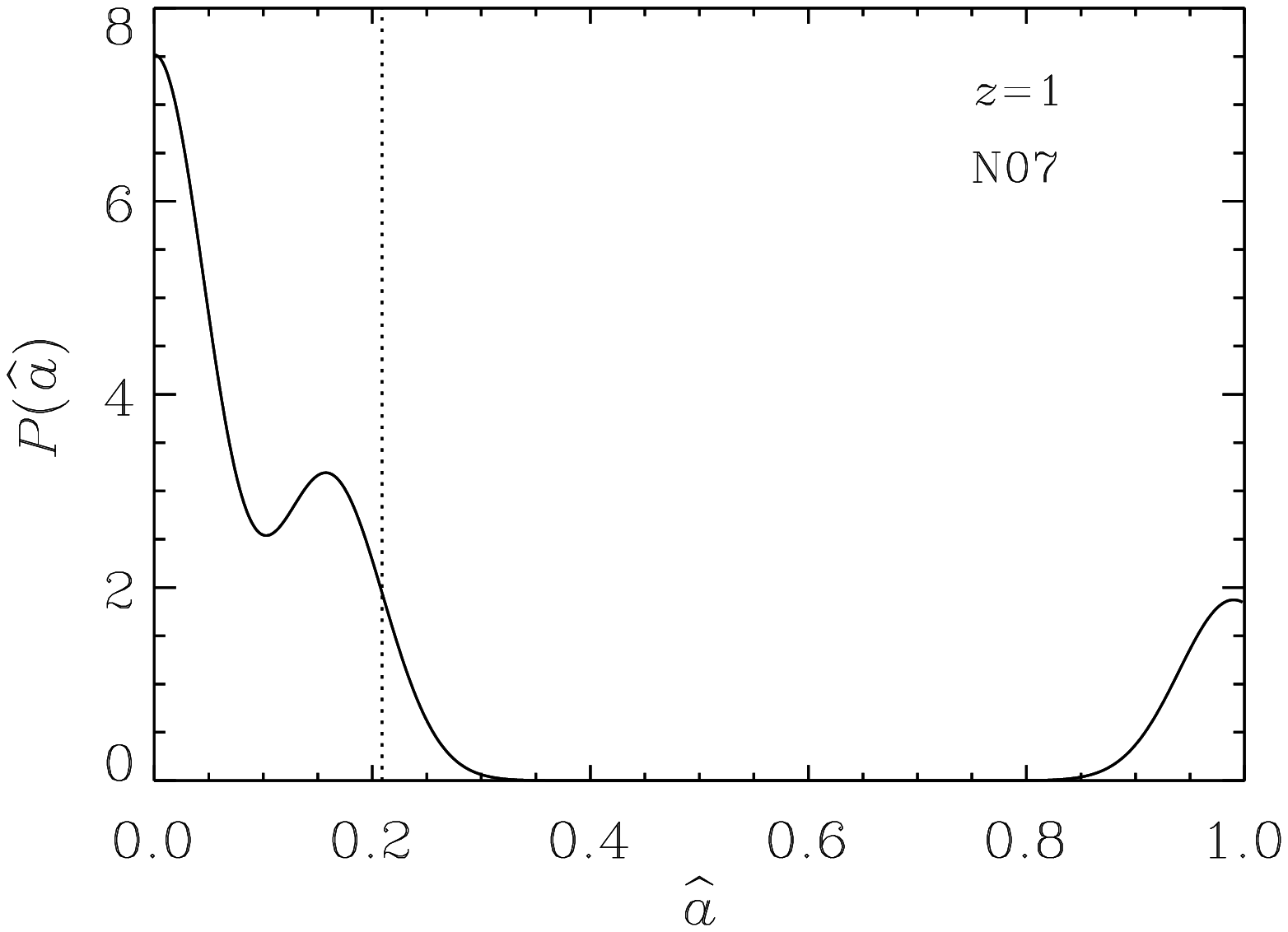, width=8cm, angle=0} }
 \hbox{ \psfig{file=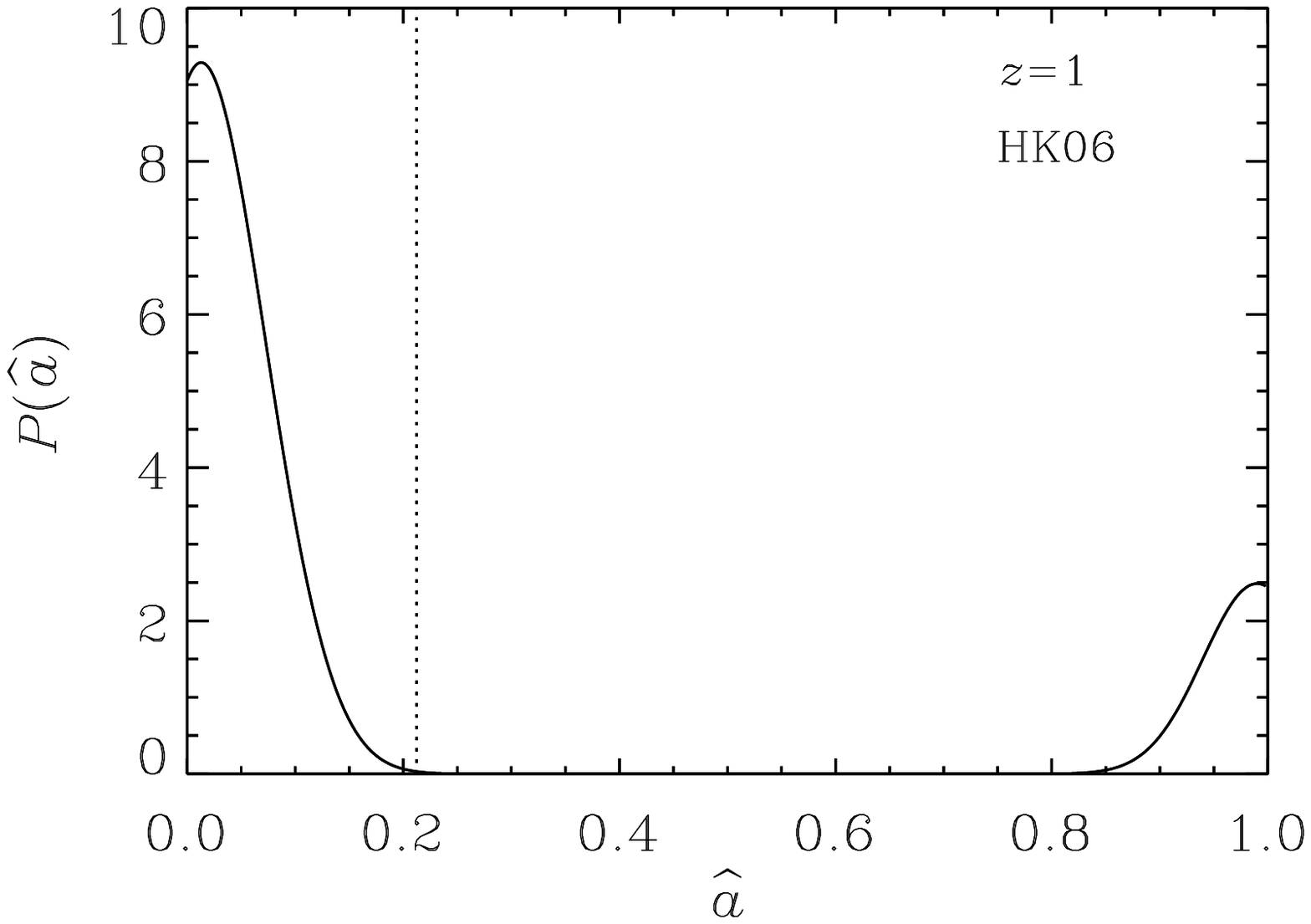, width=8cm, angle=0} \psfig{file=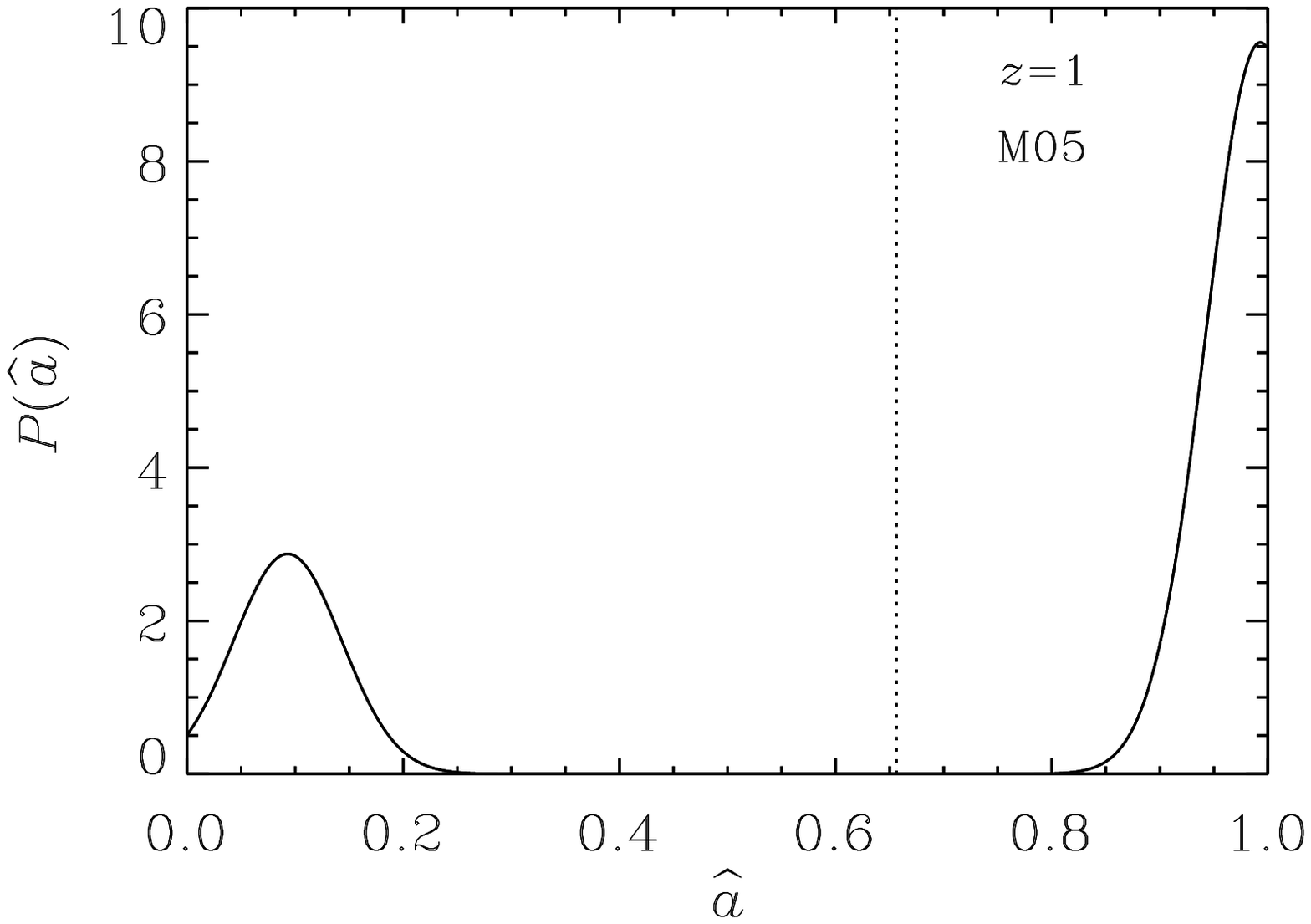, width=8cm, angle=0} }
\caption{\noindent Spin distributions at $z$$=$1 for all SMBHs. These have been obtained in an analogous way to Figure~\ref{fig:spin_dbn_z0}, only using the X-ray LF at $z$$=$1. The vertical dotted lines mark once again the mean value.  We have assumed the ADAFs not to evolve significantly up to $z$$=$1, which means we are likely to overestimate their contribution. Comparing this figure with  Figure~\ref{fig:spin_dbn_z0}, we can see that the fraction of SMBHs with a high spin decreases with increasing redshift. }
\label{fig:spin_dbn_z1}
%\end{center}
\end{figure*}

\begin{figure*}
\begin{center}
\hbox{  \psfig{file=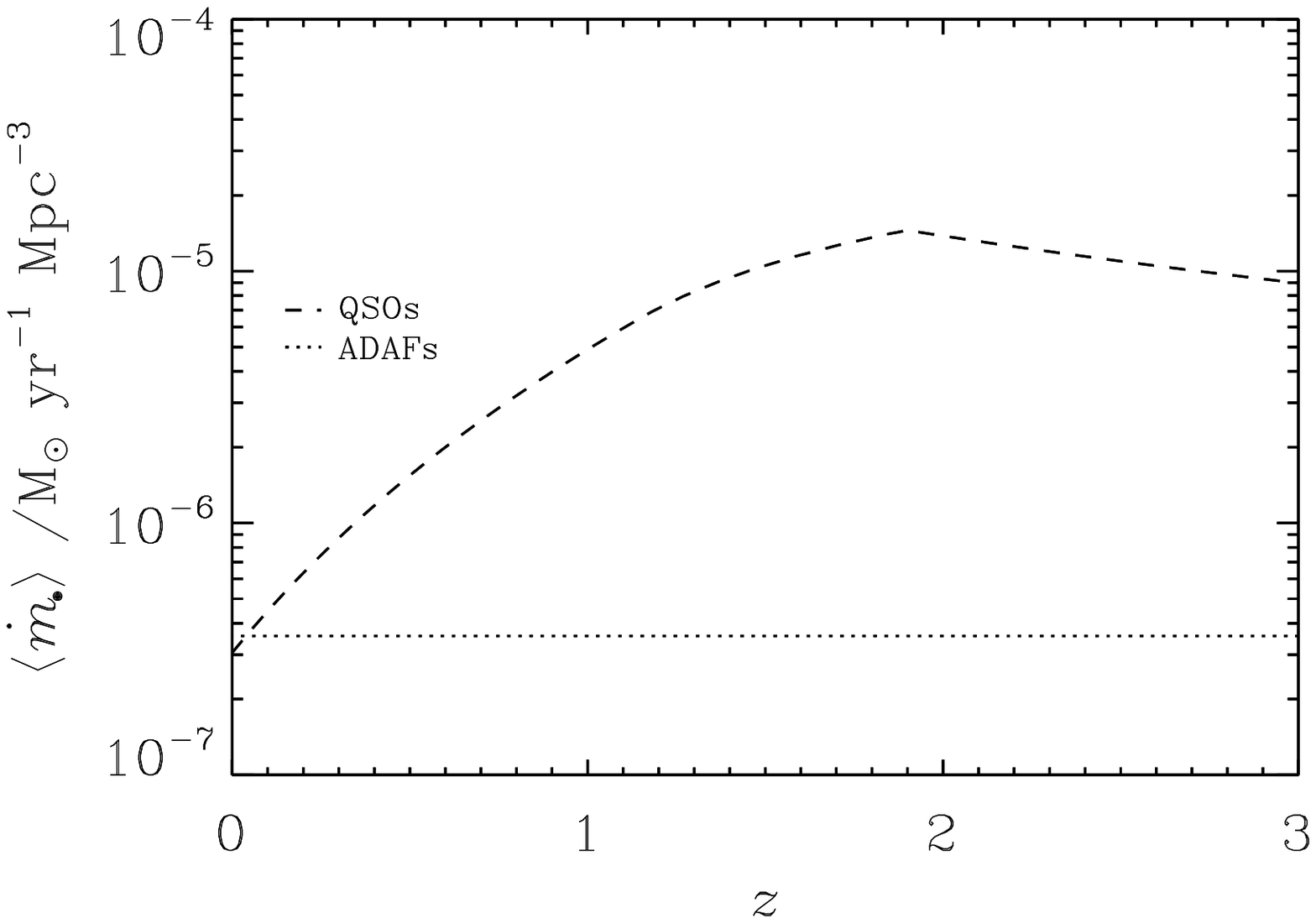, width=8cm, angle=0}  \psfig{file=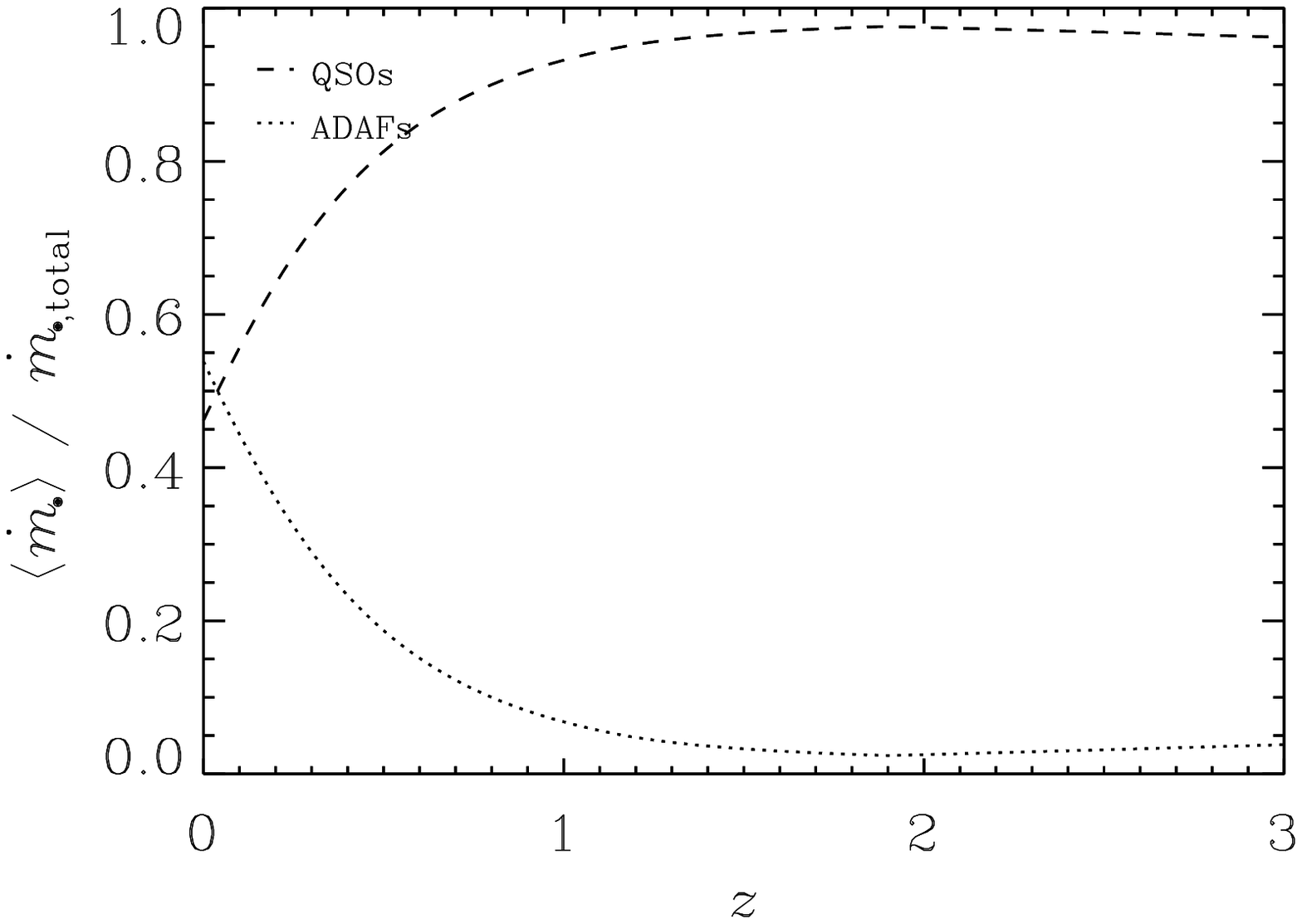, width=8cm, angle=0} }
\caption{\noindent Expectation value for the comoving accretion rate from QSOs and ADAFs (left panel) and relative fraction of the accretion rate (right panel), derived in an analogous way to Equations~\ref{eq:dbn_smbh}, \ref{eq:weights} and Equation~\ref{eq:exp_a}, but applied to $\dot{m}$ rather than $\hat{a}$. }
\label{fig:accn_phi}
\end{center}
\end{figure*}

\subsection{The comoving accretion rate of radio sources}\label{sec:accn_rate}

Applying Equations~\ref{eq:dbn_smbh}, \ref{eq:weights} and \ref{eq:exp_a} to the accretion rate,  rather than the spin, we can estimate the expectation value for the  accretion rate onto ADAFs and onto QSOs. 

Figure~\ref{fig:accn_phi} shows the evolution of the comoving accretion rate (left panel), and the fraction of the total accretion rate onto ADAFs and onto QSOs (right panel) as a function of redshift.
At $z$$=$0, approximately half of the accretion onto the most massive black holes is occurring via high accretion rates (in the form of QSOs), and the other half via low accretion rates (ADAFs). The fraction of accretion ocurring in high-accretion rate objects increases rapidly with redshift. The right panel of Figure~\ref{fig:accn_phi}, shows our estimate, although again if the mass function of SMBH decreases significantly, the curve for ADAF will overpredict the accretion rate onto low-accretion rate sources.

Our left panel of Figure~\ref{fig:accn_phi} is very similar to Figure~4 of \citet{2008MNRAS.388.1011M}. For both works, the accretion rate from QSOs is derived from the X-ray LF, so unsurprisingly the curves agree very well. 

The difference lies in the density of accretion onto ADAFs, where we have assumed a distribution of Eddington ratios for the local SMBH mass function, and assumed this to stay constant with redshift. We have used de inferred distribution of spins to produce a distribution of jet efficiencies. 
\citet{2008MNRAS.388.1011M}, on the other hand, assume a fixed fiducial jet efficiency and derive the accretion rate from the radio LF. 

Similarly, \citet{2008ApJ...676..131S} use a sample of optically- and radio-selected AGN to derive the mean jet efficiency, together with the scatter. The resulting  jet efficiency for  the  sample  is $\sim$0.01. The sample is actually the  same one used in our Section~\ref{sec:loud}, so that the same caveats mentioned in Section~\ref{sec:comparison} should be bore in mind, since the sample includes only objects which are bright enough to be detected both in the optical and radio surveys.

\section{The evolution of the cosmic spin}\label{sec:spin_his}

In this Section we use our results from Sections~\ref{sec:fit_rlf} and \ref{sec:res} to infer our best 
estimate for  the 
cosmic spin history of the most massive black holes.  Comparison of Figures~\ref{fig:spin_dbn_z0} and \ref{fig:spin_dbn_z1} shows that as the redshift increases, the fraction of SMBHs with high spin decreases.

 We stress that Figure~\ref{fig:spin_dbn_z1} assumes negligible evolution in the black hole mass function. This is probably an acceptable  approximation up to $z$$\sim$1, but at higher redshifts we expect the space density of the most massive black holes to have decreased significantly. At higher redshifts, the distribution of SMBHs will closely match that of the HEGs alone, and most of the probability density will lie  at low spins.

Before we continue, we warn of two important points.   First, it is not the intention of this paper to judge the different models of jet formation around SMBHs.   Second, as discussed earlier, the quality 
of the fit is partly determined by our choice of $f$ factor when converting jet power to radio luminosity, and a lower value would cause  the models  that currently look best to overpredict the radio LF, so that some of the less-favoured models would then provide better fits (see discussion in Section~\ref{sec:z0}).  

Hence we choose to estimate the mean cosmic spin history using two sets of efficiencies. The first uses all the models independently of whether they can reproduce the radio-loudness of quasars or the local radio LF.  The second will use only the four efficiencies that show some success at reproducing the observations, namely T10A, BB09, N07 and HK06.  We label these as the `best fitting set'. 

\begin{table}
  \begin{center}
    \begin{tabular}{cccc}
      \hline
      \hline
$\eta$ & P(\sp$\geq0.5$) & P(\sp$\geq0.5$) \\
 &  $z$$=$0&  $z$$=$1 \\
\hline
T10A & 0.14 & 0.13 \\
T10B & 0.94 & 0.78 \\
BB09 & 0.33 & 0.19 \\
N07 &  0.20 & 0.13 \\
HK06 &  0.30 & 0.17 \\
M05 &  0.80 & 0.62 \\
\hline
all   &  0.45$\pm$0.33 & 0.34$\pm$0.29 \\
best fitting &  0.24$\pm$0.09 &  0.16$\pm$0.03 \\
    \hline
      \hline
    \end{tabular}
  \end{center}
\caption{\noindent Fraction of SMBHs with  \sp$\geq$0.5  at $z$$=$0 and 1, for each of the six jet efficiencies. At the bottom we show the mean values for all the efficiencies and for the best fitting set only. }
    \label{tab:frac_0.5}
\end{table}

The cosmic spin of SMBHs evolves from   distributions dominated by low spins at high redshifts, to bimodal distributions with comparable high-spin and low-spin components at low redshifts.   It is the fraction of highly spinning black holes that increases with cosmic time.   To quantify this, we consider the fraction of SMBHs with \sp$\geq$0.5. The fractions, or probabilities of \sp$\geq$0.5, are shown for each jet efficiency in Table~\ref{tab:frac_0.5}, together with the mean fractions of all the sources, and of the best fitting set only.  If we consider all the efficiencies together, the fraction of SMBHs with \sp$\geq$0.5 increases from 0.34$\pm$0.29 at $z$$=$1 to 0.45$\pm$0.33 at $z=$0. However, we believe a more accurate result is given by the best fitting set of efficiencies, T10A, BB09, N07 and HK06: in this case the fraction of SMBHs with \sp$\geq$0.5 is 0.16$\pm$0.03 at $z$$=$1 and 0.24$\pm$0.09 at $z$$=$0. The uncertainties are due to the variations of the fraction between different efficiencies (see Table~\ref{tab:frac_0.5}).  At redshifts $>$1 we expect the fraction of SMBHs with  \sp$\geq$0.5  to be even lower than the quoted values.

Given the shape of the distribution functions, the most informative way of describing the spin history is to show the distributions and determine the fraction with high spin. A mean value does not reflect the bimodality, however, for completeness we have calculated the expectation value and its evolution, and we describe this in Sections~\ref{sec:exp_a_z} and \ref{sec:best_est}.

\subsection{The expectation value of spin with redshift}\label{sec:exp_a_z}

\begin{figure*}
%\begin{center}
\hbox{ \psfig{file=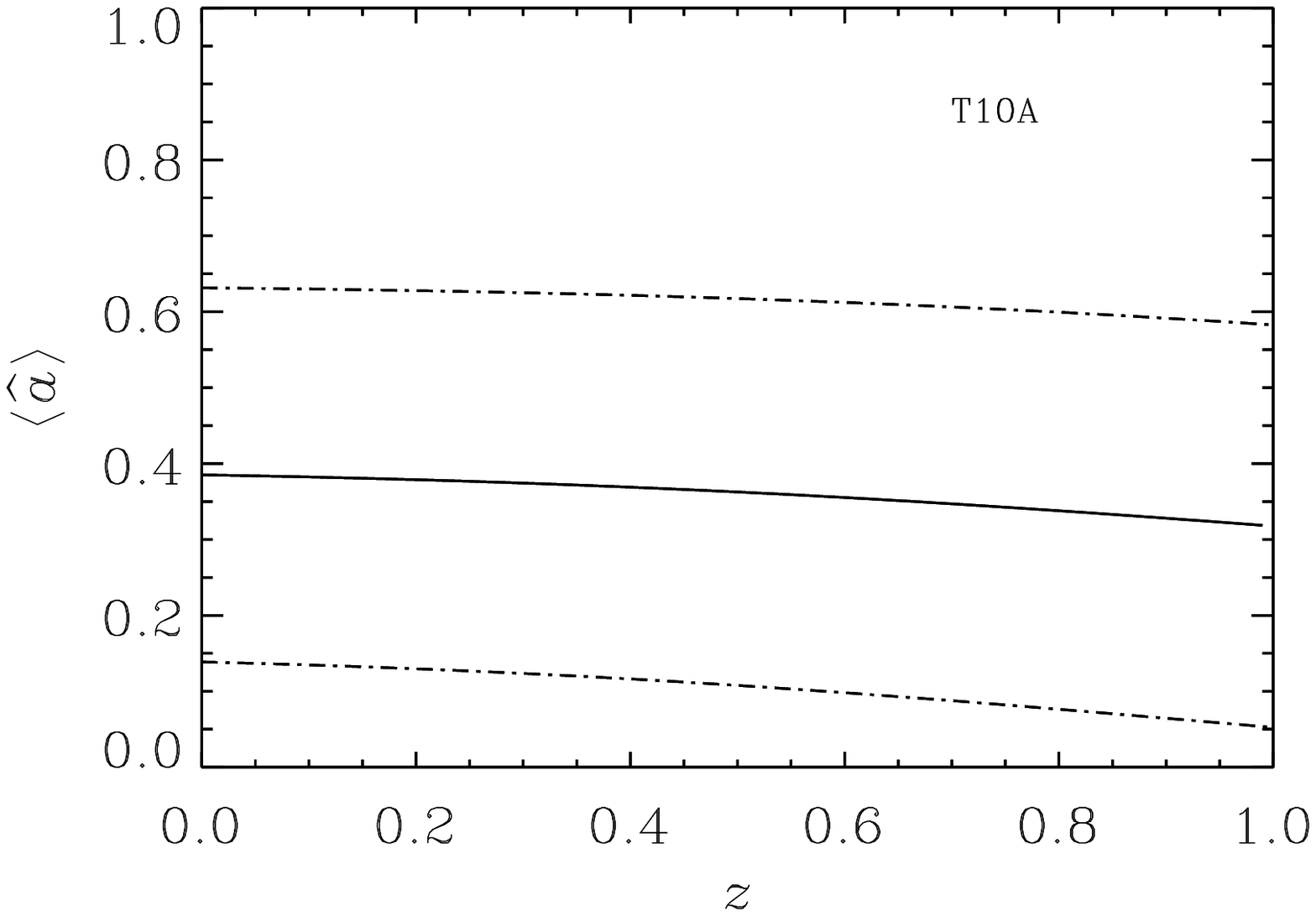, width=8cm, angle=0} \psfig{file=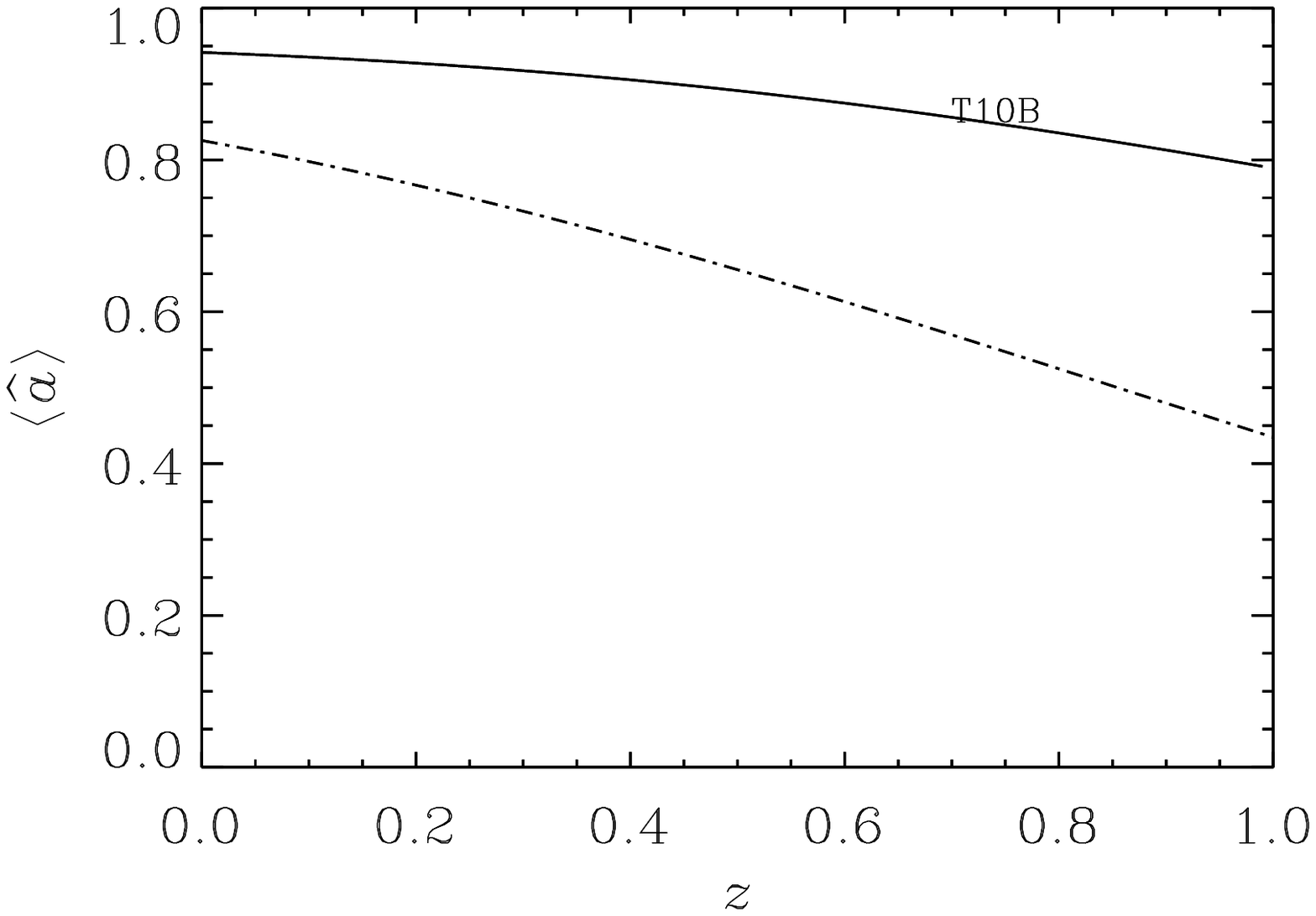, width=8cm, angle=0} }
\hbox{ \psfig{file=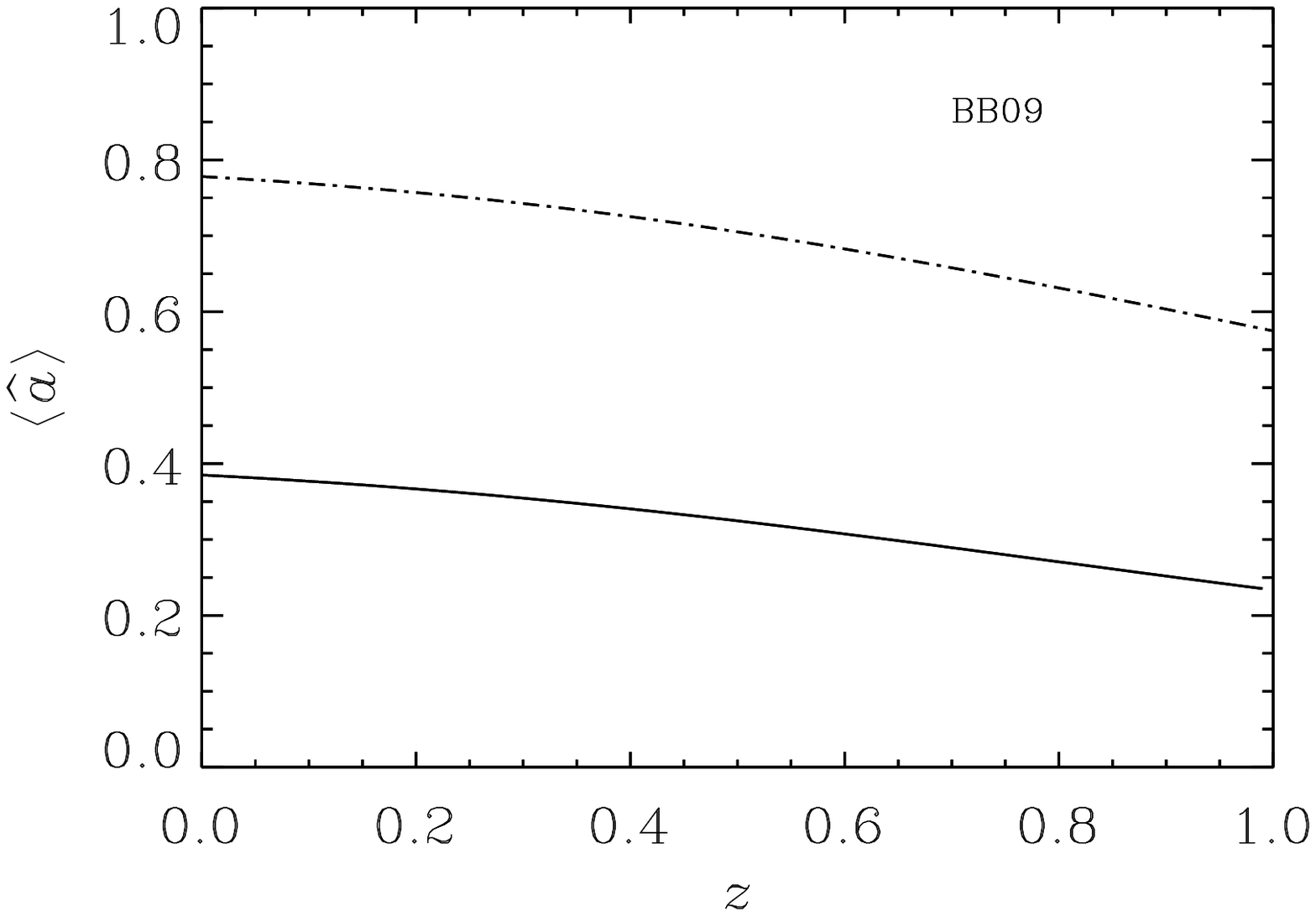, width=8cm, angle=0} \psfig{file=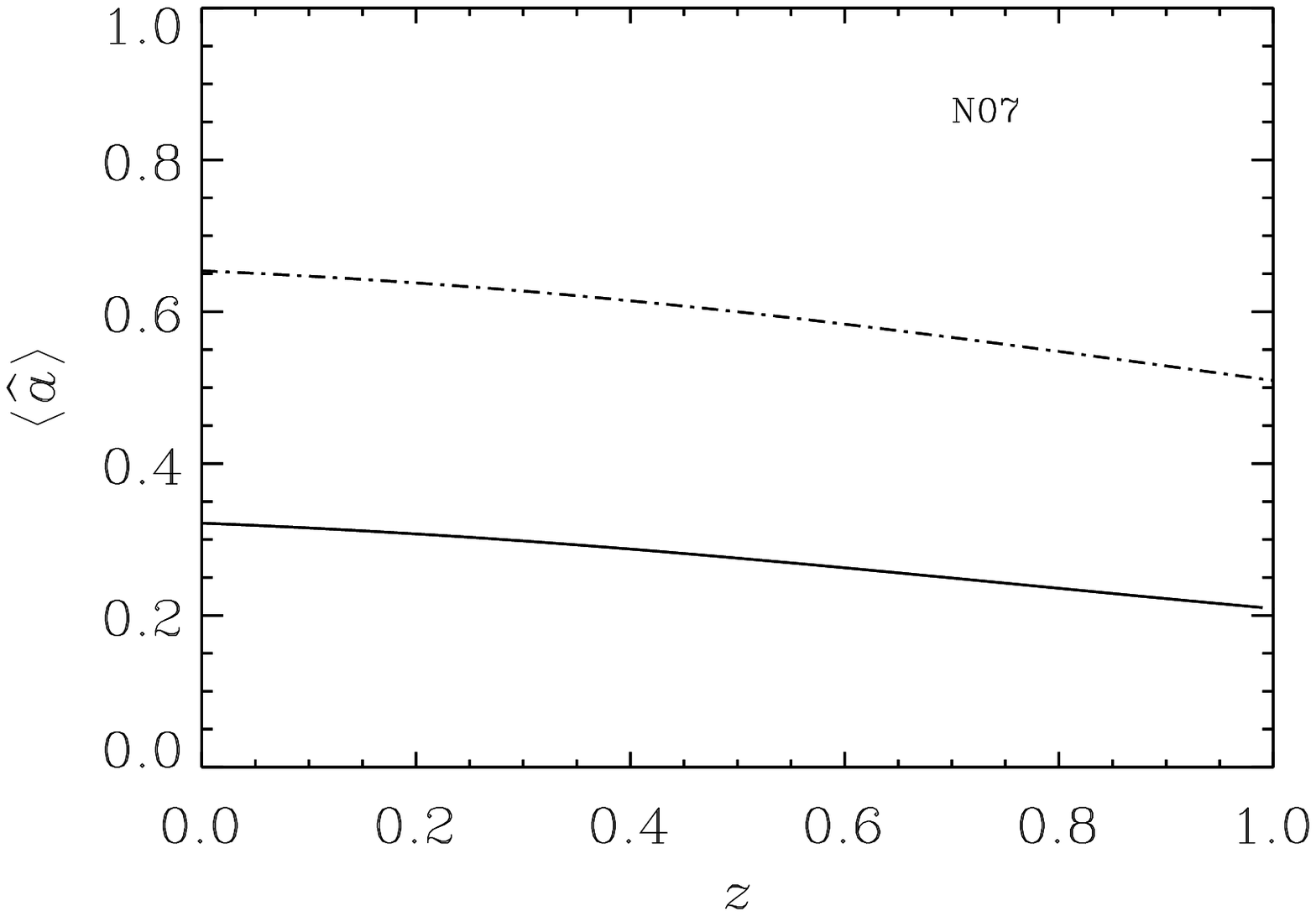, width=8cm, angle=0} }
\hbox{ \psfig{file=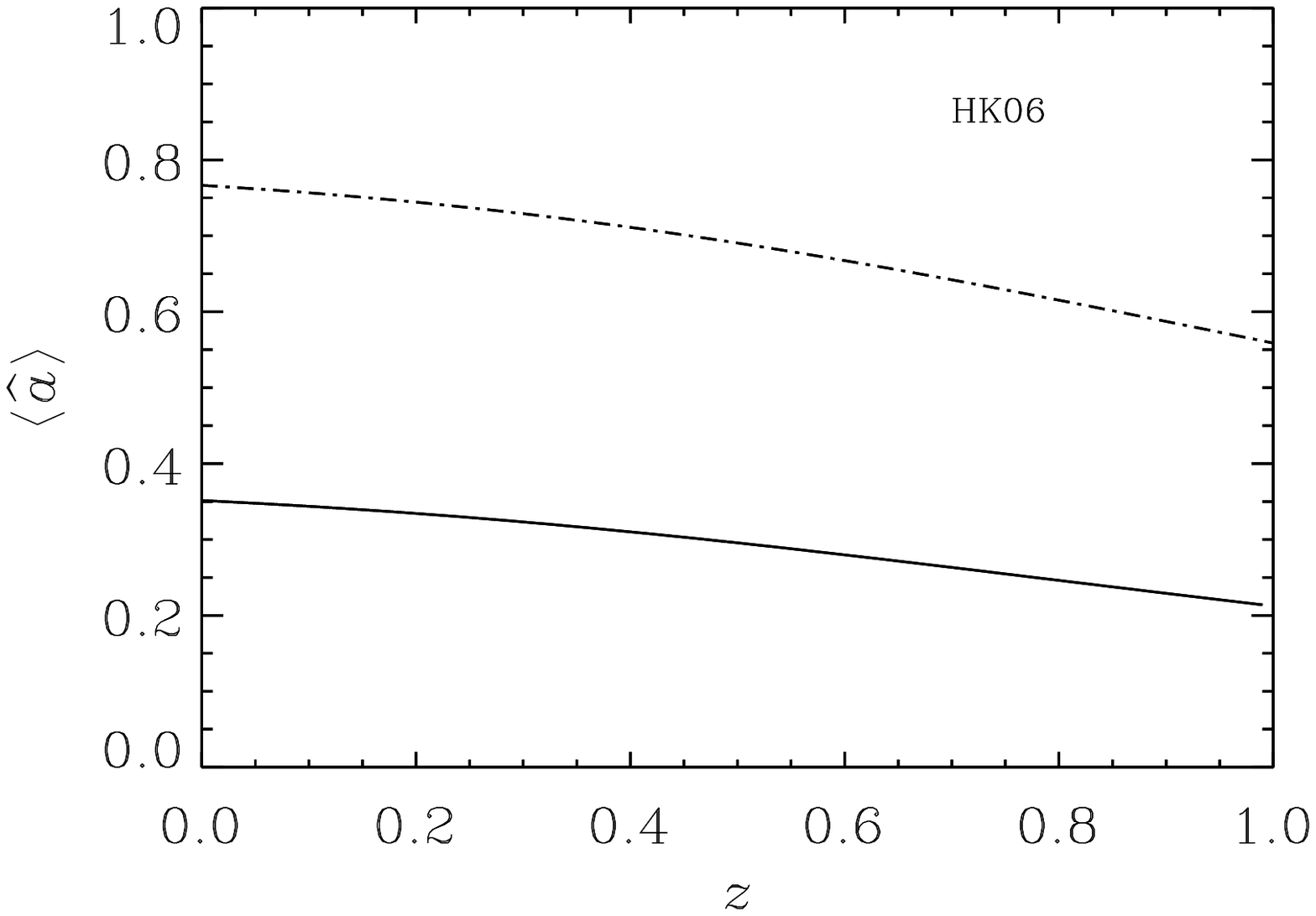, width=8cm, angle=0} \psfig{file=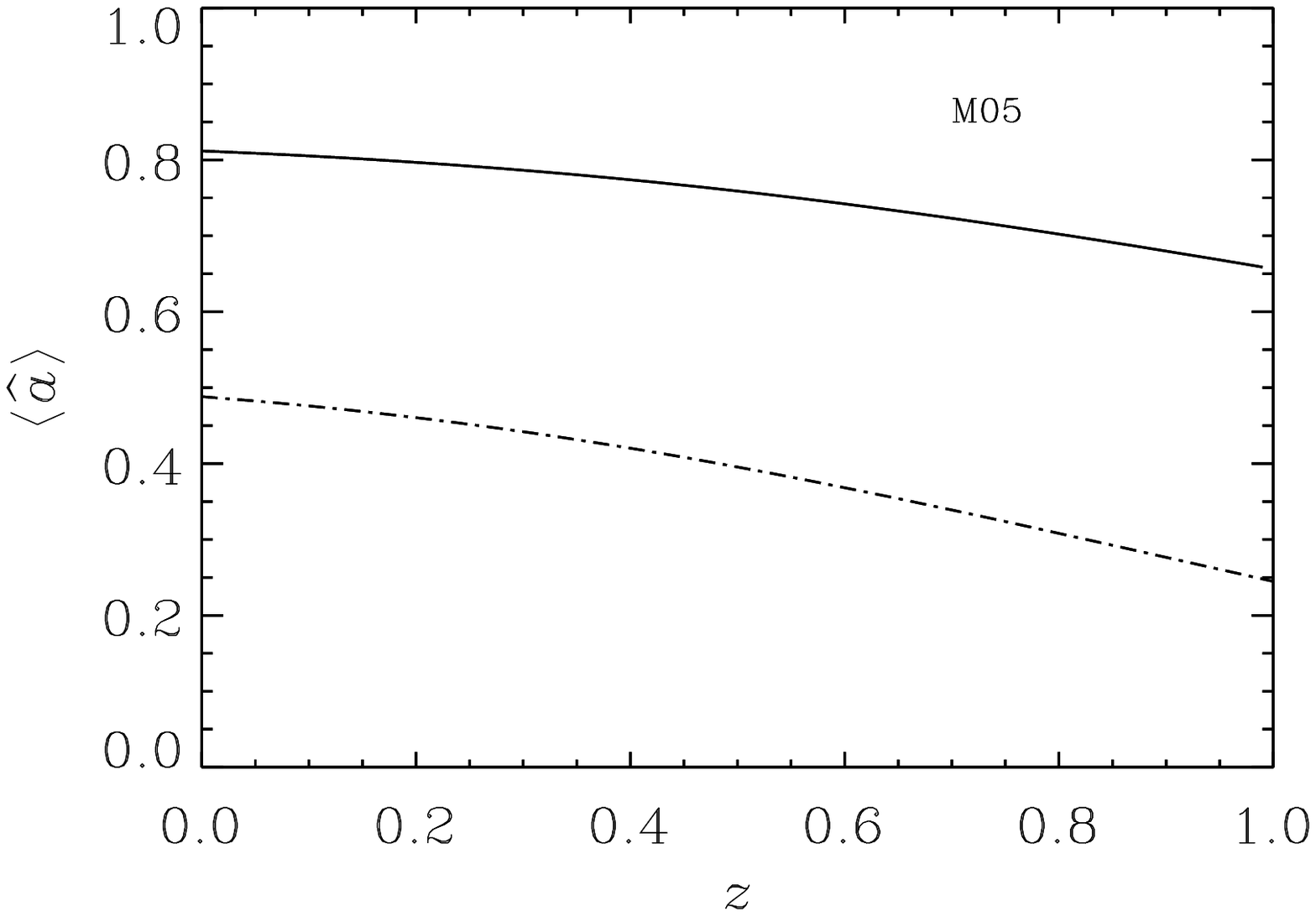, width=8cm, angle=0} }
\caption{\noindent Expectation value for the spin of SMBHs as a function of redshift (solid line),  as determined by combining the spin distributions for ADAFs and QSOs, using the X-ray LF and black hole mass function to weigh the relative contributions. The dashed dotted lines show the values of $\pm1$  standard deviations,  where the standard deviation is the square root of the variance determined from the weighted spin distribution. In the T10B, BB09, N07 and HK06, panels the -1$\sigma$ contour fall bellow \sp$=$0. In the case of M05, the +1$\sigma$ contour is higher than   \sp$=$1. }  
\label{fig:exp_a_z}
%\end{center}
\end{figure*}

Figure~\ref{fig:exp_a_z} shows the evolution of \amean\, (solid line)
together with the standard deviation (dashed lines).  The standard deviation is measured from the square root of the variance. The large value of the 
variance is caused by the bimodal nature of the spin distribution at each redshift (as shown by  Figures~\ref{fig:spin_dbn_z0} and \ref{fig:spin_dbn_z1}).

Due to the fact that the spin distributions are bimodal, the expectation value should be taken simply as a fiducial value, and the exact value is clearly uncertain. However, this expectation value evolves with redshift, and we see that it was lower at $z$$=$1  and has slightly increased to a higher value  in the present-day Universe (see Figure~\ref{fig:exp_a_z}). In Section~\ref{sec:disc} we discuss the physical implications of this evolution.

\subsection{A best estimate of the evolution of the mean spin}\label{sec:best_est}

Figure~\ref{fig:exp_a_z} shows a range of different spin histories, due to the differences
in the jet efficiencies from Figure~\ref{fig:eta}. We wish to use all the available information, in the 
form of the different jet efficiencies, for a best estimate of the spin history. We consider both the entire set of jet efficiencies, as well as considering the best-fitting set alone.

To obtain the best estimate of the spin history, we take the mean of all the results as a function of redshift. However, the individual results are not always compatible. The $N$ individual results $\hat{a}_{i}$($z$) are compatible if  $\sum_{i}\left[ \left(\hat{a}_{i}(z) -\langle \hat{a}(z) \rangle \right)/ \sigma_{i}\right]^{2}$ is $\leq$$(N-1)$, where  \amean\, is the mean from all the individual results \citep[see e.g.][]{1997upa..conf...49P}. 

 In order to  combine incompatible results, we consider the probability that each measurement of $\hat{a}_{i}$($z$) is correct, as well as the probability that it is not correct. A correct measurement is one where the quoted uncertainty is accurate, whereas an incorrect measurement is quoting uncertainties that are too small or might suffer from systematics. Incorrect measurements have error bars that are too small, hence they are incompatible. 

We follow the method describe by
\citet{1997upa..conf...49P}, and in particular their Equation~16, which gives the posterior probability for the value of \amean\, given data:

\begin{equation}
{{\rm P}(\langle \hat{a}(z)\rangle | {\rm data})} \propto \int  {\rm P({\it p})}\prod_{i}[p{\rm P}_{\rm C \it i} + (1-p){\rm P}_{\rm I \it i}]{\rm d}p
\label{eq:press}
\end{equation}

\noindent where $P_{\rm C \it i}$ and $P_{\rm I \it i}$ are the probability distribution
functions for correct and incorrect estimates, and the product $\prod_{i}$ is over the $n_{i}$ different estimates. 

The first term in the square bracket is the product of the probability $p$ that measurement $i$ is correct times the probability distribution function for that measurement (using the standard deviation of the correct measurement, $\sigma_{\rm C \it i}$). The second term is the product of the probability 1-$p$ that the measurement is incorrect. The probability distribution function $P_{\rm I \it i}$ therefore requires a standard deviation which is large enough to make the measurement correct. We label this as $\sigma_{\rm I \it i}$. The integral over d$p$ means we marginalise over both cases. 

We approximate the probability distribution functions for \amean\, as gaussians. In the case of 
$P_{\rm C \it i}$ , we assume the standard deviation $\sigma_{\rm C \it i}$ to given by the  errors shown on Figure~\ref{fig:exp_a_z}.

\begin{figure*}
\hbox{ \psfig{file=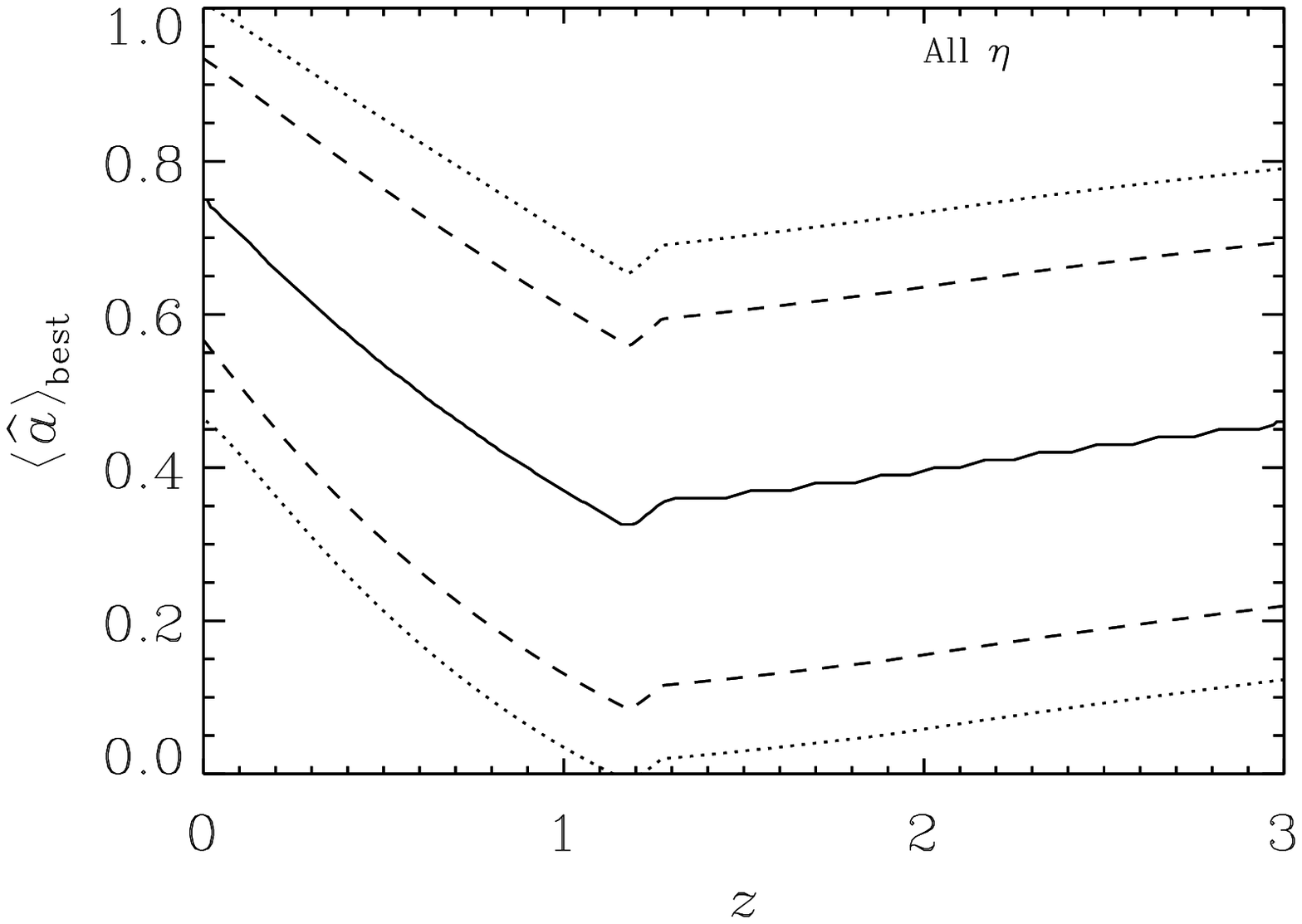, width=8cm, angle=0} \psfig{file=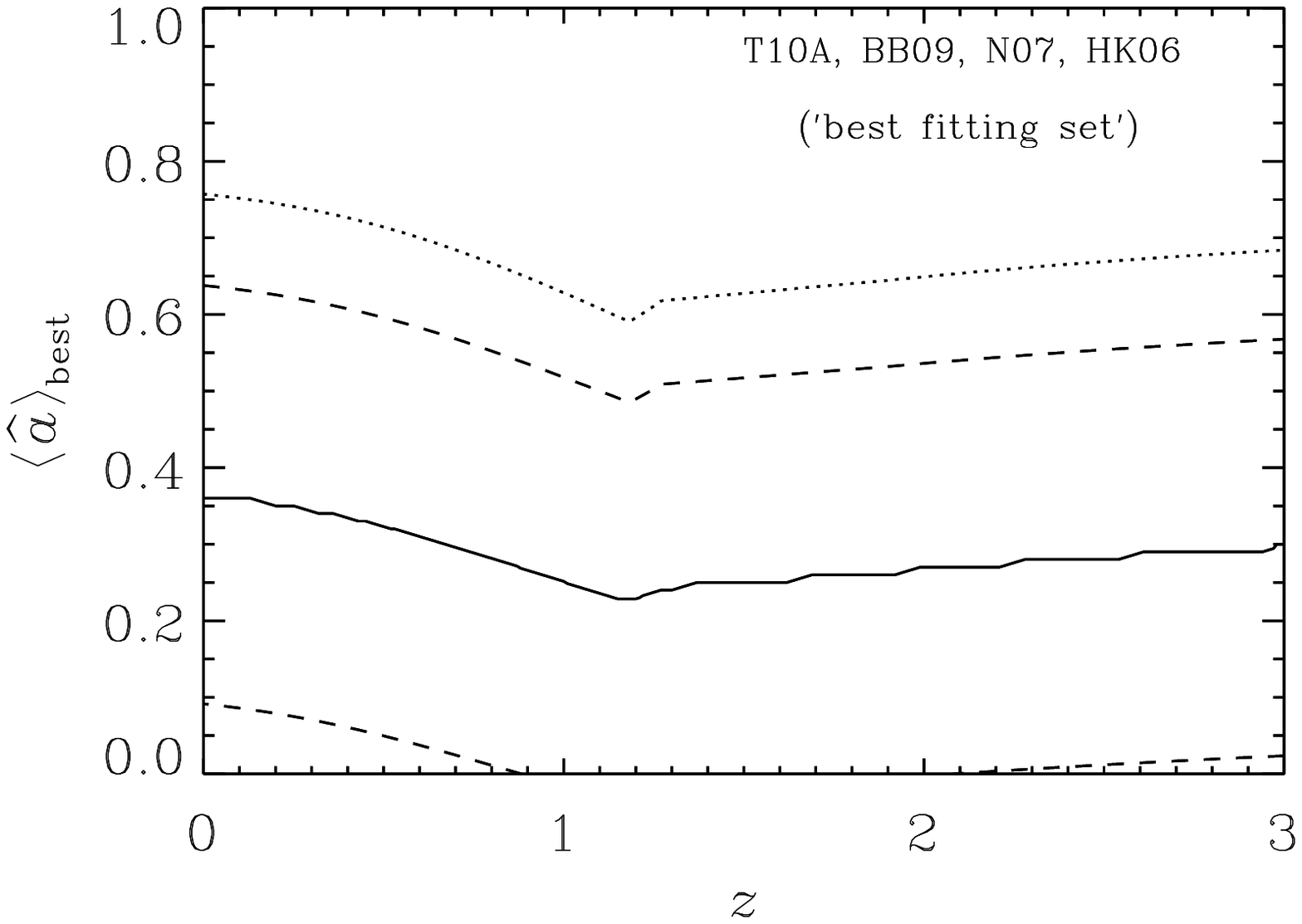, width=8cm, angle=0} }
\caption{\noindent  Best estimates of the evolution of the mean cosmic spin of SMBHs with  redshift, derived using Equation~\ref{eq:press}. The left panel combines all the results of Figure~\ref{fig:exp_a_z}, and since these are not compatible, it considers the probability that each independent estimate is correct as well as the probability that it is incorrect, and marginalises over these (see Equation~\ref{eq:press1}). The definition of compatible results is given in Section~\ref{sec:best_est}. The right panel uses the results from T10A, BB09, N07 and HK06 only (the `best fitting set'), and since these are compatible, it  considers all the independent estimates to be correct, and is simply the product of their individual posterior probability distribution functions (Equation~\ref{eq:press2}). 
In this case a more modest spin evolution is seen between $z$$=$1 and 0., and note that the -1$\sigma$ and -2$\sigma$ contours fall mostly below the \sp$=$0. Both estimates assume that the space density of low-accretion rate SMBHs, ADAFS described by the local black hole mass function, is constant with $z$. Since the ADAFs tend to have higher spin values, this approximation is likely to overestimate the mean spin at the high-redshift end ($z$$\grtsim$1). }
\label{fig:az_press}
\end{figure*}

We now consider two extreme cases:

{\it 1. There is the possibility that some of the estimates of \amean\, are indeed incorrect.} If, however, incorrect estimates were assigned a larger uncertainty, they would become correct. 
How large this new uncertainty should be is somewhat arbitrary, and might vary from model to model. Hence, we assign the same very large new uncertainty to all estimates, $\sigma_{\rm I \it i}$$=$$\sigma_{\rm I}$ =1, and within this uncertainty all our estimates agree.  We also assume a flat prior for the probability that measurements are correct, in the range 0$\leq$$p$$\leq$1, so that P($p$)$=$1. Equation~\ref{eq:press} then simplifies to:

\begin{eqnarray}
{{\rm P}(\langle \hat{a}(z)\rangle | {\rm data})} =  \nonumber \\ 
{e^{-{1\over 2}\sum_{i} \left({a(z) - \langle \hat{a}(z) \rangle_{i}  \over \sigma_{\rm C \it i} } \right)^{2}} \over 2 \left(2\pi\right)^{n_{i}\over 2}\prod_{i} \sigma_{\rm C \it i} }  +  {e^{-{1\over 2}\sum_{i} \left(a(z) - \langle \hat{a}(z) \rangle_{i}   \right)^{2}} \over 2 \left(2\pi\right)^{n_{i}\over 2} }  
\label{eq:press1}
\end{eqnarray}

{\it 2. The measurements are all correct, since their uncertainties are large, they are formally consistent with each other.} We therefore assign a delta-function prior at `certainty', P($p$)$=$$\delta$($p$$=$1) and Equation~\ref{eq:press} simplifies instead to the standard result:

\begin{equation}
{{\rm P}(\langle \hat{a}(z)\rangle | {\rm data})} =  {e^{-{1\over 2}\sum_{i} \left({a(z) - \langle \hat{a}(z) \rangle_{i}  \over \sigma_{\rm C \it i} } \right)^{2}} \over \left(2\pi\right)^{n_{i}\over 2}\prod_{i} \sigma_{\rm C \it i} }  
\label{eq:press2}
\end{equation}

Figure~\ref{fig:az_press}   shows the resulting evolution of \amean\, under these two assumptions. 
 When combining all the results of Figure~\ref{fig:exp_a_z}, some of the results are incompatible, so that we use case  {\it 1} (Equation~\ref{eq:press1}). The typical spin at $z$$=$1 is  \sp$\sim$0.35  and increases to $\sim$0.75 at $z$$=$0.  The right panel of Figure~\ref{fig:az_press} shows the resulting spin history when  combining the T10A, BB09, N07 and HK06 results only, using case {\it 2} and Equation~\ref{eq:press2}. In this case a more modest increase  is seen, from \sp$\sim$0.25 at $z$$=$1 to $\sim$0.35 at $z$$=$0.  The right panel is derived from our best fitting set, and represents our best estimate of the mean spin history.  

We stress that at all redshifts, we are assuming the population of ADAFs is well represented by the $z$$=$0 SMBH mass function convolved with the distribution of Eddington ratios discussed in Section~\ref{sec:low_acc}. This is probably a good approximation up to $z$$\sim$1,  since most of the mass of SMBHs with \mbh$\geq$$10^{8}$~\msol\, was accreted at higher redshifts. The SMBH mass function is probably similar at $z$$\sim$1 and $z$$\sim$0, but at $z$$\sim$3 it will be significantly different. The space density of high-mass SMBHs sould be much lower. Hence our inferred spin history contains too high a contribution from ADAFs at high redshifts, so that we probably overestimate the spin at $z$$\grtsim$1.

Finally, we note that although the evolution of the expectation value, \amean, appears weak, this is not necessarily the best way to illustrate the evolution in spin for such bimodal distributions.  The evolution in spin is stronger than what \amean\, would suggest:  at high redshift most of the SMBHs have the spin distribution describing the QSOs so that most sources have very low spins, with a negligible fraction having high spins. At low redshift, most of the SMBHs are described by the ADAF spin distribution, meaning that around half of the sources have high spins.

It is important to note that the evolution in the spin is driven by the disappearance of high-accretion rate, low-spin SMBHs, whose space density decreases enough that the approximately constant population of low-accretion rate, high-spin objects becomes dominant. 

\section{Comparison to other inferences on black hole spin}\label{sec:other}

The spin of black holes in AGN have been estimated in different ways, and our results do not necessarily agree with those of other groups. For example, \citet{2002ApJ...565L..75E} and \citet{2006ApJ...642L.111W}  have claimed that most SMBHs must be spinning rapidly, while more recently
\citet{2010MNRAS.406.1425F} have found no correlation between black hole spin and jet power for X-ray binaries.  We therefore compare our results to previous work and discuss the possible reasons for discrepancies.

\subsection{The mean radiative efficiency}

We begin by comparing to estimates of the spin from the mean radiative efficiency. Comparing the total energy radiated by AGN over cosmic time with the mass density of SMBHs in the local Universe can be used to estimate the mean radiative efficiency \citep{1982MNRAS.200..115S}. The mean radiative efficiency can be converted into a black hole spin using the NT73 model, but as is shown in Figure~\ref{fig:eps}, the radiative efficiency from simulations can differ from the NT73 by factors of $\sim$2, resulting in large uncertainties. 

 \citet{2002ApJ...565L..75E} used the intensity of the X-ray background and assumed the bulk of the AGN causing it to be at $z$$=$2. The authors inferred $\langle \epsilon \rangle$$\geq$0.15, suggesting the mean spin of SMBHs is $\hat{a}$$\sim$0.9. The bulk of the sources producing the X-ray backround, however, are at lower redshifts than their assumed value of $z$$=$2.  
 A more appropriate characteristic redshift would have been $z$$=$0.7-1.0 \citep[see e.g.][]{2003AJ....126..632B,2003ApJ...598..886U}, which would bring their mean 
 efficiency down to values of $\langle \epsilon \rangle$$\grtsim$0.08-0.1.

Another high value of the radiative efficiency was inferred by \citet{2006ApJ...642L.111W} using a different method. These authors compared the luminosity function of quasars to the mass function of quasars in redshift bins, and estimate $\langle \epsilon \rangle$ to be in the range 0.20-0.35, suggesting the SMBHs are spinning rapidly. 

In order to derive a radiative efficiency, \citet{2006ApJ...642L.111W}
equate the integral of the luminosity function ($\int L_{\rm bol}
\phi(L_{\rm bol}) {\rm d}L_{\rm bol}$) to the integral of the
mass-density function of quasars ($\int m_{\bullet} \phi(m_{\bullet})
{\rm d}m_{\bullet}$). However, these authors derive the mass functions
from the broad-line widths of the quasars in a magnitude-limited
sample.  Only quasars brighter than the limiting magnitude ($B$-band
magnitude of 19.4) can have their masses measured and included. Since
quasars have a range of Eddington ratios (e.g. McLure \& Dunlop 2004),
the correspondence between bolometric luminosity and black hole mass
is not one-to-one. While the luminosity integral used by
\citet{2006ApJ...642L.111W} is complete above their flux limit, the
mass integral is not. This is because not all quasars with mass above
their nominal mass lower limit are optically bright enough to be detected in
their survey so that their mass function is incomplete. The fact that
the luminosity integral is complete but the mass integral is
incomplete means that they will overestimate the amount of energy that
was radiated away. We therefore consider their estimate of the
radiative efficiency, and hence spin, as an upper limit.

Most of the recent estimates of the mean efficiency from the So{\l}tan argument typically find values of $\langle \epsilon \rangle$ in the range 0.06-0.1, with a typical value of  $\langle \epsilon \rangle$$\sim$0.07 \citep[e.g.][ see their Section~5  for a discussion  of recent results]{2004MNRAS.351..169M,2008MNRAS.390..561C,2009ApJ...690...20S,2009ApJ...692..964M}.

We can compare all the previously-mentioned estimates to our results by computing the 
luminosity-weighted mean spin:

\begin{eqnarray}
\langle \hat{a}\rangle_{\rm L}  \equiv { \int \langle \hat{a}(z) \rangle_{\rm best} \langle L_{\rm bol} \rangle {\rm d}z \over  \int  \langle L_{\rm bol} \rangle {\rm d}z}  \nonumber \\
= {\int \left[ \int \hat{a}(z) {{\rm P}(\langle \hat{a}(z)\rangle | {\rm data})} {\rm d}\hat{a}\right]\left[   \int C_{\rm X} L_{\rm X} \phi(L_{\rm X},z) {\rm d}L_{\rm X}  \right] {\rm d}z \over  \int  \int C_{\rm X} L_{\rm X} \phi(L_{\rm X},z) {\rm d}L_{\rm X}  {\rm d}z}
\label{eq:soltan}
\end{eqnarray}

For the weighting function in Equation~\ref{eq:soltan}, we use  the X-ray LF in the range 0$\leq$$z$$\leq$3, and due to the strong evolution of the X-ray LF,  practically all of the weight lies with the high redshift values of \amean, since  the bulk of the luminosity was emitted at $z$$\geq$1.  

 ${{\rm P}(\langle \hat{a}(z)\rangle | {\rm data})}$ is given by the left panel of Figure~\ref{fig:az_press}, which then yields a value of $\langle \hat{a}\rangle_{\rm L} $$=$0.40. The resultant mean radiative efficiency   $\langle \epsilon \rangle_{\rm L}$ is therefore $\epsilon$(0.40)$=$0.075. However, when using  the spin history from the right panel in Figure~\ref{fig:az_press} (the best fitting set), we infer $\langle \hat{a}\rangle_{\rm L} $$=$0.28, or  $\epsilon$$=$0.068. The mean radiative efficiency derived from our 
 spin history is therefore in excellent agreement with the estimates from the literature.

\subsection{Lack of correlation between estimated spins and jet powers in galactic black holes}\label{sec:lack}

Comparing the spin measurements of GBHs (galactic black holes) with
their radio luminosities, \citet{2010MNRAS.406.1425F} conclude that
there is evidence for an additional parameter which controls the
amount of power in jets, yet this parameter does not correlate with
the spins estimated from fitting the accretion disc or the reflection
edge.

In many aspects, AGN behave as scaled up versions of GBHs
\citep[e.g.][]{2002Natur.417..625M,2003MNRAS.345.1057M,2004A&A...414..895F,
  2006Natur.444..730M}. The lack of correlation between `radio
loudness' and spin in GBHs is  therefore a concern for our work.

It is encouraging that when plotting a measure of bolometric
luminosity versus a measure of jet power, both GBHs and SMBHs show
correlations spanning several decades, but also a scatter of 2 or 3
decades in radio luminosity
\citep[e.g.][]{2007ApJ...658..815S,2010MNRAS.406.1425F}. In this
paper, the scatter has been interpreted as arising mainly due to
variations in black hole spin \citep[see also][]{2007ApJ...658..815S},
however \citet{2010MNRAS.406.1425F} find no correlation with published
spin estimates for the GBHs.  \citet{2010MNRAS.406.1425F} conclude that the lack of correlation could be due to spin having no
effect on the observed jet power, or alternatively due to either the spin measurements
or the measurements of jet power being inaccurate.  

It is possible that at low spins,  the power from the disc jet is larger than the power from the black hole jet (e.g. the HK06, N07 and BB09 jet efficiencies in Figure~\ref{fig:eta}), complicating the interpretation. However, if spin has an observable effect on the jet power, then one would expect some correlation between spin and jet power. If the disc jet were always more important than the jet driven by the spin of the black hole, then the spin would be irrelevant when explaining observations of jets.

 Subsection~\ref{sec:jet_gbhs} discusses the uncertainties in estimating the jet power from core radio luminosities, Subsection~\ref{sec:spinx} discusses the uncertainties in the estimates of black hole spin from fitting the accretion disc luminosity or the iron reflection lines, and  Section~\ref{sec:nocorr} discusses the observed lack of correlation.

\subsubsection{Estimated jet powers}\label{sec:jet_gbhs}

 \citet{2010MNRAS.406.1425F} measure the jet powers from the core luminosities, as is usual in the GBH community. However, for AGN the core luminosity is totally incapable of predicting  the total radio luminosity \citep[see Figure~7 of][]{2009MNRAS.398..176K}, and it is the total radio luminosity that is required to trace the jet power \citep[e.g.][]{1991Natur.349..138R,1999MNRAS.309.1017W,2004ApJ...607..800B,2010ApJ...720.1066C}.

We note the cases of Cygnus X-1 and S26 (a microquasar with extended radio structure reminiscent of an FRII radio galaxy)  where the extended lobe radio luminosity is $\sim$2 decades brighter than expected from the cores \citep{2005Natur.436..819G,2010MNRAS.409..541S}, which in the case of S26 is not even detected.  
It seems that in GBHs, the core luminosity might not be  a good tracer of the total jet power either  \citep[see also e.g.][for the case of GRS~1915+105]{2004ApJ...612..332K}. 

 To complicate things further, the discovery of  a one-sided bow-shock in Cygnus X-1 \citep{2005Natur.436..819G}, suggests the
environments of the GBHs can be much less homogeneous than those of AGN, on the respective scales of their jets.

The extended jets will not be detectable unless they are seen to do work on  the interstellar medium (as is only the case in the Northern lobe of Cygnus X-1). This suggests accurate determination of the total jet power in GBHs might suffer from even larger uncertainties than in the AGN case (Section~\ref{sec:obs}).

\subsubsection{Estimated spins}\label{sec:spinx}

The spin estimates used by  \citet{2010MNRAS.406.1425F} are  obtained either from modelling the thermal continuum from the
accretion disc \citep{1997ApJ...482L.155Z} or from the iron fluorescence reflection lines \citep{1989MNRAS.238..729F}. 

The spin measurements from the iron reflection (K-$\alpha$ and L) lines assume that the red wing
of the line is due to general-relativistic  effects in the vicinity of a black hole. From the 
line width, the characteristic radius where most of the reflection is occurring is estimated. 

The disc-fitting methods model the spectrum of the source as continuum thermal emission originating in an accretion disc. The free parameters include the mass of the black hole and the spin, which determines the truncation radius,  
the accretion rate as well as the tilt angle of the accretion disc.

The two methods  rely on identifying a
characteristic radius. This radius can be where the continuum emission truncates or 
where the bulk of the reflection originates. If the observed characteristic radius corresponds
closely to the ISCO, which depends on the spin parameter in the Kerr metric, then 
the characteristic radius yields an estimate of the spin.  

However, if the characteristic radius does not correspond to the ISCO, then the resulting spin estimate will be incorrect. One concern is therefore whether significant radiation or reflection can occur at radii smaller than the ISCO, so that assigning the characteristic radius to the ISCO underestimates the ISCO and therefore overestimates the spin. 

Several authors have studied this possibility, with varying conclusions, and we summarise here
 recent progress.  \citet{2010A&A...521A..15A} discuss how different methods (disc-fitting, iron  lines) will probe different characteristic radii. They argue that for low Eddington ratios, $\lambda$$<$0.3, implying thin discs,  the approximation that these radiation or reflection edges will correspond to the ISCO is valid, but that this approximation will break down for the higher Eddington ratios that produce thicker discs. The results of \citet{2010ApJ...718L.117S} for LMC X-3 support this: analysing data that span 26 years, and using only the data for periods during which the Eddington ratio is 0.05$\leq$$\lambda$$\leq$0.3, they find an inner disk radius that is constant to within 4-6\%.

 The simulations by \citet{2008ApJ...687L..25S}, \citet{2008ApJ...675.1048R} and \citet{2010MNRAS.408..752P} also support the view that for thin discs ($h/r$$\lesssim$0.1) the deviations from NT are small. 

The simulations of \citet{2008MNRAS.390...21B} and \citet{2009ApJ...692..411N} suggest that deviations from the NT73 model are particularly important at low-spins. \citet{2008MNRAS.390...21B} also discuss the possible dependence of this on $h/r$, and note that  height in their simulations is  $\sim$0.06-0.2. 

For moderately thick discs  ($h/r$$\sim$0.2), the simulations by  \citet {2009ApJ...706L.246F}  
found that when the tilt angle was zero, the dependence of the characteristic radius with spin resembled closely that of the NT73 model. However, with tilt angles of only 15 degrees, the characteristic radius is constant with spin, and 
any attempt to infer the spin from the characteristic radius will severely underestimate the spin \citep{2009ApJ...706L.246F}.

Another potential problem can occur if the accretion disc truncates at a larger radius than the ISCO, for example if the accretion rate is too low to sustain an optically-thin disc in the innermost region, and the accretion flow becomes an ADAF instead. In this case,  the radius of the ISCO will be overestimated, meaning the inferred value of the spin will be a lower limit. Indeed, in Cygnus X-1, the fits to the thermal continuum and the K-$\alpha$ line suggest the truncation radius is at a distance $\sim$2-3$\times$ larger than the ISCO \citep{1999MNRAS.309..496G,2008PASJ...60..585M}\footnote{ For another example where the truncation radius might larger, see the study of XTE J1650-500 by \citet{2006MNRAS.367..659D}.}.

For one of the sources with a measured high spin \citep[\sp$=$0.94][]{2009ApJ...697..900M}, a subsequent analysis by \citet{2010MNRAS.407.2287D} has found that the extremely broad line is artefact due to the brightness of the source causing `Pile-up' of photons\footnote{Pile-up occurs when more than one photon reaches the CCD in between two read outs. The local excess in charge will appear larger, hence a higher energy will be erroneously assigned to the photon. However, if the excess is above a cut-off threshold, the flux from photons will not be counted at all. See e.g. \citet{1999A&AS..135..371B} or \citet{2010MNRAS.407.2287D} for more details.}. 
The authors find that after taking this effect into account the line is narrower than previously determined, corresponding to a characteristic radius that does not require high spin and might even imply a truncated disc. 

Indeed,
we note that spin measurements of the same GBH from fitting the disc
continuum and from the fitting the reflection components do not always
agree within the quoted uncertainties \citep[see Table~1 of ][]{2010MNRAS.406.1425F}. Some GBHs have had their spin estimated both from from disc-fitting and from iron lines, yet the two estimates do not always agree within their formal uncertainties. 

From disc-fitting alone,  the reported measurements for GRS1915+105 from disc-fitting range from 0 to 0.998 \citep[see references in][]{2010MNRAS.406.1425F}. More recently, \citet{2009ApJ...706...60B} have estimated the spin again from disc-fitting and from the K-$\alpha$ reflection line, and obtain spin values of  \sp$=$0.56 and 0.98 respectively. \citet{2009ApJ...706...60B} use observations of GRS1915+105 during a low-hard state only, with $\lambda$$<$0.3, and argue that some of the discrepancy with other groups  could be partly due to observing at different states \citep[e.g.][ observed the source when $\lambda$$>$0.3]{2006MNRAS.373.1004M}.

Finally, an additional complication arises from the presence of absorbing material 
along the line of sight and near the black hole. Photoelectric absorption from 
gas with a range of ionisation parameters can lead to a spectral shape similar to a 
red wing, and some authors suggests that the broad reflection lines can be fitted without requiring relativistic blurring \citep[e.g.][]{2006MNRAS.367..659D,2008A&A...483..437M,2009A&ARv..17...47T,2009PASJ...61.1355M,2010MNRAS.401..411Y}.  This is an area of intense debate, where other authors suggest that relativistic blurring is indeed required to explain the lines \citep[e.g.][]{2009MNRAS.397L..21R,2009Natur.459..540F,2010MNRAS.401.2419Z}.

\subsubsection{Lack of correlation between spins and jet powers}\label{sec:nocorr}
 
 To summarise, while measurements from the reflection lines of disc fitting offer an exciting and promising method to determine black hole spin, the accuracy of these measurements is still being debated.  Our primary concern is that the characteristic radius inferred by these measurements might not always correspond to the ISCO. These concerns apply to fitting both the continuum-fitting and the reflection line methods.

 This is not a criticism on our part, since the spin estimated from $\eta$ in our work is also hugely uncertain and
 Figure~\ref{fig:eta} shows the different jet efficiencies we have
 used: it is clear that they vary enormously.  Converting between jet powers and 
radio luminosities is also highly uncertain in GBHs as well as in AGN (see  Sections~\ref{sec:obs} and \ref{sec:jet_gbhs}). An additional consideration to bear in mind is that measures of AGN jet power from the total radio luminosities are actually derived from the total energy deposited over typically $\sim$$10^{7}$ years, so that they are only time-averaged values. 
 
Given the uncertainties we do not think that the observed lack of 
 correlation 
 between  spins and radio luminosities found by 
\citet{2010MNRAS.406.1425F} is at odds with our assumption that spin is a monotonically-increasing function of spin. The uncertainties in the spin estimates and the estimates of the jet power are still too large. We do not think that the current evidence is strong enough to conclusively rule out the spin paradigm.  

We stress that there is strong evidence amongst GBHs for the jet power to correlate with the state of the accretion flow \citep[e.g.][]{2004MNRAS.355.1105F,2007A&ARv..15....1D}.  We are not arguing here that such effects are not important in AGN. 

Amongst AGN, however, huge differences 
in (time-averaged) jet power are also observed with no noticeable  difference in the accretion flow \citep[reflected, for example, by the shape of the SED or the Eddington ration, e.g.][]{1994ApJS...95....1E,2007ApJ...658..815S}. It is this difference that requires an additional hidden variable such as spin.

Finally, we note also that winds with similar escape velocities have been detected in radio-loud and radio-quiet AGN \citep[][2010b]{2010A&A...521A..57T}. Their results are consistent with both populations having similar underlying distribution, but  the radio-loud sample is much smaller (5 objects) than the radio-quiet sample (42), and unsurprisingly the latter shows a larger spread in velocities, with maximum values $\sim$0.3$c$. These winds are likely to be powered by accretion close to the Eddington limit \citep{2003MNRAS.345..657K}, rather than jets.

\section{Discussion}\label{sec:disc}

We have used an assumption where the power available for the production of jets is a fraction of the accreted energy, $\eta$. We have used the values for $\eta$ given by recent theoretical and simulation work, summarised in Figure~\ref{fig:eta}.  This approximation can explain the radio loudness of quasars as a dependence on the spin of the accreting SMBH as well as explaining several observed properties of radio-selected AGNs.

In Section~\ref{sec:fit_rlf} we use three parametrisations for the spin distribution of high- and low-accretion rate SMBHs, and use the radio LFs of HEGs and LEGs together with a bayesian model-selection method to select the most appropriate parametrisation. A two-gaussian model is chosen,
ahead of power-law distributions or single-gaussian models.  This two-gaussian distribution of spins can explain reasonably well the radio LF of HEGs and LEGs individually.

It is interesting that both the high-accretion rate regime (in the form of the upper envelope for the quasars in Section~\ref{sec:loud}), as well as the low-accretion rate regime (the ADAF-LEGs in Section~\ref{sec:fit_rlf}) require a mechanism to produce very high jet efficiencies, $\eta_{\rm max}$$\sim$1. This seems to be independent of the accretion mode, and black hole spin provides one possible mechanism for such high jet efficiencies, as shown by some of the models in Figure~\ref{fig:eta}.

For the two-gaussian distribution, the best-fit parameters for the LEGs-ADAFs choses  two gaussians of similar amplitude, one with minimum spin and the other with maximum spin. This suggests that there are similar numbers of high-spin and low-spin SMBHs amongst the SMBHs with low-accretion rates.  The best-fit distribution for the HEGs-QSOs is also found to have a  bimodal distribution, but this time with a very low amplitude for the high spin peak. Hence, most of  the  SMBHs with  high-accretion rates have low spins.

Since we infer a bimodal distribution for spins, which is heavily favoured over smooth power-law distributions, our model predicts a dichotomy in the radio-loudness of quasars. However, detecting such a dichotomy might prove difficult due to source-to-source variations in the conversion from jet power to radio luminosity density and from bolometric to monochromatic luminosity, as well as due to observational selection effects. 

We are able to make falsifiable predictions about the contribution of HEGs and LEGs to the  radio LF. At $z$$=$0, we predict that the radio LF of LEGs is dominated by black holes with \mbh$\grtsim$$10^{8}$~\msol\, down to radio luminosity densities of $\sim$$10^{21}$~\whzsr\, at 1.4~GHz. 
At $z$$=$1 and above a certain luminosity density, which we estimate to correspond to $\sim$$10^{26}$~\whzsr\, at 1.4~GHz, we expect all the radio sources to be composed entirely of HEGs. Below this value, we expect a rapidly increasing number of LEGs.  The appearance of a large population of HEGs amongst the radio sources is due to the the strong increase of the comoving accretion rate onto SMBHs.

As discussed in Section~\ref{sec:spin_his}, the evolution in cosmic spin is most notable in the fraction of SMBHs with high spins. This fraction is low at high redshifts, and increases towards lower redshifts.  The evolution in spin is really driven by the gradual disappearance of a high-accretion rate but low-spin population, so that the low-accretion rate and high-spin population becomes gradually dominant.    

\begin{figure*}
\hbox{ \psfig{file=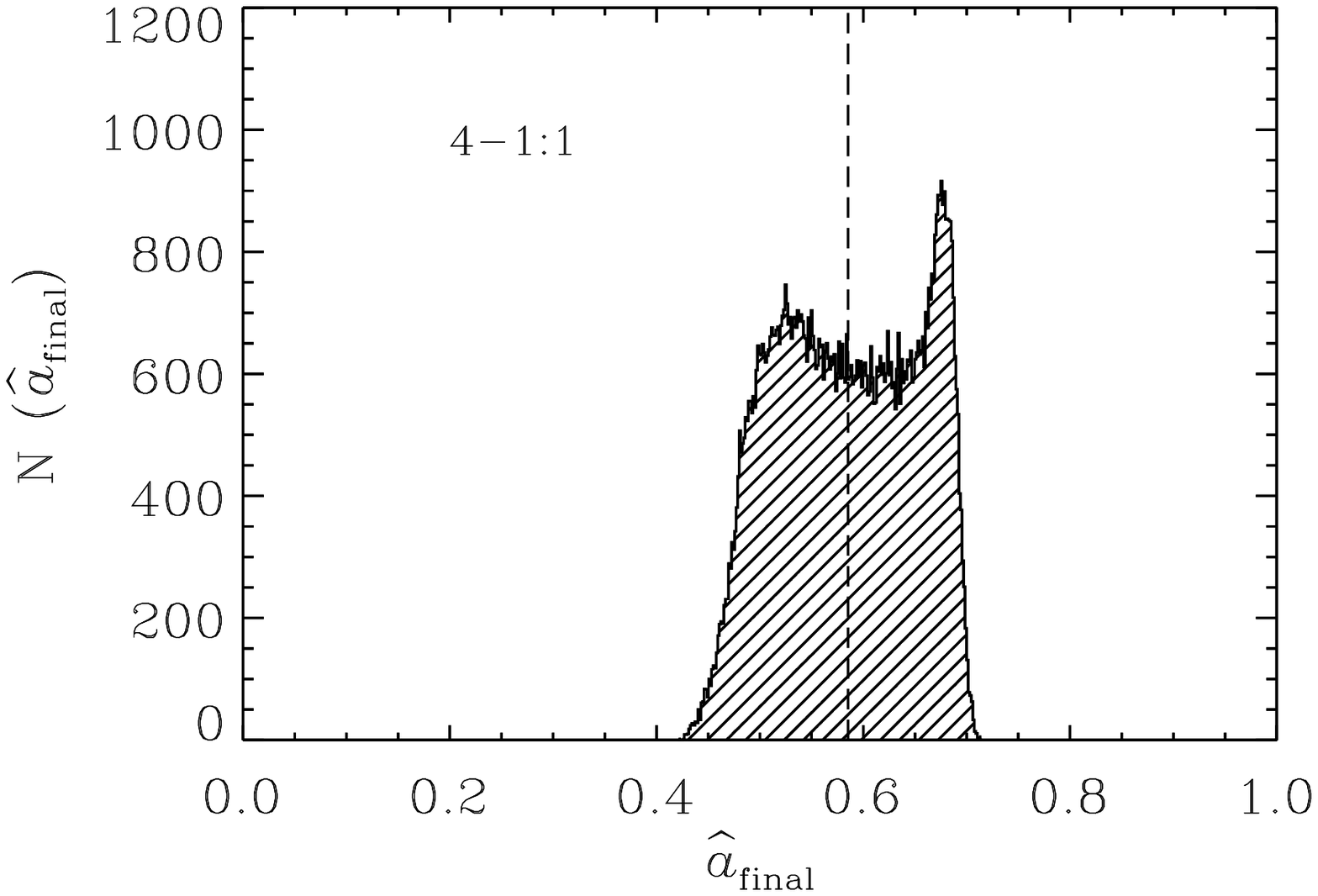, width=8cm, angle=0}\psfig{file=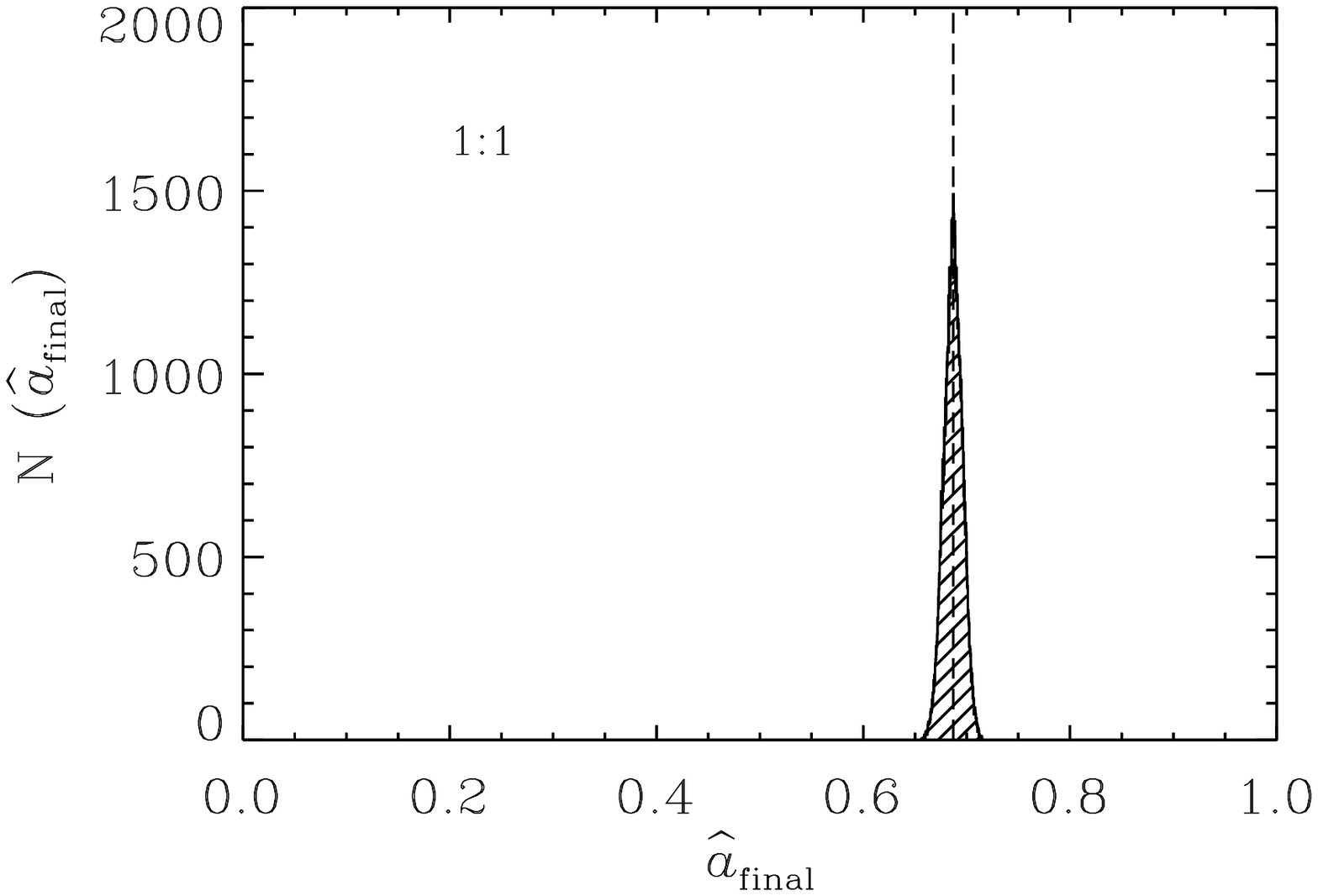, width=8cm, angle=0}}
\caption{\noindent Distribution of final black hole spins from a Montecarlo simulation of $10^{5}$ 
SMBHs undergoing one major merger each. We use the analytic formula of \citet{2008PhRvD..78d4002R},  which requires the initial masses and spins of both black holes, as well as three angles describing the relative orientations with each other of the angular momentum vectors of each black hole and the total angular momentum vector (for a total of 7 free parameters).  The three angles have flat  probability distribution functions between  0 and 2$\pi$. 
The two initial spins are allowed any values between 0 and 0.1, while the three angles any value between 0 and 2$\pi$. The smaller mass is always fixed at a value of 1. In the left-hand panel,  the larger mass is drawn from a flat distribution within the range 1-4. We are therefore  representing what is normally considered as a major merger.  In the right-hand panel, the larger mass is fixed at 1, representing equal mass mergers. The vertical dashed line represents the mean.  The panels illustrate that high mean spin values, $\langle \hat{a} \rangle$$=$0.6-0.7 can be reached by single-event major mergers. Allowing  the initial spins to have values   0$\leq$\sp$\leq$1 instead, with a flat distribution in this range, leads to much broader distributions of the final spins, but with essentially the same mean values.   }
\label{fig:bhmerg}
\end{figure*}

Combining the results from the different  jet efficiencies, with the best-fitting parameters, and the evolution of the X-ray LF, we are able to estimate the expectation value for the spin as a function of redshift for  black holes with \mbh$\grtsim10^{8}$~\msol: We find marginal evidence for evolution in the spin distributions. If we use all the efficiencies, we find the fraction of SMBHs with  \sp$\geq$0.5 increases from 0.34$\pm$0.29 at $z$$=$1 to 0.45$\pm$0.33 at $z=$0. If only the best fitting set is used, this fraction is found to be 0.16$\pm$0.03 at $z$$=$1 and 0.24$\pm$0.09 at $z$$=$0.

The mean cosmic spin increases from $\hat{a}$$\sim$0.35 at $z$$\sim$1 to $\hat{a}$$\sim$0.75 at $z$$\sim$0. However, when using the three best  fitting jet efficiencies only, the evolution is found to be more modest: from \sp$\sim$0.25 to \sp$\sim$0.35 in the same redshift interval. This second set of values represents our best estimate. 

The luminosity-weighted spin comes out to be $\hat{a}_{\rm L}$$=$0.40 for all efficiencies and  $\hat{a}_{\rm L}$$=$0.28 for the best fitting set. This predicts the mean radiative efficiency to be $\langle \epsilon \rangle$$=$0.075  (or $\langle \epsilon \rangle$$=$0.068 when using the best fitting set). These quantities can be compared to independent constraints from the So{\l}tan argument, and are in excellent agreement with them. 

Overall, we infer a very modest evolution of the mean spin, although given the bimodal distributions a mean value is not the best description of the evolution of the spin. Indeed, the spin distributions evolve from distributions dominated by a single low spin component at high redshift, towards bimodal distributions at low redshift.  
In addition, as stressed in Section~\ref{sec:spin_his}, we are probably overestimating the high-spin contribution at $z$$\grtsim$1. 

Perhaps counter-intuitively, the epoch of lowest black hole spin corresponds to the epoch of highest accretion. 
The evolution of \amean\, inferred here, however,  is consistent with the picture of `chaotic accretion': matter from the host galaxy falls into the quasar from random directions and with different orientations of the angular momentum vector \citep[see ][]{2008MNRAS.385.1621K}.

Depending on the angle between the angular momentum vectors of the black hole and the accretion disc, these two will either align or counter-align. During approximately half of the time the accreted matter has the opposite direction of angular momentum to the black hole \citep{2005MNRAS.363...49K,2006MNRAS.368.1196L}. Co-rotating orbits are physically closer to the event horizon of the hole than counter-rotating ones. Therefore co-rotating matter increases the angular momentum by a smaller amount than the decrease brought by counter-rotating matter.   Chaotic accretion is expect to lead to SMBHs with low spins \citep[\sp$\sim$0.1][]{2008MNRAS.385.1621K}, and our result is in agreement with spun down AGNs at high redshift, when the mean accretion rate was higher.

For accretion to significantly affect the spin of a black hole, the increase in mass must be significant \citep{1970Natur.226...64B}.  We approximate this as the condition

\begin{equation}
{\Delta m_{\bullet} \over m_{\bullet}} \sim 1,
\label{eq:deltam}
\end{equation}

\noindent which, if not met, implies no significant decrease in  \sp\,  due to chaotic accretion.

\begin{figure}
\begin{center}
\psfig{file=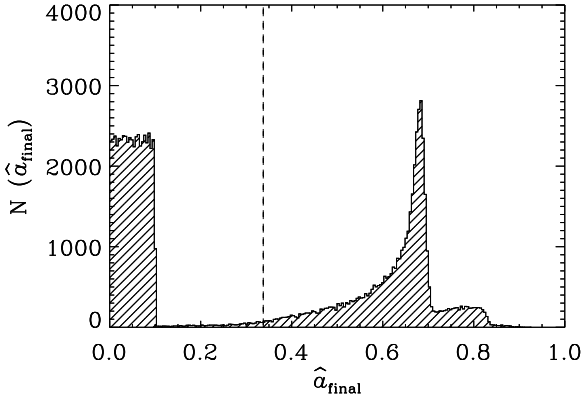, width=8cm, angle=0}
\caption{\noindent  Simulation of final black hole spins after $10^{5}$ SMBHs undergoing a variable number of mergers. The number of mergers undergone  is drawn from a  Poisson function with mean 0.7 \citep[following the results of][]{2010ApJ...719..844R}. The final spin is calculated 
using the analytic formula of \citet{2008PhRvD..78d4002R} (see caption to Figure~\ref{fig:bhmerg}). We have assumed flat distributions for all the two black hole masses, the two black hole spins and the three angles.  In this simulation, the black hole masses are assumed to be a constant fraction $1.4\times10^{-3}$ of the galaxy stellar masses \citep{2004ApJ...604L..89H}. These stellar masses are allowed to have masses in the range $5\times10^{10}-4\times10^{11}$~\msol, and are drawn at random from a Schechter function with a break at mass $1.4\times10^{11}$~\msol, and low-mass end power law index of -1.2 \citep{2001MNRAS.326..255C}. The initial spins of the SMBHs are drawn from a flat probability distribution between \sp$=$$0-0.1$, and the spin of the coalesced SMBH is tracked (later incoming SMBHs are also assumed to have an initial spin in the range 0-0.1). The mergers are assumed to occur isotropically, and the final black hole spins are again inferred using the analytic formula of  \citet{2008PhRvD..78d4002R}. The distribution shows 3 obvious signals: the sources having undergone zero mergers are gathered around $\langle \hat{a} \rangle$$=$0-0.1, those having merged once peak around 0.65, and those having merged twice extend up to 0.85. The resulting  expectation value of \sp$=$0.35 is marked by the vertical dashed line.  }
\label{fig:merg_his}
\end{center}
\end{figure}

The higher mean spin at lower redshifts suggests major mergers of SMBHs are driving the increase in spin.  Galaxy major mergers are expected to lead to major mergers of the central SMBHs. During the coalescence of two black holes of similar mass, the orbital angular momentum will contribute significantly to the final angular momentum of the coalesced black hole. Hence,  major galaxy mergers  are expected to lead to central SMBHs with a moderately high spin value \citep[e.g.][]{2003ApJ...585L.101H,2008PhRvD..78d4002R}.  Minor black hole mergers, on the other hand, tend to lead to low values of spin in the coalesced hole \citep[see ][]{2003ApJ...585L.101H}. Values of $\langle \hat{a} \rangle$$\sim$0.7 can be reasonably achieved by SMBH major mergers, but values $\sim$1 are very hard to reach.

In the low-redshift Universe, the most massive SMBHs, which dominate our estimate of \amean, are found in the most massive galaxies \citep[with bulge masses $\grtsim10^{11}$~\msol, ][]{2004ApJ...604L..89H}.  From mass-selected samples of galaxies and using the two-point correlation function, \citet{2010ApJ...719..844R}  suggest that since $z=1.2$, galaxies more massive than $1\times10^{11}$~\msol\, have undergone, on average, 0.7 mergers involving two progenitor galaxies more massive than $5\times10^{10}$~\msol.

To illustrate the effect of every SMBH undergoing one major merger we have carried out a simple simulation. Figure~\ref{fig:bhmerg} shows the distributions of final spins from two Montecarlo realisations of $10^{5}$ black hole single-event mergers. Once again, the amount of accretion after the merger is assumed to be small (Equation~\ref{eq:deltam}), so that the change in spin after the merger due to accretion is negligible. 

In the left-hand panel, the mass of the more massive black hole is allowed any value in the range 1-4$\times$ larger than the smaller black hole with a uniform distribution. The distribution therefore represents the final product of isotropic  single-event major mergers, the distribution is centred around $\langle \hat{a} \rangle$$=$0.58 and has a full width zero intensity of about 0.3. The right-hand panel represents equal-mass mergers only  (1:1 mass ratio), it is centred around $\langle \hat{a} \rangle$$=$0.69 and is very narrow, with a full width zero intensity of about 0.05. The narrow distribution is partly due to the narrow initial spins, in the range 0-0.1. A broader distribution of initial spins would also lead to a broader distribution of final spins yet the mean would remain unchanged.

We have carried out a second simple simulation, shown in   Figure~\ref{fig:merg_his}. Here,  $10^{5}$ black holes have each suffered a random number of mergers. All initial SMBH spins are drawn from a flat distribution between 0 and 0.1, reflecting the low spin values expected from chaotic accretion. After the merger, they are assumed to accrete a small enough fraction of their own mass that the spin is not significantly changed (the condition in Equation~\ref{eq:deltam} is not met).   The exact number  of mergers each black hole will undergo is determined by drawing from a Poisson distribution with mean 0.7, reflecting the mean number of mergers the host galaxy has experienced since $z\sim1$ \citep{2010ApJ...719..844R}. The stellar masses of the bulges of the host galaxies are drawn from a Schechter function, and the masses of the merging  black hole  are assumed to be a fixed fraction of this mass \citep[1.4$\times10^{-3}$ of the stellar mass,][]{2004ApJ...604L..89H}. The final spin is estimated using the formula given by \citet{2008PhRvD..78d4002R},  where we have randomised the orbital parameters (the orientations of each angular momentum component with respect to each other) as well as the initial black hole spins. 

In this simulation,   50\% of the SMBHs do not undergo any mergers, and they retain the small initial spin (the flat component around $\hat{a}$$=$0-0.1). A significant fraction undergo 1 or 2 mergers (35\% and 12\% respectively, peaking at 0.65 and 0.8), and a small  fraction undergo three or more mergers (3\%, causing the high-spin tail). The distribution shows a sharp peak around \sp$=0.65$, but the expectation spin is only $\langle \hat{a} \rangle=$0.35 (vertical dashed line in Figure~\ref{fig:merg_his}), in very good agreement with our results from the best fitting set. This low mean value is due to the large fraction of black holes that have not undergone any mergers and have kept their original spins (0-0.1). 
The mean from this simple simulation is slightly lower than our estimate from fitting the radio LF of HEGs and LEGs, but the distribution is  qualitatively similar.

As discussed by \citet{2010ApJ...719..844R}, the mass limit of their survey requires both progenitor galaxies to have masses $\geq5\times10^{10}$~\msol. Hence major mergers involving one progenitor with a mass $<5\times10^{10}$~\msol\, are missed out, and their estimate is a lower limit on the mean number of major mergers experienced by galaxies with mass $\grtsim10^{11}$~\msol.  Indeed, studying the projected axial distribution of galaxies more massive than $10^{11}$~\msol,  \citet{2009ApJ...706L.120V}  have argued that they must have all undergone a major merger.  In addition, these mergers are expected to have been relatively gas-poor, otherwise the projected axial distributions would be more similar to discs   \citep{2006MNRAS.372..839N,2009ApJ...706L.120V}.   Hence the mean of 0.7 mergers per galaxy, on average,  might be an underestimate.

However, Figure~\ref{fig:merg_his} does predict an approximately bimodal distribution, qualitatively similar to what we infer at $z$$=$0 (see Section~\ref{sec:res} and in particular Figure~\ref{fig:spin_dbn_z0}), with reasonably similar expectation values.  Our results suggest the high-spin end to peak around \sp$\sim$0.80-0.99, rather than 0.7, but otherwise the distribution Figure~\ref{fig:merg_his}  is similar to the distributions inferred for the LEGs. We do note that achieving the high spins we infer in our bimodal distributions, \sp$\sim$1, remains very difficult.

The galaxy mergers driving the spin-up  do not need to be completely `dry' \citep[e.g.][]{2005AJ....130.2647V}: provided that after the merger,  the condition in Equation~\ref{eq:deltam} is not met, the spin of the SMBHs will be predominantly determined by the coalescence of the two SMBHs as no significant spin-down will occur.  
Given a black hole mass  of $10^{8}$~\msol\, needs to accrete a mass $\sim10^{8}$~\msol\, to spin down, and assuming a fraction 1.4$\times10^{-3}$ of the galaxy's gas makes it to the black hole \citep[e.g.][]{2004ApJ...604L..89H}, spinning down the most massive black holes requires $\sim$$10^{11}$~\msol\, of molecular gas. This is a very high gas mass, much higher than found in low-redshift elliptical galaxies \citep{1991ApJ...379..177L,2002AJ....124..788Y}.

Another key issue is that our estimate of \amean\, is dominated by  black holes with masses $\grtsim$$10^{8}$~\msol, which reside in bulges with stellar masses $\grtsim$$10^{11}$~\msol, and these have generally elliptical morphologies. Thus, we have indirectly selected galaxies with elliptical morphologies, which is a proxy for selecting galaxies that have undergone at least one major merger.  It is the fact that elliptical galaxies have undergone major mergers that allows their central SMBH to have high spin values. This point was noted by \citet{1992ersf.meet..368S} and related arguments have also been put forward by \citet{1995ApJ...438...62W}, \citet{2007ApJ...658..815S} and \citet{2007ApJ...667..704V}.

\begin{figure}
\begin{center}
 \psfig{file=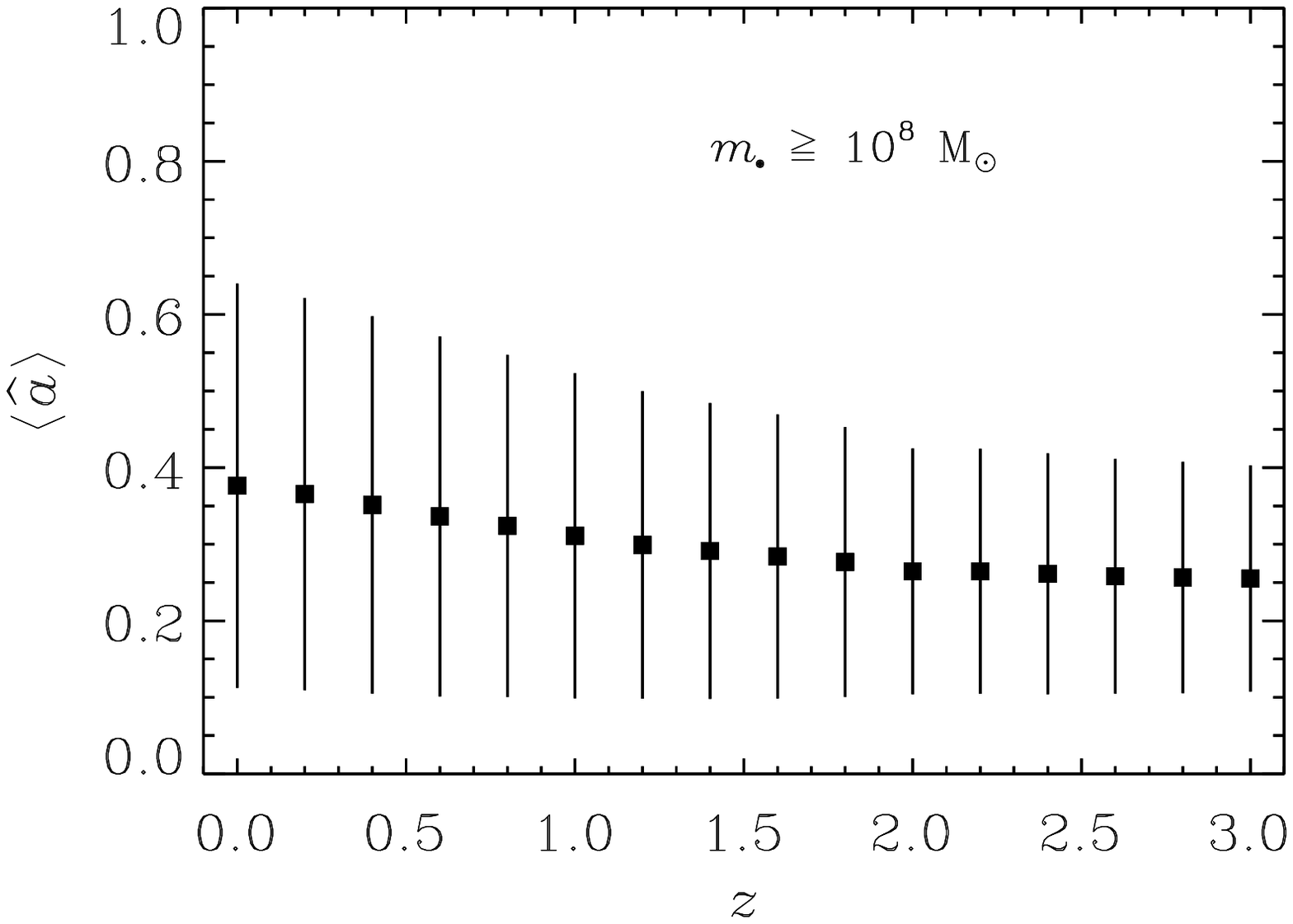, width=8cm, angle=0} 
\caption{\noindent Evolution of the mean spin as a function of redshift for SMBHs with \mbh$\geq10^{8}$~\msol, from the simulation of  \citet{2011MNRAS.410...53F}.  The squares mark the mean  value of the spin in each redshift bin, with the vertical bars representing the standard deviation, rather than error bars. This incorporates mergers of SMBH, using the prescription of  \citet{2008PhRvD..78d4002R} as well as a prescription for `chaotic accretion' \citep{2008MNRAS.385.1621K}. The behaviour of the spin in the simulation is extremely similar to our inference of the evolution from observations (Figure~\ref{fig:az_press}). }
\label{fig:fanidakis}
\end{center}
\end{figure}

Figure~\ref{fig:merg_his} is only a simple simulation, and  detailed studies of the cosmic evolution of SMBH spins from mergers can be found, for example, in \citet{2007ApJ...667..704V}, \citet{2008ApJ...684..822B}, \citet{2009MNRAS.395..625L}.  The simulations carried out by   \citet{2011MNRAS.410...53F} incorporate a semi-analytic model of the accretion onto SMBHs in the centres of galaxies, and in this simulation mergers of galaxies also cause mergers of the central SMBHs. The simulation tracks down the spin of the central SMBHs, and assumes chaotic accretion spins the black holes down.  Figure~\ref{fig:fanidakis} shows the resulting spin history for black holes with \mbh$\geq 10^{8}$~\msol\, (N. Fanidakis, 2010, priv. comm.).  

The behaviour seen in  Figure~\ref{fig:fanidakis}  is  almost identical to our Figure~\ref{fig:az_press}. This similarity extends also to the standard deviations,  reflecting the broad distribution of SMBH spins at each redshift. The value of \amean\, from the simulation is also in good quantitative agreement with our values at $z$$\grtsim$1.

This line of argument suggests that their lower-luminosity AGN in spiral galaxies will have SMBHs with low spin values \citep{2007MNRAS.377L..25K}. At first sight, this would appear  in contradiction with  the spin estimate for the nearby AGN MCG-6-30-15\footnote{MCG-6-30-15 contains a SMBH with mass $\sim$2$\times 10^{6}$~\msol, accreting in the analogue to the high or very high state of GBHs. In fact, it is probably accreting close to the Eddington ratio, with $\lambda$$\grtsim$0.8 \citep{2005MNRAS.359.1469M,2008A&A...492...93B}.}, \sp$=$0.98, from a broad iron line \citep{2006ApJ...652.1028B}. However, the host galaxy of this source is actually an S0 \citep{2000ApJS..128..139F}, so the high spin could have been acquired through merging.   In addition, the caveats discussed in Section~\ref{sec:lack} should be kept in mind.

Major mergers are also expected to happen at high redshift ($z\grtsim1$), but higher gas masses in the host galaxies imply higher accretion rates which will lead to a spin down (Equation~\ref{eq:deltam} will be satisfied).  
Hence, while the mechanism to spin SMBHs up (major merging) is occuring both at high and low redshifts, the increase in spin at low redshifts can be attributed to a gradual disappearance of the  `braking' mechanism (chaotic accretion of a significant mass).

If high-redshift quasars have a low mean spin due to chaotic accretion, it implies they have had time to spin down. The timescale for spin down is related to the timescale for alignment \citep[$\sim10^{7}$~yr for \sp$\sim$1, Equation~15 of][] {2006MNRAS.368.1196L} since after a few alignments and anti-alignments the spin will have decreased to $\sim$0.1. 
AGNs with a higher spin value are expected to have a longer alignment timescale \citep {2006MNRAS.368.1196L}, so jets from high-spin SMBHs will be more stable than those of SMBHs with low spin, implying powerful observed  FRIIs could live for a few $\times10^{7}$~yr \citep[indeed the ages estimated by][]{1995MNRAS.277..331S,1999MNRAS.309.1017W,2000AJ....119.1111B} and will be rapidly spinning.  Given that the most powerful radio-loud quasars typical jet advance speeds  are$\sim0.05c$ \citep{1995MNRAS.277..331S}, and assuming a maximum age of $5\times10^{7}$~yr, the largest jets are expected to be $\sim$8~Mpc at the most, in excellent agreement with observations \citep[e.g.][]{1999AJ....117..677B}.

Our bimodal distribution of spins suggests  that in the local Universe, the SMBHs with mass $\grtsim10^{8}$~\msol\, will be able to produce jets with only a modest rate of accretion ($\lambda$$\ll$1). It is therefore no surprise that approximately 30\% of local massive ellipticals have jets with  moderate radio luminosities \citep[e.g.][]{2007MNRAS.375..931M}. 

Our estimate of \amean\, is not in disagreement with the  high values of \sp$\sim$0.9 inferred for the SMBHs of low-redshift elliptical galaxies (see N07).   The authors inferred the spin from the correlation between Bondi accretion rate and jet power for nearby elliptical galaxies found by \citet{2006MNRAS.372...21A}, and using an ADAF model for the production of jets around rotating SMBHs. The sources in the sample of \citet{2006MNRAS.372...21A} were selected to show jet-induced cavities in the hot gas imaged using X-rays. Hence, the sample is biased towards elliptical galaxies with jets, which in our picture and that of N07 corresponds to high spin. 
Although our mean spin in the local Universe is typically lower than 0.9, approximately half of the SMBHs accreting as ADAFs are predicted to have near-maximal spins (Section~\ref{sec:fit_rlf}), hence there is no conflict between our results and those of N07. 

Finally, we speculate how our inferred cosmic spin history can be explained in the context of galaxy groups and clusters. We have argued that SMBH major mergers caused by galaxy major mergers will spin up the black holes.  For major mergers to occur efficiently, there must be a high density of massive galaxies, yet the translational velocities of the galaxies must not exceed the escape velocities of the galaxies themselves, which correspond to $\sim$200-400~\kms\, for massive galaxies hosting black holes of mass $\grtsim10^{8}$~\msol\, \citep{2000ApJ...539L...9F}.

The ideal environment for mergers therefore corresponds to groups or to  the outskirts of low-redshift clusters of galaxies. At $z$$\sim2$, the clusters of galaxies have not yet virialised fully, and the characteristic velocities of galaxies in these high-redshift protoclusters will not exceed this velocity significantly. Hence, mergers are expected to be efficient, and the mean spin of SMBHs will be initially high due to mergers.  However, at $z$$\sim2$ large reservoirs of gas will also be present, allowing the SMBHs to accrete large masses. 

The most massive SMBHs will reside at the centres of these protoclusters: with black hole masses $\sim10^{9}$~\msol, $\lambda\grtsim0.1$ and initially \sp$\sim0.6$ due to mergers, these central SMBHs will produce powerful FRII jets. The  injection of energy from these FRII jets  into the intracluster medium will prevent further cooling of the gas cutting down the supply of gas for future star-formation and SMBH growth in  galaxies in the protocluster \citep[e.g.][]{1995MNRAS.276..663B,2004MNRAS.355L...9R}. 

The SMBHs in the central dominant galaxies 
might well retain their high spin from the high-redshift epoch, $z\sim2$, provided that the total mass accreted after their last major merger is small compared to the black hole mass,  \mbh$\sim10^{9}$~\msol. Indeed the little evolution observed amongst the high mass end of the stellar mass function, with stellar mass $>$$3\times10^{11}$~\msol, suggests little merging activity below $z$$\sim$1 for the most massive galaxies \citep{2010MNRAS.402.2264B}. 

The bulk of the accretion is, however, in the form of radio-quiet quasars (\mbh$\sim10^{8}$~\msol, $\lambda\grtsim0.1$ and \sp$\sim$0), whose spin at high redshift is low due to prolonged chaotic accretion. Presumably these objects have not yet undergone a major merger or have been able to accrete their own mass, $\Delta m_{\bullet}/ m_{\bullet} \grtsim 1$,  to spin themselves down to \sp$\sim$0.  The number of such objects is large compared to the number of $\sim10^{9}$~\msol\, SMBHs in future central dominant galaxies, which brings the mean spin down. 

At $z<1$ mergers of massive galaxies will still occur, particularly in environments with velocities $\sim$300~\kms, corresponding to group environments and the outskirts of clusters. These mergers will spin up the SMBHs. The  gas in the host galaxies will have been stripped by ram pressure, while  the intracluster gas is hot and will not resupply the galaxies. With little gas in the host galaxy, the braking mechanism, chaotic accretion of a significant mass,  disappears.

Future episodic accretion onto SMBHs with  \mbh$=10^{8}-10^{9}$~\msol\, will be modest and will not affect the spin. It will lead to intermitent FRI-ADAF activity, rather than to radio-quiet quasars. Occasionally a major accretion episode might occur, creating sources such as Cygnus~A, but such episodes will be very rare.

 \section{Acknowledgements}

 We warmly thank Philip Best for providing, prior to publication, the classification of the radio luminosity function into HEGs and LEGs, which we used in Section~\ref{sec:fit_rlf}.  We are also very grateful to Alexander Tchekhovskoy for deriving jet efficiencies for his model (Section~\ref{sec:jet_eff}), as well as for many clear explanations regarding his work. We also thank warmly Nikolaos Fanidakis for providing us with the spin distribution of SMBHs from his simulation, used in Figure~\ref{fig:fanidakis}, as well as to Michele Cirasuolo for providing the data for the quasars used in Figure~\ref{fig:cira}.  We are also grateful to Andrew Benson for clarifications regarding his model. We thank Chris Done, Ryan Houghton, Thomas Mauch, Lance Miller, Danail Obreschkow, Brian Punsly, Aday Robaina and Stanislav Shabala for useful discussions. We gladly thank the anonymous referee for criticism that significantly improved the paper. A.M.-S. gratefully acknowledges a
 Post-Doctoral Fellowship from the United Kingdom Science and
Technology Facilities Council, reference ST/G004420/1. This effort was partly supported by the
European Community Framework Programme 6, Square Kilometre Array
Design Studies.

\bibliographystyle{bibstyle}

\label{lastpage}

\end{document}